\documentclass[a4paper]{ar-1col}
\usepackage[utf8]{inputenc}
\usepackage{natbib}
\usepackage{graphicx}
\usepackage{lscape}
\usepackage{amsmath}
\usepackage{amssymb}
\usepackage{color}
\definecolor{darkblue}{cmyk}{1.0,0.4,0,0.5}
\definecolor{darkred}{rgb}{0.6,0.0,0.0}
\definecolor{titlecolor}{cmyk}{0.4,0.4,0.4,0}
\definecolor{headcolor}{cmyk}{0,1.0,1.0,0.30}
\definecolor{shadecolor}{cmyk}{0.03,0.03,0.12,0.0}
\definecolor{fignumcolor}{cmyk}{1.0,0.4,0,0.5}
\definecolor{marginrulecolor@val}{cmyk}{0,0,0,0.30}
\definecolor{textboxcolor@val}{cmyk}{0.12,0.04,0.08,0.0}

\usepackage{hyperref}
\hypersetup{
    colorlinks=true,
}

\newcommand{\apj}{ApJ}
\newcommand{\apjl}{ApJL}
\newcommand{\apjs}{ApJS}
\newcommand{\aap}{A$\&$A}

\newcommand{\apss}{APSS}

\newcommand{\araa}{ARA$\&$A}

\newcommand{\mnras}{MNRAS}
\newcommand{\pasj}{PASJ}
\newcommand{\prd}{Phys. Rev. D}
\newcommand{\prl}{Phys. Rev. Lett.}
\newcommand{\aj}{AJ}
\newcommand{\nat}{Nature}

\newcommand{\physrep}{Physics Reports}
\newcommand{\pasp}{PASP}
\newcommand{\ssr}{{Space~Sci.~Rev.}}%

\newcommand{\pasa}{PASA}

\begin{document}

% Page header
\markboth{Klessen \& Glover}{The First Stars}

% Title
\title{The first stars: formation, properties, and impact}
\author{Ralf S.\ Klessen$^{1,2}$, Simon C.\ O.\ Glover$^{1}$
\affil{$^{1}$Universit\"{a}t Heidelberg, Zentrum f\"{u}r Astronomie, Institut f\"{u}r Theoretische Astrophysik, Albert-Ueberle-Str.\ 2,  69120 Heidelberg, Germany. \\ emails: klessen@uni-heidelberg.de, glover@uni-heidelberg.de}
\affil{$^2$Universit\"{a}t Heidelberg, Interdisziplin\"{a}res Zentrum f\"{u}r Wissenschaftliches Rechnen, Im Neuenheimer Feld 225, D-69120 Heidelberg, Germany.}
}

\date{\today}

%Abstract
\begin{abstract}
The first generation of stars, often called Population III (or Pop III), form from metal-free primordial gas at redshifts $z \sim 30$ and below. They dominate the cosmic star formation history until $z \sim 15 -20$, at which point the formation of metal-enriched Pop II stars takes over. We review current theoretical models for the formation, properties and impact of Pop III stars, and discuss existing and future observational constraints. Key takeaways from this review include the following:
\begin{itemize}
\setlength{\leftskip}{-5.5mm}
\item Primordial gas is highly susceptible to fragmentation and Pop III \\
stars form as members of small clusters with a logarithmically flat \\ 
mass function. 
\item Feedback from massive Pop III stars plays a central role in regulating \\ 
subsequent star formation, but major uncertainties remain regarding 
\\ its immediate impact.
\item In extreme conditions, supermassive Pop III stars can form, reaching \\ 
masses of several $10^5\,$M$_\odot$. Their remnants may be the seeds of the \\
supermassive black holes observed in high-redshift quasars.
\item Direct observations of Pop III stars in the early Universe remain \\
extremely challenging. Indirect constraints from the global 21~cm \\
signal or gravitational waves are more promising.
\item Stellar archeological surveys allow us to constrain both the low-mass \\ 
and the high-mass ends of the Pop III mass distribution. Observations \\
suggest that most massive Pop III stars end their lives as core-collapse \\
supernovae rather than as pair-instability supernovae.
\end{itemize}
\end{abstract}

%Keywords, etc.
\maketitle

%Table of Contents
{
  \hypersetup{linkcolor=black}
  \tableofcontents
}

\hypersetup{
   linkcolor={darkred},
   citecolor={darkblue},
   urlcolor={darkblue}
}

\section{Introduction}
From studying the cosmic microwave background (CMB), we know our Universe started out very simple. It was by and large homogeneous and isotropic, with small fluctuations that can be described by linear perturbation theory. In stark contrast, the Universe today is highly structured on a vast range of length and mass scales. In the evolution towards increasing complexity, the formation of the first stars marks a primary transition phase of cosmic evolution. Their light ends the so-called `dark ages', and they play a key role in cosmic metal enrichment and reionization, thereby shaping the Universe at large and the present-day galaxy population. The study of stellar birth in the early Universe is therefore important for many areas of modern astronomy and astrophysics but is also a relatively young field of research. Only with the advent of advanced numerical methods and powerful supercomputers did a comprehensive modeling of early star formation become feasible. As a consequence, there is still considerable debate about the physical processes that govern stellar birth at high redshifts and the overall properties of the first stars. This review aims at providing an overview of the current state of the field.

The first generation of stars, the so-called Population III (or Pop III) build up from truly metal-free primordial gas. Second generation stars, sometimes termed early Pop II stars, form from material that has been enriched from the debris of the first stars. Unlike the very first stars, 
which have not yet been directly detected, members of the second generation have been found and characterized in surveys looking for extremely metal-poor stars in our Milky Way and neighboring satellite galaxies. In concert with observational data at high redshift, this allows us to  constrain the properties of genuine Pop III stars. Initially, primordial star formation was proposed to be very simple, governed by well defined initial conditions provided by Gaussian fluctuations of the cosmic density field. These were thought to result in the build-up of solitary high-mass stars.  %\citep[e.g.][]{Omukai2001, Abel2002, Bromm02, oshea2007a}. 
This simple picture has undergone dramatic revisions.  We now understand that fragmentation is a wide-spread phenomenon during first star formation, %\citep[e.g.][]{clark2011a, greif2012a}, 
and that Pop III stars form as members of multiple stellar systems with separations as small as the distance between the Earth and the Sun.
%\citep[see, e.g.][]{turk2009a, clark2011b, greif2011b, smith2011a, stacy2013a, latif2013c, latif2014b}. 
Studies that include radiative feedback, %\citep{hirano2014a, hirano2015a, hosokawa2016a},
magnetic fields, 
%\citep[e.g.][]{machida2006a, machida2008b, schleicher2009a, schleicher2010c,  sur2010a, sur2012a, turk2012a, schober2012c, schober2012b, bovino2013a,latif2016a}, 
dark matter annihilation, 
%\citep[e.g.][]{smith2012a,  stacy2014b}, 
as well as the primordial streaming velocities %\citep[e.g.][]{tseliakhovich2010a, greif2011a, maio2011b, stacy2011a} 
add to this complexity. All of these processes are relevant and need to be included in any realistic model. There is agreement now that primordial star formation is just as dynamic and complicated as stellar birth at the present day.
%% MRNT
\begin{marginnote}[]
\entry{Pop III stars}{Population III stars form from truly metal-free primordial gas in the high-redshift Universe. }
\end{marginnote}
%% MRNT
\begin{marginnote}[]
\entry{Pop II stars}{Population II stars build up from metal-enriched material. Early Pop II stars constitute the second stellar generation in cosmic history. }
\end{marginnote}
% MRNT
\begin{marginnote}[]
\entry{Pop I stars}{Population I stars, such as the Sun, are relatively young and metal rich. They are found in the thin disk of the Milky Way and other spiral galaxies. }
\end{marginnote}

We structure our review as follows: First we provide in Section \ref{sec:basics} an overview of all relevant physical processes that contribute to first star formation. This includes an introduction of the cosmological model that constitutes the basis of our analysis, together with the equations that govern the evolution of the cosmic fluid. We end this part with an account of the instabilities that lead to gravitational collapse and subsequently to stellar birth, and we stress the importance of the thermodynamic properties of the star-forming gas. In Section \ref{sec:PopIII-SF} we discuss the critical mass for collapse in the primordial Universe and describe the evolution of the cosmic star-formation rate density. We then discuss in detail the standard Pop III formation pathway in isolated pristine halos. We also account for more complex scenarios in which further physical processes add to the complexity of the problem. Finally, we turn our attention to the most extreme physical conditions that result in the formation of supermassive stars and lead to the supermassive black holes we observe in quasars at high redshift. The impact of stellar feedback is the focus of Section \ref{sec:feedback}. We first look at the consequences of the radiative and the mechanical energy and momentum input from young massive stars on their immediate birth environment and then consider their influence on neighboring halos and on the larger-scale intergalactic gas. We also discuss chemical feedback of Pop III stars, which drives the transition to Pop II star formation and governs the early metal enrichment of the Universe. A critical assessment of possible observational probes of primordial star formation is presented in Section \ref{sec:obs-probes}, covering measurements at high redshift as well as multi-messenger data from the local Universe. Finally, we conclude and summarize in Section \ref{sec:summary}.  

Our focus here is to provide a comprehensive overview of the developments and successes in the field during the past one or two decades. For further reading, including a more historic account, we refer to the reviews by \citet{Barkana2001}, \citet{Bromm04},  \citet{Glover2005},  \citet{Yoshida2012}, and \citet{Bromm2013},  as well as to the book by \citet{Loeb2010}.  A good overview with a specific focus on astrochemistry  is provided by \citet{Glover2013}, and an account of important numerical aspects of high-redshift star formation is given by \citet{Greif15}. For further reading on the transition to the second generations of stars and build up of the first galaxies we recommend \citet{bromm2011a} or the textbook by \citet{Loeb2013}. In order to calibrate our understanding of primordial star formation with what we know about stellar birth at present days, we also refer to the reviews by \citet{Maclow2004}, \citet{Mckee2007}, \citet{Zinnecker2007},  \citet{Krumholz2015}, or \citet{Klessen2016} on different aspects of this subject.

\section{Important physical concepts}
\label{sec:basics}
In this Section, we introduce the basic physical concepts needed to understand the formation of the first stars. We begin with the cosmological model that forms the foundation of all further considerations, introduce the main equations governing the evolution of the cosmic fluid, and study the criteria for the onset of gravitational instability in the early Universe and consequently for the formation of the first stars.

\subsection{Cosmological model} 
\label{sec:cosmological-model}
Throughout this review, our approach is based on the standard $\Lambda$CDM model \citep[e.g.][]{Peebles93, Bullock17} in which the Universe consists of matter in form of cold dark matter and baryons, radiation,  as well as dark energy. 
In this model, the Hubble parameter $H(a) = \dot{a}/a$, which describes the expansion of the Universe, evolves as
\begin{equation}
\dfrac{H^2(a)}{H^2_0} = \dfrac{\Omega_{r}}{a^4} + \dfrac{\Omega_{\rm m}}{a^3} + \dfrac{\Omega_{\Lambda}}{a^{3(1+w)}} ~~~~~~~~\mathrm{with}~~~~~~~ 1 = \Omega_{\rm r} + \Omega_{\rm r} + \Omega_{\Lambda}\;, 
\label{eq:Hubble}
\end{equation}
where $a(z) = (1+z)^{-1}$ is the cosmic scale factor and $z$ is the redshift.
At the present day, i.e.\ for $a=1$, the densities of radiation, matter and dark energy with respect to the critical density are $\Omega_{\rm r} \approx 10^{-5}$, $\Omega_{\rm m} = 0.31$ and $\Omega_{\Lambda} = 0.69$ \citep{Planck18VI}. The parameter $w$ relates the pressure and energy of the dark energy component ($p = w \rho c^{2}$); here, we assume that it is constant in time and equal to -1, as appropriate for the classical cosmological constant.
The current value of the Hubble parameter $H_0$ is $\sim 70 \,\mathrm{km}\,\mathrm{s}^{-1}  \mathrm{Mpc}^{-1}$; for historical reasons, this is usually expressed as $H_0 = 100 \, h\, \mathrm{km}\,\mathrm{s}^{-1}  \mathrm{Mpc}^{-1}$ with $h\approx 0.7$. Note that there remains some debate regarding the precise value of $H_{0}$ \cite[for a comprehensive overview, see][]{DiValentino21}.

\subsection{Evolution of the cosmic fluid} 
\label{sec:evolution-of-cosmic-fluid}
In order to understand primordial star formation we need to identify and characterize the first regions in the Universe that decouple from the cosmic expansion and contract under their own gravitational attraction. Observations of the CMB show that the Universe started out extremely simple. The initial density distribution was isotropic and almost perfectly homogeneous with spatial fluctuations, $\delta(\mathbf{x}) = \rho(\mathbf{x})/\langle \rho \rangle -1$, of order $10^{-5}$ on large scales with respect to the average background density $\langle \rho \rangle$. In the dark matter, these fluctuations are present on all scales and grow due to their own self-gravity. In the linear regime, the evolution of $\delta$ is governed by the equation
\begin{equation}
\dfrac{\partial^2 \delta}{\partial t^2} + 2H \dfrac{\partial \delta}{\partial t} =  4 \pi G \langle \rho \rangle  \delta\;.
\label{eq:growth}
\end{equation}
This corresponds to the famous \citet{Jeans02} equation describing the evolution of self-gravitating isothermal gas spheres, written in the limit of zero pressure, and with an additional drag term due to the expansion of the Universe. 
In the $\Lambda$CDM model, the initial density fluctuations are well characterized as a Gaussian random field, and so their statistical properties are completely determined by the power spectrum of the field, which has been inferred with high precision from the observed temperature fluctuations in the CMB \citep{Planck20V}. Consequently, the physical properties of the Universe at a redshift $z \sim 1000$, and hence the initial conditions for cosmic structure formation, are extremely well constrained.

\subsection{Gravitational instability}
Whereas (cold) dark matter can be considered a pressure-less zero-temperature fluid, this is not the case for the baryonic component of the Universe. On large scales, the gravity of the dark matter dominates, and the evolution of the baryons is very similar to that of the dark matter. On small scales, the effects of gas pressure become increasingly important. In the classical \citet{Jeans02} stability analysis, we can identify a critical mass scale $M_{\rm J}$ separating the gravity-dominated and pressure-dominated regimes. Perturbations with $M > M_{\rm J}$ undergo gravitational collapse, whereas those with $M < M_{\rm J}$ do not. In the simple case of an isothermal gas sphere, we have
\begin{equation}
M_{\rm J} \;=\; \dfrac{\pi^{5/2}}{6}\left( \dfrac{1}{G}\right)^{3/2} \rho^{-1/2} c_{\rm s}^{3} \;\approx\; 50 \,{\rm M}_{\odot} \;\mu^{-2} \left(\frac{n}{1 \, {\rm cm}^{-3}}\right)^{-1/2} \left(\frac{T}{1 \, {\rm K}}\right)^{3/2}\;, 
\label{eq:Jeans-mass}
\end{equation}
% python: 3.14159**(2.5) / 6.0 * (1.38e-16 / 6.67e-8)**(1.5) / (1.0 * 1.67e-24)**(2) / 2e33
where $n$ is the number density of H nuclei and $c_{\rm s}$ is the sound speed, which are related to $T$ and $\rho$ by $\rho = \mu m_{\rm H} n $ and $T =  \mu m_{\rm H} c^2_{\rm s} / k_{\rm B}$.
%% MRNT
\begin{marginnote}[]
\entry{Jeans instability}{~~~ Instability that leads to the collapse of self-gravitating isothermal spheres if gravity dominates over gas pressure.}
\end{marginnote}
%% MRNT
Information about the chemical makeup of the gas is encoded in the weight factor ${\mu} = (1 + 4 \chi)$ with $\chi$ being the ratio of He to H atoms by number. For primordial gas, $\chi = 0.079$ and so ${\mu} = 1.32$.
In this convention, atomic hydrogen at one particle per cubic centimeter corresponds to $n = 1\,$cm$^{-3}$, whereas fully molecular hydrogen with one particle per cm$^{3}$ yields $n = 2\,$cm$^{-3}$. Note that the chemical weight factor $\mu$ could simply be replaced by the mean molecular weight if we take the total particle number density in Equation~\ref{eq:Jeans-mass}, rather than just the number density of hydrogen atoms.
Finally, we note that a similar expression to Equation~\ref{eq:Jeans-mass} can be derived by considering the growth of plane-wave baryonic perturbations in the linear regime \citep[see e.g.][]{1999coph}. More generally, an expression differing only by a small numerical factor can be derived by comparing the gravitational and sound-crossing timescales or the gravitational and thermal energy of the perturbation. 
\begin{marginnote}[]
\entry{Primordial gas}{~~~~~ Big Bang nucleosynthesis produces mostly H and $^4$He with mass fractions of 0.76 and 0.24, respectively, and trace amounts of $^2$H, $^3$He, and $^7$Li at the level of $10^{-5}$ to $10^{-10}$.}
\end{marginnote}

The critical mass for collapse can also be derived by considering the two competing timescales in the problem: 
\begin{equation}
\tau_{\rm ff} = \left( \dfrac{3 \pi }{ 32 G \rho}\right)^{1/2} \mbox{~~~(free-fall time)~,~~~~~} \tau_{\rm sound} = \dfrac{R}{c_{\rm s}} \mbox{~~~(sound-crossing time)~.}
\label{eq:free-fall-and-sound-crossing-time}
\end{equation}
The dynamical or free-fall timescale, $\tau_{\rm ff}$ expresses the characteristic duration of gravitational collapse in the absence of pressure, whereas the sound crossing timescale, $\tau_{\rm sound} = R/c_{\rm s}$ denotes the time it takes to communicate pressure gradients across a fluctuation of size $R$. If $\tau_{\rm ff} < \tau_{\rm sound}$ the system is unstable against contraction, and if $\tau_{\rm sound} < \tau_{\rm ff}$ then pressure gradients are able to provide stability against gravitational attraction. Note that in equilibrium, these two timescales are the same. 
Note also that there is a third relevant timescale to consider, which is the age of the Universe. It is reasonably well approximated as the inverse of the Hubble parameter, $\tau_{\rm H} = 1/H(z)$, tracing the cosmic expansion history. Even if $\tau_{\rm ff} < \tau_{\rm sound}$, if both numbers are larger than $\tau_{\rm H}$ there is not enough time for collapse to progress to sufficiently large densities for star formation to set in. 

For a collapsing sphere, we can estimate the associated accretion rate from the Jeans mass and the free-fall time $\tau_{\rm ff} = \left(3 \pi / 32 G \rho\right)^{1/2}$ via:
\begin{equation}
\dot{M} = f \dfrac{M_{\rm J}}{\tau_{\rm ff}} \approx f \dfrac{c_{\rm s}^3}{G}\;,
\label{eq:accretion-rate}
\end{equation}
where $f$ can take on values between $\sim 1$ and several tens, depending on the initial density profile and the ratio between $M$ and $M_{\rm J}$. Note, in this simple approximation, the accretion rate only depends on the gas temperature.  For further discussion, see \citet{Whitworth1985} and references therein.
Additional physical processes can also be accounted for in our definition of the Jeans mass, most easily by defining and using an effective sound speed in place of $c_{\rm s}$. For example, 
when the gas is turbulent on scales much smaller than the dynamical scales of interest \citep{Chandrasekhar1951A, Chandrasekhar1951B, VonWeiz1951}, we can simply add the velocity dispersion $\sigma$ to the sound speed. 
A similar role has been ascribed to magnetic fields $B$, in which case we include a contribution from the Alfv\'{e}n velocity, $v_{\rm A} = B/(8\pi\rho)^{1/2}$ \cite[e.g.][]{Federrath2012}, leading to 
\begin{equation}
c_{\rm s, eff} = \left(c_{\rm s}^2 + \sigma^2 + \dfrac{1}{2} v^2_{\rm A} \right)^{1/2}\;.
\label{eq:effective-sound-speed}
\end{equation} 
We  revisit these aspects later, in Section \ref{sec:alternative-PopIII-pathways}.

\subsection{Impact of thermodynamics} 
\label{sec:TD}
The above considerations demonstrate the importance of the thermodynamic response of the evolving system (see also Section \ref{sec:initial-collapse}). 
As an illustration, look at the simplified case of the gas following an effective polytropic equation of state, 
\begin{equation}
p \propto \rho^{\gamma_{\rm eff}}   ~~~\mbox{with index}~~~ \gamma_{\rm eff} = 1 + d \ln T/ d \ln \rho\;,
\label{eq:polytropic-EOS}
\end{equation}
where the $\gamma_{\rm eff}$ is the result of the competition between various heating and cooling mechanisms \citep[e.g.][]{Omukai2005, Klessen2016}. From Equation~\ref{eq:Jeans-mass} we see that 
\begin{equation}
M_{\rm J} \propto \rho^{\frac{3}{2}(\gamma_{\rm eff}-\frac{4}{3})}\;.
\label{eq:Jeans}
\end{equation}
If $\gamma_{\rm eff} > {4}/{3}$ the Jeans mass increases during the contraction and eventually becomes comparable with the mass of the system (including both dark matter and baryons). In this case pressure forces will stop further collapse. For adiabatic gas with $\gamma_{\rm eff} = 5/3$, as appropriate for monoatomic gas without internal degrees of freedom, this will happen long before stellar densities are reached. In order for stars to form, the gas must be able to radiate energy away during its collapse, so that 
$\gamma_{\rm eff}$ remains below the critical value. 
\begin{marginnote}[]
\entry{Equation of state}{~~~ It relates thermodynamic state variables such as pressure, density, temperature or internal energy, and is needed to turn the equations of hydrodynamics into a closed and thus solvable system. }
\end{marginnote}

The importance of cooling can be assessed by comparing the cooling time,
\begin{equation}
\tau_{\rm cool} = \dfrac{1}{\gamma - 1} \dfrac{n_{\rm tot} k T}{\Lambda(T,n)}\;,
\label{eq:cooling-time}
\end{equation}
with the free-fall timescale, $\tau_{\rm ff}$. Here  $n_{\rm tot}$ is the total number density of particles, $k$ the Boltzmann constant, $T$ the temperature, and $\Lambda$  the cooling rate. If $\tau_{\rm cool}  < \tau_{\rm ff}$ then the gas can cool rapidly, and collapse proceeds roughly on a free-fall timescale. However, if $\tau_{\rm ff} < \tau_{\rm cool}$ then the gas quickly becomes pressure supported, and the contraction slows down and proceeds quasi-statically for a duration of order of $\tau_{\rm cool}$. 
As before, if $\tau_{\rm H}$ is the shortest timescale, we can consider the system as being stable. 
We note that the situation is typically more complicated than this simple analysis suggests, as the thermodynamic response of the gas may vary with time as density, temperature, and chemical composition evolve, and it may depend on location, specifically on the proximity to sources of stellar feedback, as discussed in Section \ref{sec:feedback}.

\subsection{Instability of rotationally-supported systems}
\label{sec:Toomre}
In rotationally-supported systems, such as protostellar accretion disks or spiral galaxies,  
the criterion for gravitational instability takes a slightly different form, owing to the stabilizing effect of shear. 
For infinitely thin disks, this was investigated by \citet{Toomre1964}, who derived the following criterion for instability: 
\begin{equation}
Q = \dfrac{\kappa c_{\rm s}}{\pi G \Sigma} \lesssim 1\;. 
\label{eq:Toomre}
\end{equation}
Here $\Sigma$ and $\kappa$ are the surface density and epicyclic frequency, respectively. For systems in Keplerian rotation, we have $\kappa = \Omega$, where $\Omega$ is the rotational frequency \citep{Kratter2016}. This approach can be extended to thick disks with multiple components \citep{Rafikov2001, Elmegreen2002, Romeo13} by introducing appropriate correction factors to Equation~\ref{eq:Toomre}. 
%% MRNT
\begin{marginnote}[]
\entry{Toomre instability}{~~~~~ Instability occuring in differentially rotating self- gravitating thin disks if  gravity  dominates over pressure and rotational shear.}
\end{marginnote}

There are two main pathways towards disk fragmentation. First, in the absence of accretion onto the system, \citet{Gammie01} argue that even an initially stable disk will become unstable if the cooling timescale, $\tau_{\rm cool}$, is shorter than the orbital time scale $\tau_{\rm orbit} = 1/ \Omega$.
Second, in the presence of accretion from the surrounding gas envelope, if the mass load onto the disk exceeds its capability to transport material inwards, 
$\Sigma$ increases beyond the critical value and the disk becomes unstable. As we see later, this latter scenario is commonly encountered when studying Pop III accretion disks.

\section{Pop III star formation}
\label{sec:PopIII-SF}
We begin our discussion of Pop III star formation with a critical review of their birth  environment, then turn our attention to the most likely formation pathway, and speculate about alternative scenarios. In the most extreme cases this can lead to the formation of supermassive stars as possible progenitors of supermassive black holes. 

\subsection{Critical mass for collapse and cosmic star-formation rate density}
\label{sec:minimum-mass}
The formation of the first stars in the Universe occurs in regions where the cosmic fluid fulfils two conditions. First, it needs to decouple from the global expansion and begin to contract due to the self-gravity of dark matter. We call a region where this happens a dark matter halo. Second, the gas within the dark matter halo needs to be able to cool and go into run-away collapse to dramatically increase the baryon to dark-matter ratio and eventually reach stellar densities. As we discuss in more detail later, halos with virial temperatures above $T_{\rm vir} \sim 8000\,$K can cool initially via Lyman-$\alpha$ emission from atomic hydrogen, whereas lower mass halos with lower $T_{\rm vir}$ depend on cooling from H$_{2}$. For historical reasons, the latter type of halo is often referred to as a `minihalo'. In the current $\Lambda$CDM paradigm, gravitationally bound objects form in a hierarchical fashion with smaller objects forming first, implying that minihalos are the first sites in which Pop III stars can potentially form.

\subsubsection{Simple models for the critical mass} 
\label{sec:simple-models}
Calculations of the growth of density perturbations in an expanding Universe \citep[e.g.][]{Barkana2001} show that substantial baryonic overdensities can only develop in halos with masses above a critical mass given by
\begin{equation}
M_{\rm crit} \approx 5 \times10^3 {\rm M}_{\odot} \left( \dfrac{\Omega_{\rm m}h^2}{0.143} \right)^{-1/2} \left( \dfrac{\Omega_{\rm b}h^2}{0.022} \right)^{-3/5} \left( \dfrac{1+z}{10} \right)^{3/2}\;.
\label{eq:Jeans-halo}
\end{equation}
This mass scale is very similar to the Jeans mass (Equation \ref{eq:Jeans-mass}) of fluctuations with a density close to the mean density of the Universe and a temperature that declines adiabatically as the Universe expands, and holds for $z \lesssim 100$ when gas is no longer thermally coupled to the CMB by Compton scattering.  However, the development of a baryonic overdensity is a necessary but not sufficient condition for the formation of Pop III stars. In addition, as we have argued above, the gas must also be able to cool. This requirement yields a much higher critical mass. For instance, using a very simple model for the behavior of gas in a high-redshift minihalo \citet{Glover2013} derives a value
\begin{equation}
M_{\rm crit} \approx 1.4 \times 10^6 {\rm M}_{\odot}  \left( \dfrac{\Omega_{\rm m}h^2}{0.143} \right)^{-1/2}  \left( \dfrac{\mu}{1.32} \right)^{-3/2} \left( \dfrac{1+z}{10} \right)^{-3/2}\;,
\label{eq:cool-halo}
\end{equation}
where we encounter again the chemical weight factor $\mu$ introduced in Equation~\ref{eq:Jeans-mass}. 

\subsubsection{Models with more physics included} 
\label{sec:more-complex-models}
Clearly, Equations~\ref{eq:Jeans-halo} and \ref{eq:cool-halo} are substantial simplifications. For example, the latter assumes that the chemical evolution of the gas takes place at constant density and temperature, and neither equation accounts for the fact that conditions in the Universe change during the time that it takes for a perturbation to grow into the non-linear regime. This can lead to significant deviations from the prescriptions above (see e.g.\ \citealt{Gnedin98} or  \citealt{Naoz07} for more complex models). To capture the non-linear evolution and to account for the complex interplay between the different physical processes involved requires us to resort to numerical simulations. 
Processes that must be accounted for include the build-up of a cosmic Lyman-Werner (LW) or X-ray radiation background as star formation sets in \citep[e.g.][see also Sections \ref{sec:LW} and \ref{sec:x-rays}]{Gnedin00} and the existence of relative streaming velocities between dark matter and baryons \citep[][see also Section \ref{sec:more-fragmentation}]{Tseliakhovich10, Tseliakhovich11}. 
%% MRNT
\begin{marginnote}[]
\entry{LW and UV photons}{Lyman-Werner photons in the energy range $11.2\,{\rm eV} \le h\nu < 13.6\,{\rm eV}$ can photodissociate H$_2$ molecules. UV photons with $h\nu \ge 13.6\,{\rm eV}$ can photoionize H.  }
\end{marginnote}
%% MRNT
\begin{marginnote}[]
\entry{Streaming velocity}{Second-order cosmic perturbation theory predict a relative motion between dark mass and baryons, which decreases linearly as the Universe expands.  }
\end{marginnote}

\begin{figure}[htbp]
\begin{center}
\includegraphics[width=9cm]{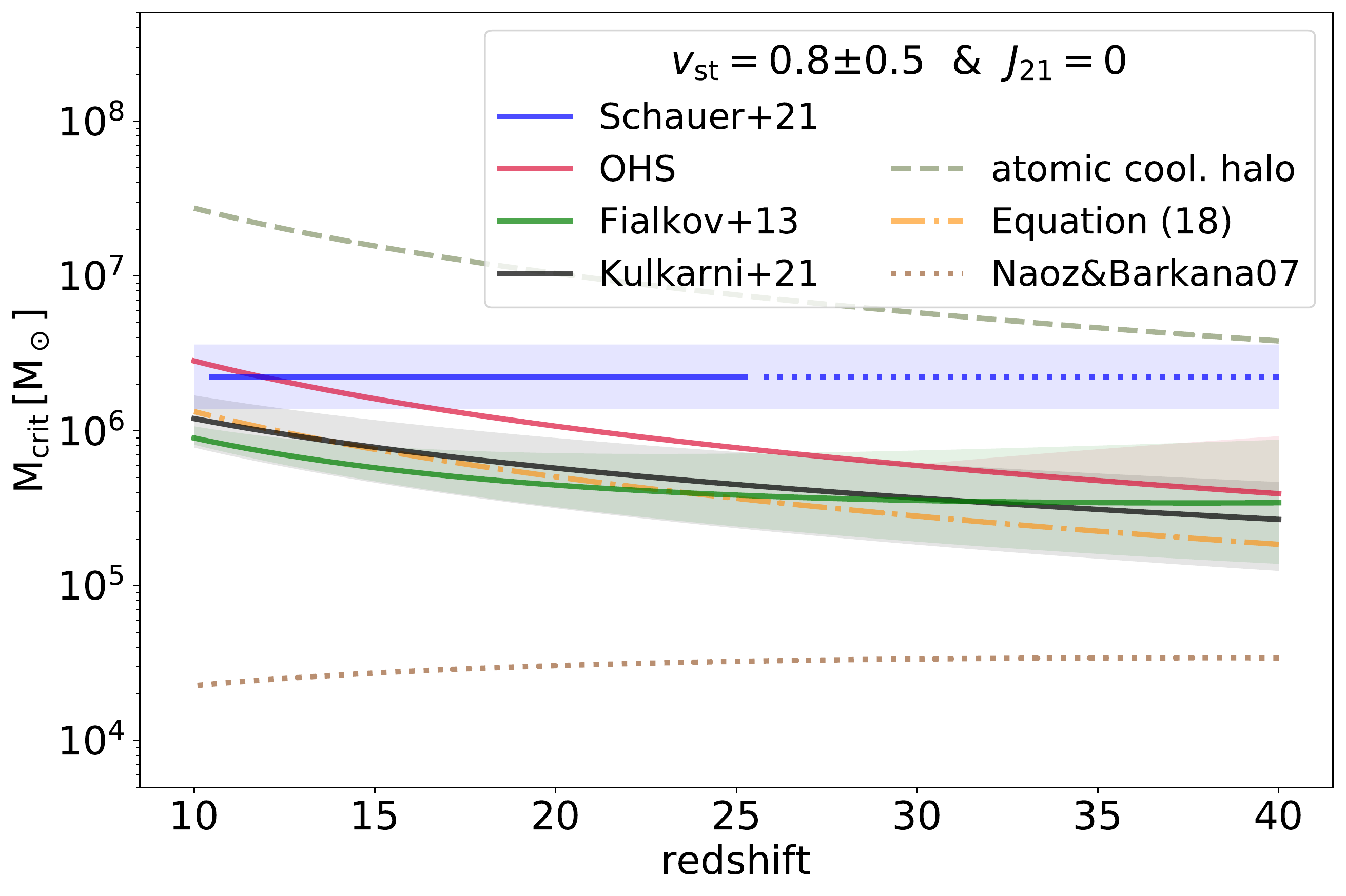}
\caption{Critical mass for fluctuations in the cosmic fluid for collapse to set in. The solid lines show results from models that take the streaming velocity into account, adopted from \citet{Schauer21}, \citet{Kulkarni21}, and also models OHS and F13 from \citet{Hartwig22} and \citet{Chen22}, which are derived from \citet{Oshea08}, \citet{Hummel12}, and \citet{Stacy2011} and  from \citet{Fialkov12, Fialkov13}, respectively. The fit from \citet{Schauer21} is based on simulation data at $z < 25$ and we indicate our extrapolation of it to higher redshifts with a dotted line. The lines are computed for the most probable value of the streaming velocity, $v_{\rm st} = 0.8$ (in units of the root-mean-squared value of the streaming velocity; see \citealt{Schauer19}) and the shading indicates the effect of varying this value by $\pm 0.5$. For further comparison, we show an estimate of $M_{\rm crit}$ in the absence of streaming based on our Equation~\ref{eq:cool-halo} (dash-dotted line). The lower envelope (dotted line) shows the minimum halo mass required for a baryonic overdensity to develop, as computed by 
\citeauthor{Naoz07}~(2007; see their Equations 17 and 18). The upper
envelope (dashed line) is given by the critical mass for atomic cooling halos (see e.g.\ Equation~1 of \citealp{Hummel12} or Equation~15 of \citealt{Hartwig22}). 
}
\label{fig:Mcrit}
\end{center}
\end{figure}

The latter results from second-order cosmological perturbation theory and therefore is often neglected in a purely linear analysis. Prior to recombination, baryons are tightly coupled to photons resulting in a standing acoustic wave pattern \citep{Sunyaev1970} and consequently in oscillations between baryons and dark matter  with relative velocities of about $30\,$km$\,$s$^{-1}$ and coherence lengths of  of a few Mpc in comoving units at $z \approx 1000$ \citep{Silk1968}. After recombination, baryons  are no longer tied to photons, their sound speed drops to $\sim 6\,$km$\,$s$^{-1}$, and the velocity with respect to the dark matter component becomes supersonic with Mach numbers of ${\cal M} \approx 5$ \citep{Tseliakhovich10}. As the Universe expands, the relative streaming velocity decays linearly and reaches  $\sim 1\,$km$\,$s$^{-1}$ at $z \approx 30$, which is comparable to the virial velocity of the first halos to cool and collapse \citep{Fialkov12}. Simulations that include streaming velocities  suggest that their presence reduces the baryon overdensity in low-mass halos, delays the onset of cooling, and leads to a larger critical mass for collapse to set in \citep{Greif2011, Stacy2011, Maio2011, Naoz2012, Naoz2013, Oleary2012, Latif2014Stream, Schauer17a,Nakazato2022}. They may also have substantial impact on the resulting $21\,$cm emission \citep{Fialkov12, McQuinn2012,Visbal2012}. Increasing the LW background intensity also increases $M_{\rm crit}$, as discussed in more detail in Section~\ref{sec:LW}.

Altogether, there remains significant uncertainty about the most appropriate value of $M_{\rm crit}$ with different simulations reaching different conclusions, depending on the numerical resolution adopted and the number of physical processes included \citep[see, e.g.][in addition to the above]{Yoshida2003, Latif19, Kulkarni21}. We illustrate some of these variations in Figure \ref{fig:Mcrit}, where we plot predictions for $M_{\rm crit}$ from \citet{Schauer21}, \citet{Kulkarni21} and from models OHS and F13 in \citet{Hartwig22} and \citet{Chen22}, which themselves are derived from the prescriptions of  \citet{Oshea08}, \citet{Hummel12}, and \citet{Stacy2011} and from \citet{Fialkov12, Fialkov13}, respectively. For further details, see Appendix A.2 of  \citet{Hartwig22}.
The curves shown are computed in the absence of LW background radiation ($J_{21} = 0$), and for a streaming velocity $v_{\rm st} = 0.8$ (in units of the root-mean-squared value), which is the most likely value to be encountered in the Universe \citep{Schauer19, Schauer21}. For a discussions of the large-scale impact of LW feedback and the associated uncertainties, we again refer  to Section \ref{sec:LW}. 
For completeness, we also plot the estimate for $M_{\rm crit}$ with $v_{\rm st} = 0$ given by Equation~\ref{eq:cool-halo}, the minimum mass required for the development of a baryonic overdensity computed by \citet{Naoz07}, and the critical mass for an atomic cooling halo 
(taken from \citealt{Hummel12} and \citealp{Hartwig22}). These systems have a virial temperature of about $10^4\,$K, and their thermodynamic properties are dominated by Lyman-$\alpha$ cooling rather than H$_{2}$ cooling  \citep{Sutherland93}.
Halos of this mass and above are able to cool and collapse even in the presence of a very strong LW radiation background \citep{Oh2002} or a high streaming velocity \citep{Schauer19}, and so this curve constitutes the upper envelope of all models for $M_{\rm crit}$ considered here.  

\begin{figure}[htbp]
\begin{center}
\vspace*{-0.5cm}
\includegraphics[width=9.3cm]{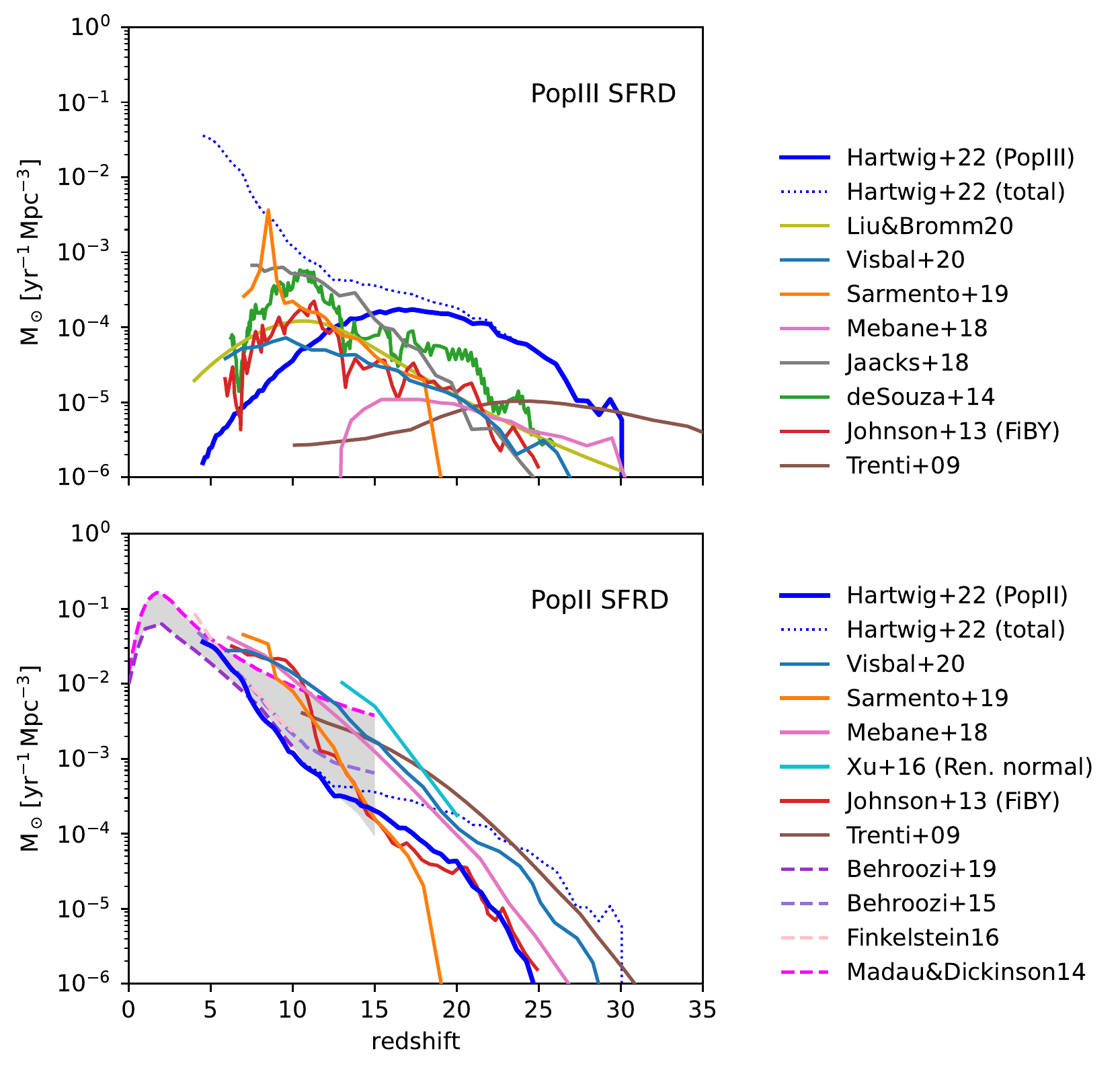}
\caption{Star formation rate density for Pop III and Pop II stars at $z > 5$ from different studies, including numerical simulations \citep{Johnson2013, deSouza14, Xu16a, Jaacks18b, Sarmento19, Liu2020} and semi-analytical models \citep{Trenti09, Mebane18, Visbal20, Hartwig22, Magg22}. For comparison, we also provide the observational constraints from \cite{Madau14-review}, \citet{Behroozi15}, \citet{Finkelstein16}, and \citet{behroozi19} with the gray shaded area indicating the proposed range.}
\label{fig:SFRD}
\end{center}
\end{figure}

\subsubsection{Cosmic star-formation rate density} 
\label{sec:csfrd}
In the $\Lambda$CDM model, Pop III star formation begins at a redshift $z \gtrsim 30$ in rare high-sigma fluctuations. The rate increases as collapse becomes possible in more and more halos and reaches a peak at redshifts $z \sim 15 - 20$. Although the overall cosmic star formation rate continues to increase \citep{Madau2014}, the rate at which  metal-free Population III stars form declines again, because regions in the Universe that have not been  enriched by supernova ejecta from massive stars become increasingly rare. This is the transition to the birth of (slightly) metal-enriched Population II stars, which we discuss in Section \ref{sec:transition-PopIII-to-PopII}. Different models for the star formation rate density (SFRD) as function of redshift are depicted in Figure \ref{fig:SFRD}. The scatter gives a feeling for the current uncertainties in determining the onset and evolution of high-redshift star formation in both numerical simulations \citep{Johnson2013, deSouza14, Xu16a, Jaacks18b, Sarmento19} and semi-analytical models \citep{Trenti09, Mebane18, Visbal20, Hartwig22}. For comparison, we also provide the observational constraints from \cite{Madau14-review}, \citet{Behroozi15}, \citet{Finkelstein16}, and \citet{behroozi19} with the total range indicated in gray, although we caution that the James Webb Space Telescope (JWST) will soon improve the situation for  $z \gg 5$.

\subsection{Standard Pop III formation pathway}
\label{sec:standard-scenario-of-PopIII-formation}
We refer to stellar birth in pristine halos containing zero-metallicity gas, which has not been affected by stellar feedback from neighboring halos, as the standard Population III formation pathway. For early studies in this field see \citet{Yoneyama1972}, \citet{Silk1977a, Silk1977b, Silk1977c}, \citet{Hutchins1976}, \citet{Carlberg1981}, \citet{Kashlinsky1983}, \citet{Palla1983}, or \citet{Stahler1986, Stahler1986b}. Here we focus on the current state of affairs, and briefly mention alternative scenarios in Section \ref{sec:alternative-PopIII-pathways}.

\subsubsection{Initial collapse phase}
\label{sec:initial-collapse}
% {\sl (450 words)}
As discussed above, the ability of the gas in a halo to collapse and form stars depends on its ability to cool. All main cooling processes in zero metallicity gas are related to hydrogen, either in atomic or molecular form. At high temperatures, collisions can populate the excited electronic states of H which then de-excite by emitting Lyman series photons. This process is often referred to simply as Lyman-$\alpha$ cooling and is most efficient around temperatures of $\sim 10^4\,$K. To reach lower temperatures, molecular hydrogen is needed. The lightness of the H$_{2}$ molecule and its lack of a dipole moment conspire to render it an ineffective coolant at very low temperatures: its lowest energy radiative transition is the $J = 2 \rightarrow 0$ transition in its vibrational ground-state, with an energy that corresponds to a temperature of 
$\sim 512\,$K. The high-velocity tail in the thermal Maxwell-Boltzmann velocity distribution allows the gas to cool below this value, but only down to about $200\,$K (see \citealt{Greif15}). This is often termed the Pop III.1 formation pathway \citep[e.g.][]{Mckee2008, Clark11b}. The temperature can drop even further, if cooling by deuterated hydrogen (HD) takes over \citep{Nagakura2005}. HD has a non-zero dipole moment and its lowest energy transition is between the $J=1$ to $J=0$ rotational levels, corresponding to a temperature of $\sim 128\,$K. This is sometimes called Pop III.2 formation pathway, and in practice becomes relevant only in regions with enhanced fractional ionization, for example in very massive or in externally irradiated halos (see Sections \ref{sec:alternative-PopIII-pathways} and \ref{sec:small-scale-impact}).

For a full account of primordial chemistry and the associated cooling and heating processes, we refer to the reviews by \citet{Glover2005, Glover2013},  or \citet{Bovino19}. Here we focus on the most essential concepts. At low densities, H$_2$ has two main formation pathways. First, it can form by a two-stage reaction pathway involving the H$^{-}$ ion as an intermediate step \citep{Mcdowell1961, Peebles1968}. The reactions are:
\begin{eqnarray}
{\rm H} + {\rm e}^- &\rightarrow& {\rm H}^- + \gamma\;,\\
{\rm H}^- + {\rm H} &\rightarrow& {\rm H}_2 + {\rm e}^-\;.
\label{eq:H-}
\end{eqnarray}
Second, there is also a contribution from a similar reaction pathway involving H$_{2}^{+}$ as an intermediary molecule \citep{Saslaw1967}:
\begin{eqnarray}
{\rm H} + {\rm H}^+ &\rightarrow& {\rm H}_2^+ +\gamma\;,\\
{\rm H}_2^+ + {\rm H} &\rightarrow& {\rm H}_2 + {\rm H}^+\;.
\label{eq:H2+}
\end{eqnarray}
Both reaction pathways require the gas to be partially ionized, and the amount of H$_{2}$ that can form via these routes is limited by the slow formation rates of the H$^{-}$ and H$_{2}^{+}$ ions and the recombination of the gas. Typically, the final molecular fraction is around $10^{-3}$. At high particle densities above $\sim 10^9\,$cm$^{-3}$, a third process becomes important. It is the three-body reaction   \citep{Palla1983}:
\begin{eqnarray}
{\rm H} + {\rm H} + {\rm H} &\rightarrow& {\rm H}_2 + {\rm H} \;.
\label{eq:3H}
\end{eqnarray}
As a result all atomic hydrogen is converted into H$_2$ once particle densities of $n \approx 10^{11}\,$cm$^{-3}$ are reached. However, as collapse proceeds and the temperature exceeds values of $\sim 2000\,$K at densities of above $10^{13}\,$cm$^{-3}$ the H$_2$ molecules become collisionally dissociated and the gas eventually turns atomic again.

\begin{figure}[ht]
%\begin{center}
 \setlength{\unitlength}{1cm}
 \begin{picture}(15,10)
   \put(-2.6,0){\includegraphics[width=8.0cm]{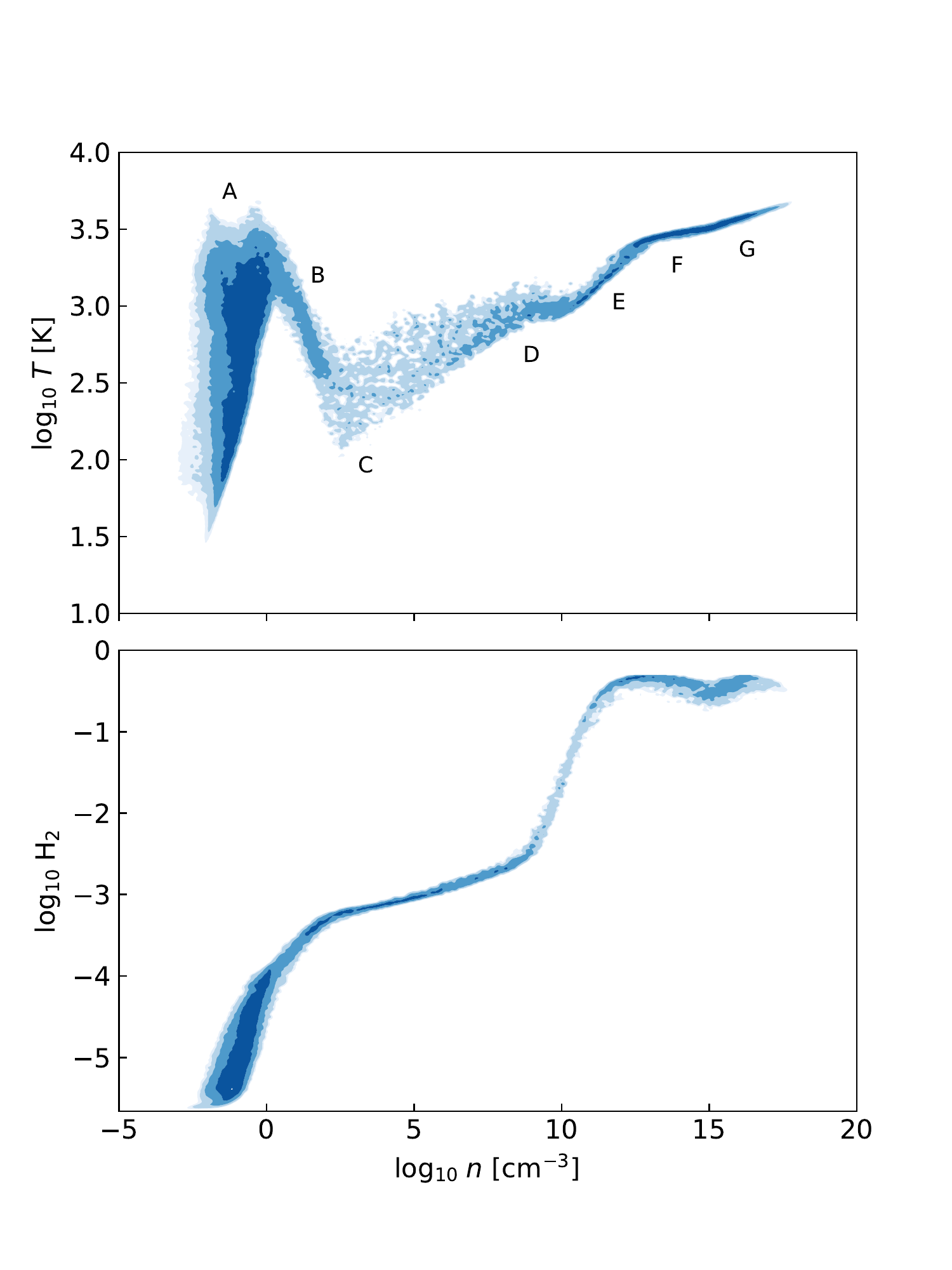}}
   \put( 5.4,0){\includegraphics[width=8.0cm]{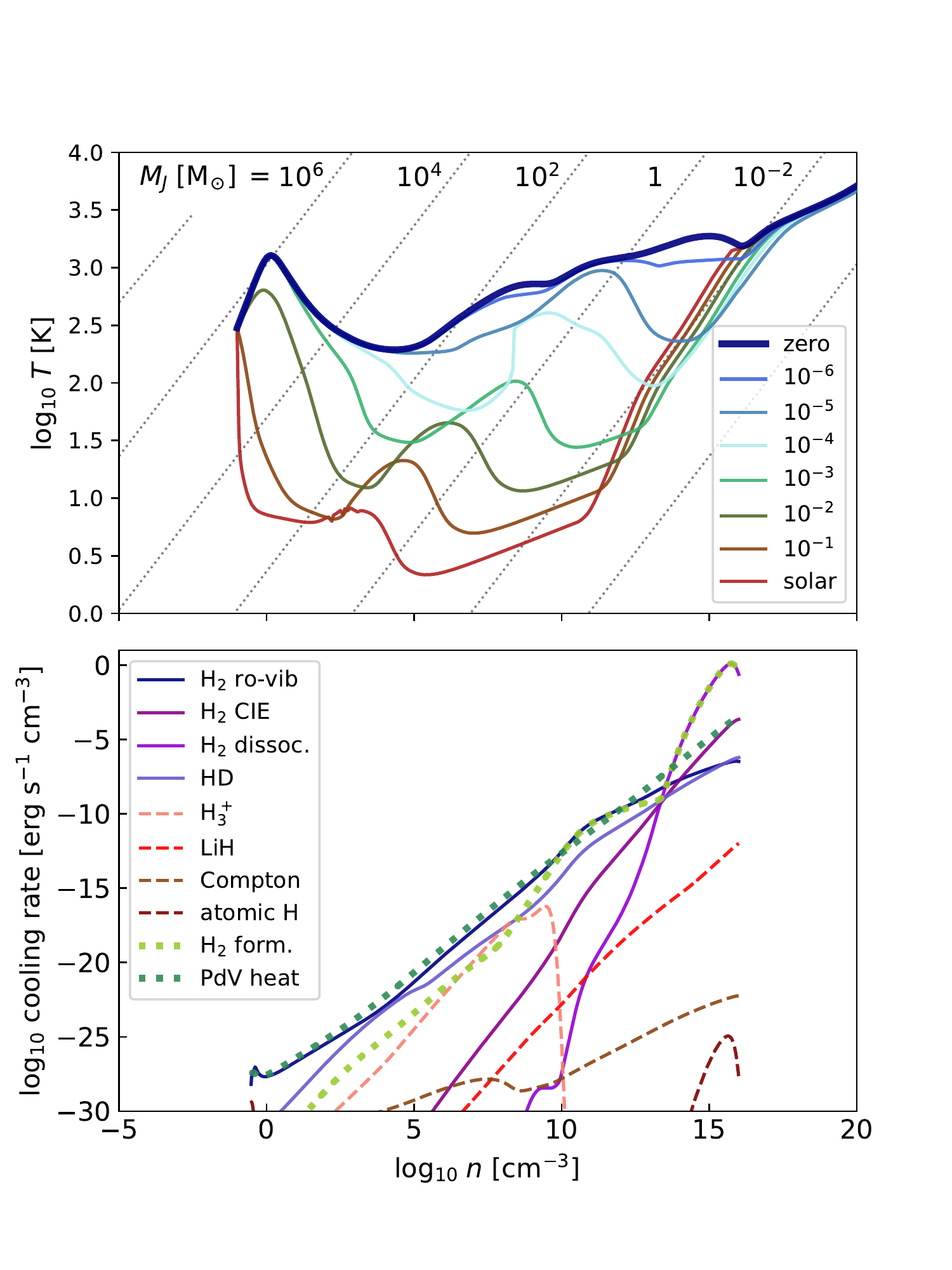}}
%   \put(-2.6,0){\includegraphics[width=8.0cm]{Figures/fig-logT-logn}}
%   \put( 5.4,0){\includegraphics[width=8.0cm]{Figures/fig-EOS-cooling}}
 \end{picture}
%\end{center}
 \caption{The left side shows the relation between temperature $T$ and hydrogen number density $n$ during the initial collapse phase (top) and the corresponding fraction of molecular hydrogen (bottom). The data are taken from a resimulation of a halo from \citet{Schauer21} provided by L.\ Prole. The right side (top) displays the  relation between log$_{10} T$ and log$_{10} n$, corresponding to the effective equation of state of the gas, for different metallicities, ranging from a purely primordial composition (dark blue, $Z = 0$) to the solar value (red, $Z = Z_{\odot}$) in steps of ten. The data are computed with a one-zone astrochemical model following the collapse of an isolated system \citep[for full details, see][]{Omukai2005, Omukai10}. The dashed lines indicate constant Jeans mass for atomic gas (Equation \ref{eq:Jeans-mass}). The right side (bottom) illustrates the main heating and cooling processes relevant for primordial gas. Cooling rates associated with hydrogen are depicted with solid lines, other cooling processes with dashed line, and the two main heating mechanisms with dotted lines. The data for the top right plot are provided by K.\ Omukai and the figure is reproduced with permission of ApJ.}
\label{fig:initial-collapse}
\end{figure}

\pagebreak
With the appropriate chemical network and the corresponding heating/cooling functions included in cosmic structure formation calculations we can follow the initial collapse phase from the cosmic mean up to the formation of the first hydrostatic core in the halo center. An example of this behavior is provided at the left side of Figure \ref{fig:initial-collapse}. The data are taken from a resimulation of a halo studied by \citet{Schauer21} at a time just before the formation of the first protostar. We show the effective equation of state, i.e.\ the relation between temperature $T$ and number density $n$ of hydrogen atoms (top), as well as the corresponding fraction of molecular hydrogen (bottom). The labels in the $\log_{10} T - \log_{10} n$ plot indicate key phases of the initial collapse: {(A)} As the gas begins to flow into the potential well of the dark matter halo, it is compressionally heated to the virial temperature of the system. At the same time the H$_2$ fraction increases from $< 10^{-5}$ to $\sim 10^{-3}$, which is sufficient for the gas to go into a run-away cooling phase {(B)} and brings it down to the minimum temperature of $\sim 200\,$K {(C)}. As more gas flows into the halo center, eventually the potential becomes dominated by the gas rather than by dark matter. The contraction proceeds and the heat provided by $PdV$ work begins to dominate again over cooling. As a consequence,  the gas temperature rises to about $1000\,$K at $n \approx 10^{9}\,$cm$^{-3}$ (D). At this stage, three-body H$_2$ formation (Equation \ref{eq:3H}) becomes important  and the gas quickly becomes fully molecular {(E)}. Below densities $n \approx 10^{13}\,$cm$^{-3}$ the main cooling process is the (forbidden) ro-vibrational line emission from H$_2$. However, at larger $n$ two additional cooling processes become important {(F)}. The first is collision-induced emission (CIE), which occurs when two molecules come close to each other. Van der Waals forces can then induce a temporary dipole which allows for efficient dipole emission during the interaction time interval \citep{Omukai1998, Ripamonti2004}. The second one is associated with the collisional dissociation of  H$_2$, which sets in at temperatures around $2000\,$K and quickly dominates the overall cooling behavior. As the density increases further {(G)},  more and more H$_2$ molecules are destroyed. This also implies that more hydrogen atoms become available again for H$_2$ formation through the three-body process and the associated energy release starts to dominate the overall heating rate over $PdV$ work. This continues until the molecular gas is largely depleted and the collapse becomes almost adiabatic at $n\approx 10^{20}\,$cm$^{-3}$.

For comparison, we provide at the right side of Figure \ref{fig:initial-collapse} (top) the relation between log$_{10} T$ and log$_{10} n$ from a one-zone astrochemical model describing the time-evolution of the central region during the initial collapse \citep{Omukai1998} It essentially depicts the effective equation of state. We see that the  polytropic index $\gamma_{\rm eff}$ varies considerably across the different regimes introduced above. Data are taken from \citet{Omukai2005, Omukai10} and describe the thermodynamic response of gas across a wide range of metallicities. We start with a purely primordial composition ($Z = 0$), indicated by the dark blue line, and show the results for log$_{10}\,Z/Z_{\odot} =$ -6, -5, $\dots$ up to the solar value ($Z = Z_{\odot}$) in red. For comparison, we also show lines of constant Jeans mass following Equation (\ref{eq:Jeans-mass}) for atomic gas.

For a more quantitative assessment of the most relevant heating and cooling processes during the initial collapse phase of gas in primordial halos, we also plot the corresponding rates as a function of the hydrogen number density, using data taken from the one-zone model of \citet{Glover2009}. Here, solid lines depict cooling processes associated with molecular hydrogen: ro-vibrational line emission from H$_2$ and HD, collision-induced emission (CIE) of H$_2$ at high densities and collisional dissociation of H$_2$. The cooling processes associated with other species are less relevant. Using dashed lines, we list cooling from H$_3^+$, LiH, Compton scattering, and H. In the absence of stellar feedback only two heating processes are important (dotted lines): $PdV$ work dominates at densities below $n \approx 10^{13}\,$cm$^{-3}$, and the latent heat released by H$_2$ formation is relevant at higher densities.

We note that over the roughly 18 orders of magnitude in density covered in Figure \ref{fig:initial-collapse}, the temperature of the primordial gas only varies by a factor of 25 or so. Overall the gas roughly exhibits an effective index of $\gamma_{\rm eff} \approx 1.08$, which is close to the isothermal value of $\gamma_{\rm eff} =1$.  This has important consequences for the level of fragmentation in primordial gas and it is essential to understand the dynamical evolution of the accretion disk that inevitably builds up around the central object (see Section \ref{sec:disks-and-fragments}). We also note that Figure~\ref{fig:initial-collapse} only considers compressional and chemical heating. The situation may change if radiative feedback from newly formed stars  (Section \ref{sec:small-scale-impact}) or the possible energy input from turbulent dissipation or from dark matter annihilation is considered (Section \ref{sec:less-fragmentation}).

\begin{figure}[tp]
\resizebox{18cm}{9cm}
{
\setlength{\unitlength}{1cm}
\begin{picture}(18,9)(5,-0.5)
\put(0.0,6.0){\includegraphics[width=2.5cm,height=2.5cm]{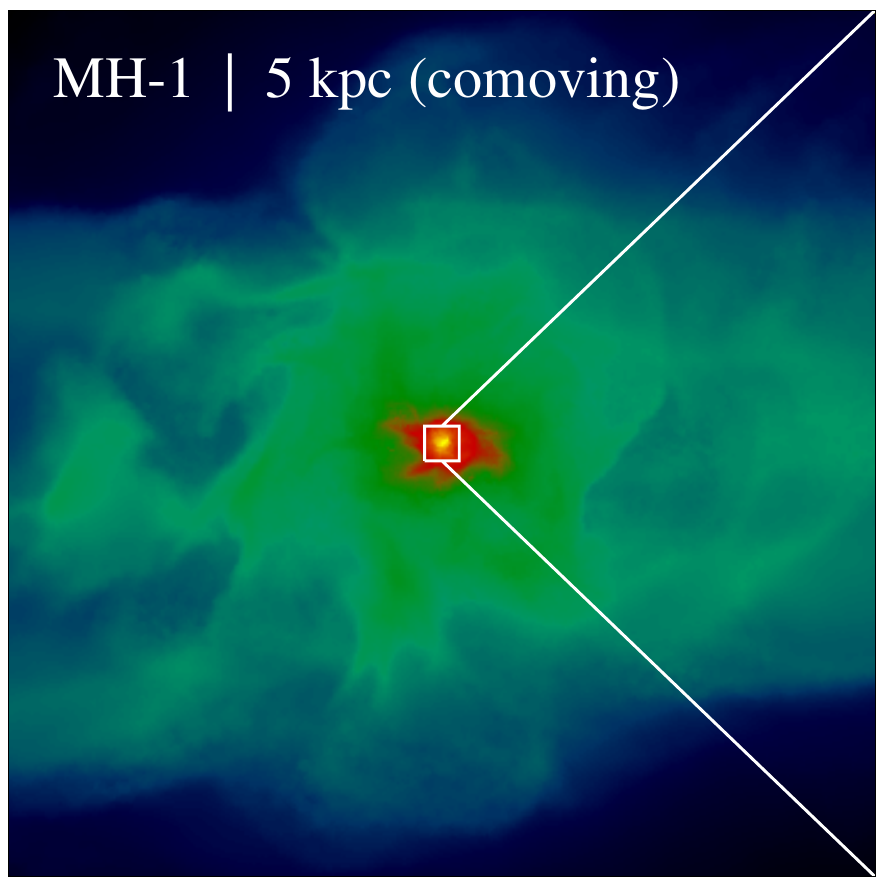}}
\put(2.5,6.0){\includegraphics[width=2.5cm,height=2.5cm]{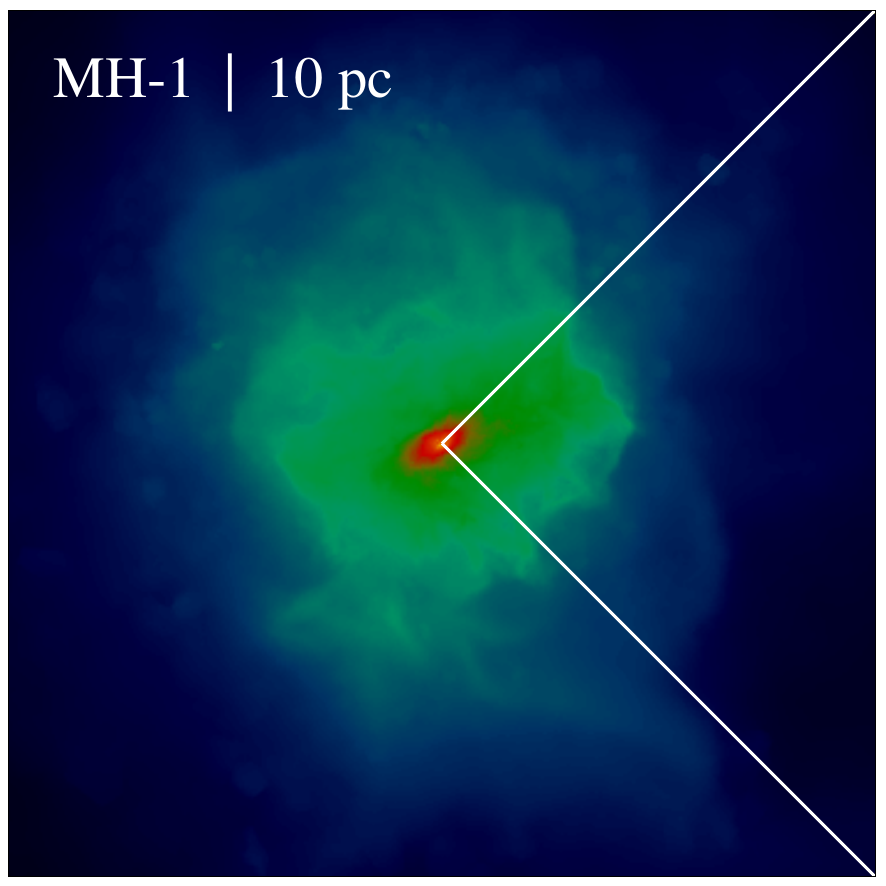}}
\put(5.0,6,0){\includegraphics[width=2.5cm,height=2.5cm]{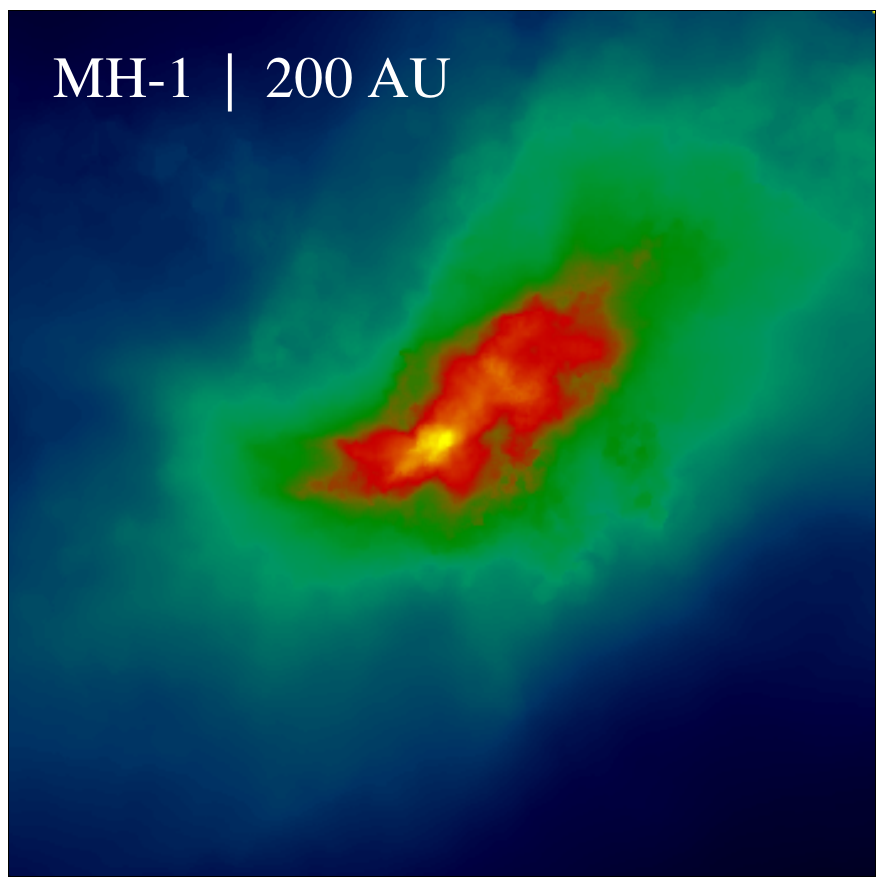}}
\put(0.0, 3.5){\includegraphics[width=2.5cm,height=2.5cm]{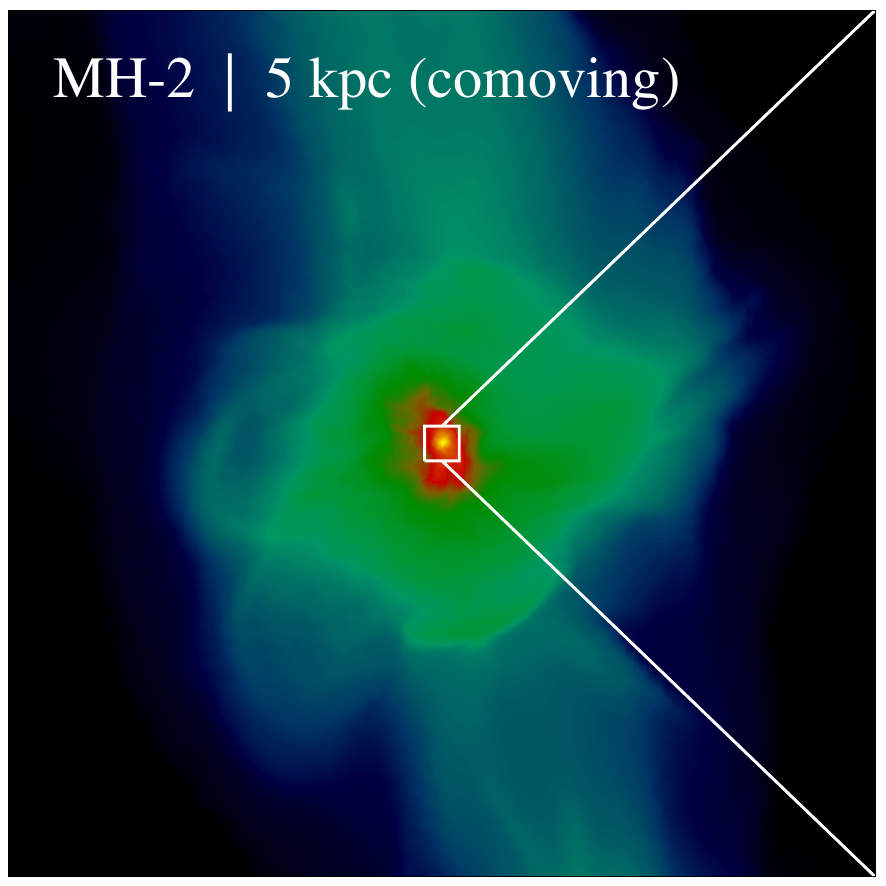}}
\put(2.5, 3.5){\includegraphics[width=2.5cm,height=2.5cm]{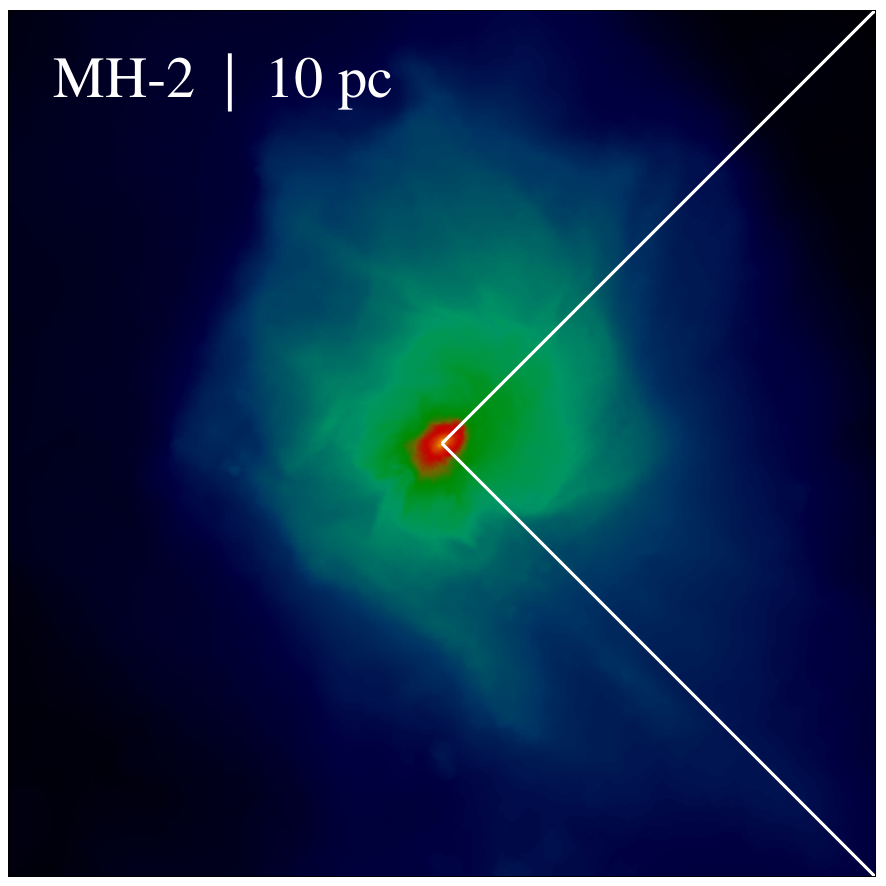}}
\put(5.0, 3.5){\includegraphics[width=2.5cm,height=2.5cm]{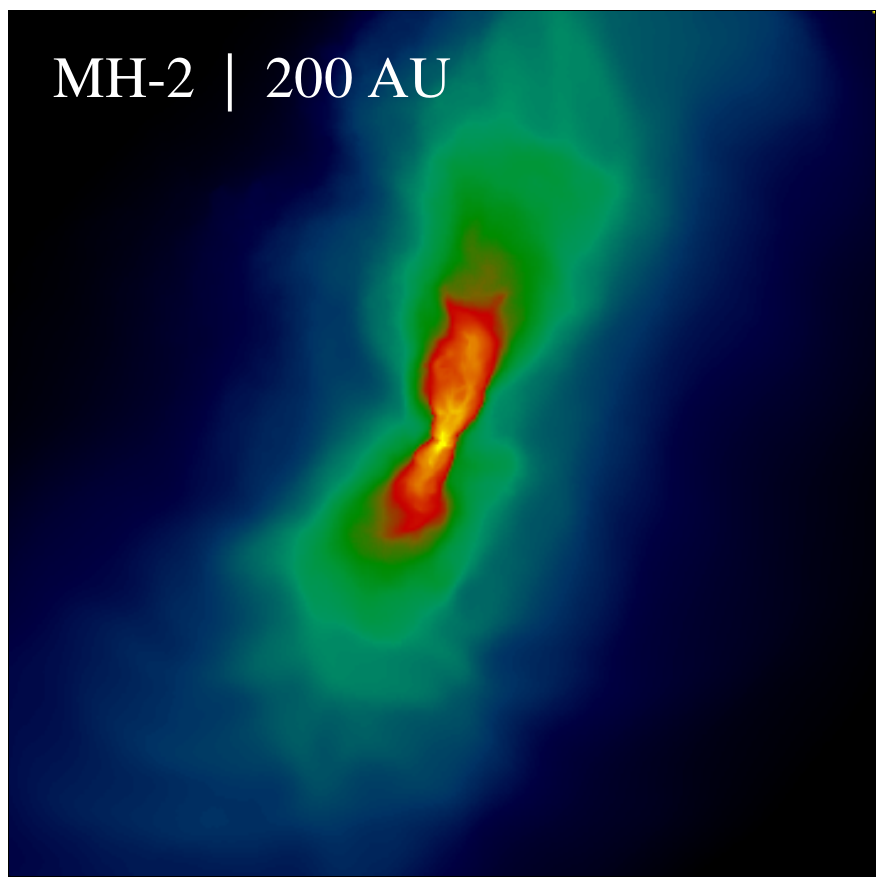}}
\put(0.0, 1.0){\includegraphics[width=2.5cm,height=2.5cm]{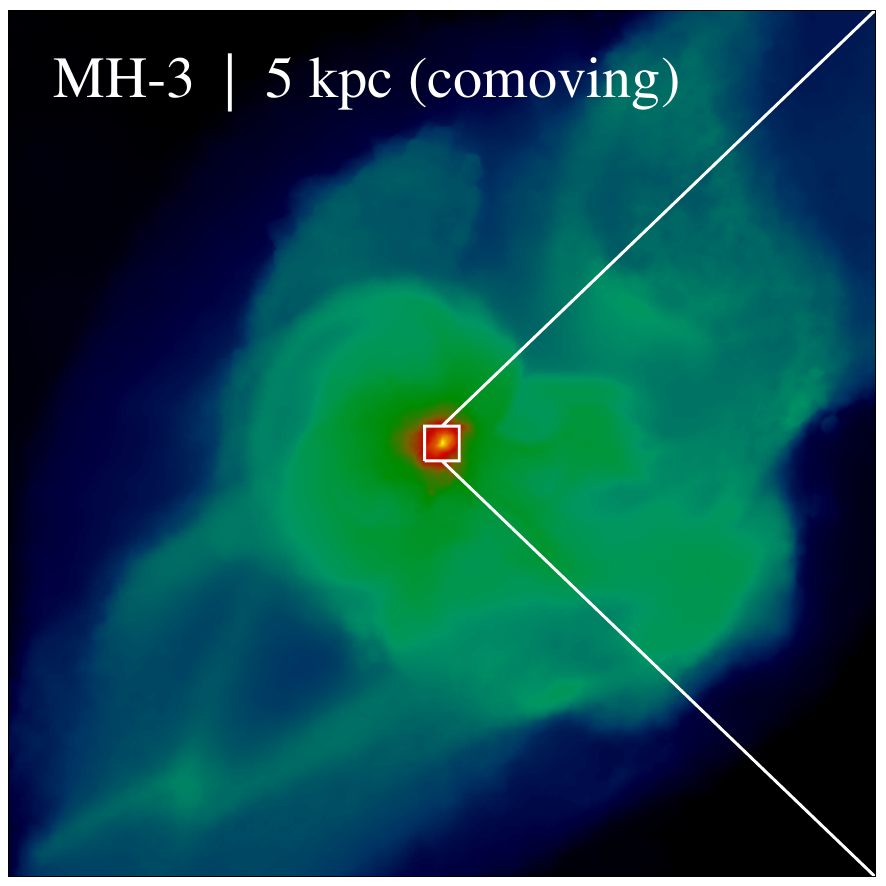}}
\put(2.5, 1.0){\includegraphics[width=2.5cm,height=2.5cm]{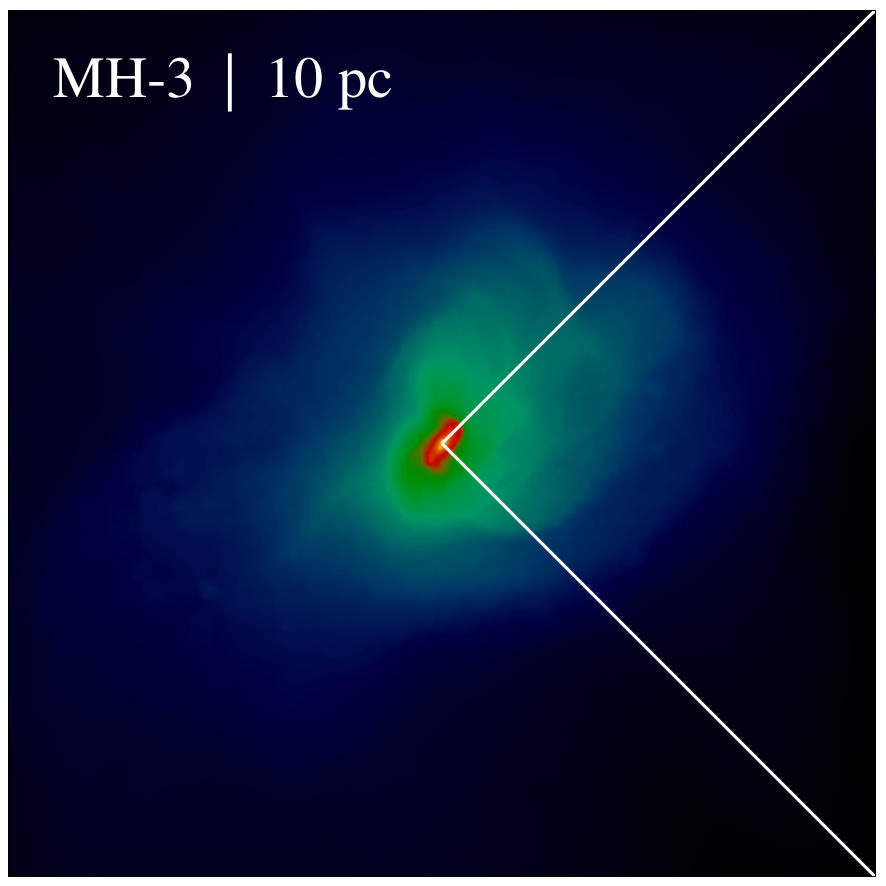}}
\put(5.0, 1.0){\includegraphics[width=2.5cm,height=2.5cm]{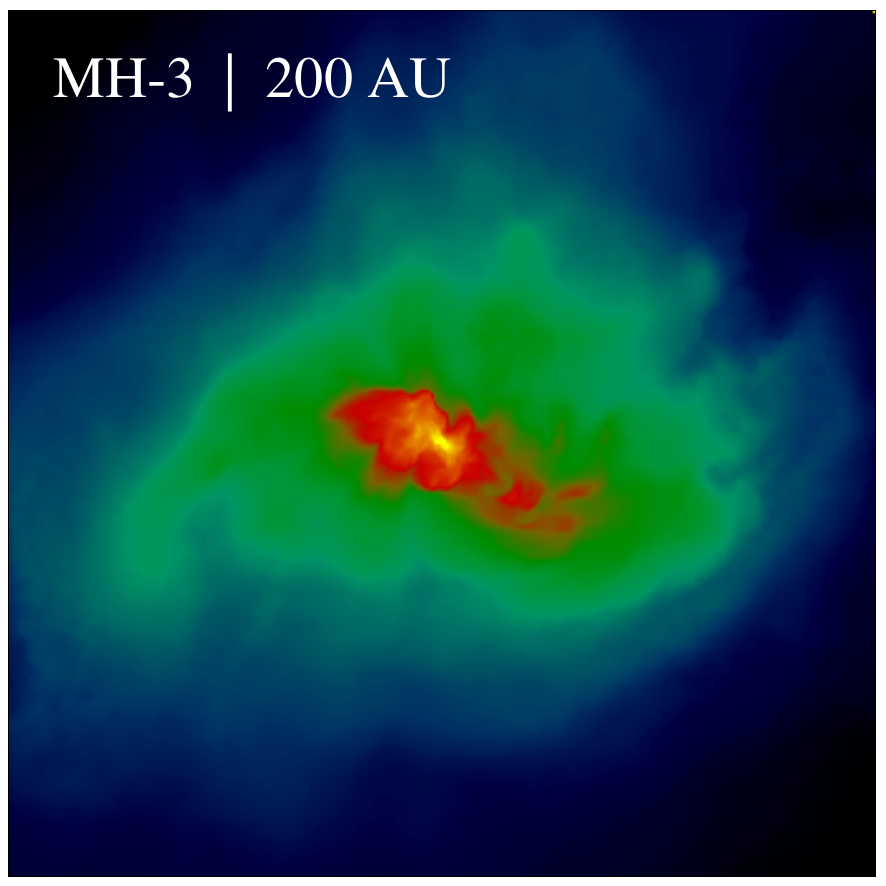}}
\put(0.0, 0.1){\includegraphics[width=2.5cm,height=0.84cm]{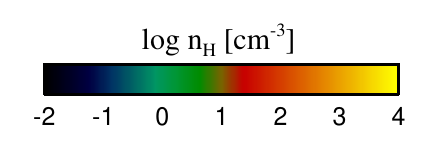}}
\put(2.5, 0.1){\includegraphics[width=2.5cm,height=0.84cm]{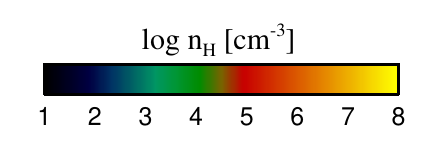}}
\put(5.0, 0.1){\includegraphics[width=2.5cm,height=0.84cm]{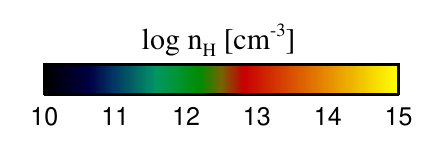}}
\put(10.0, 0.5){\includegraphics[height=8cm]{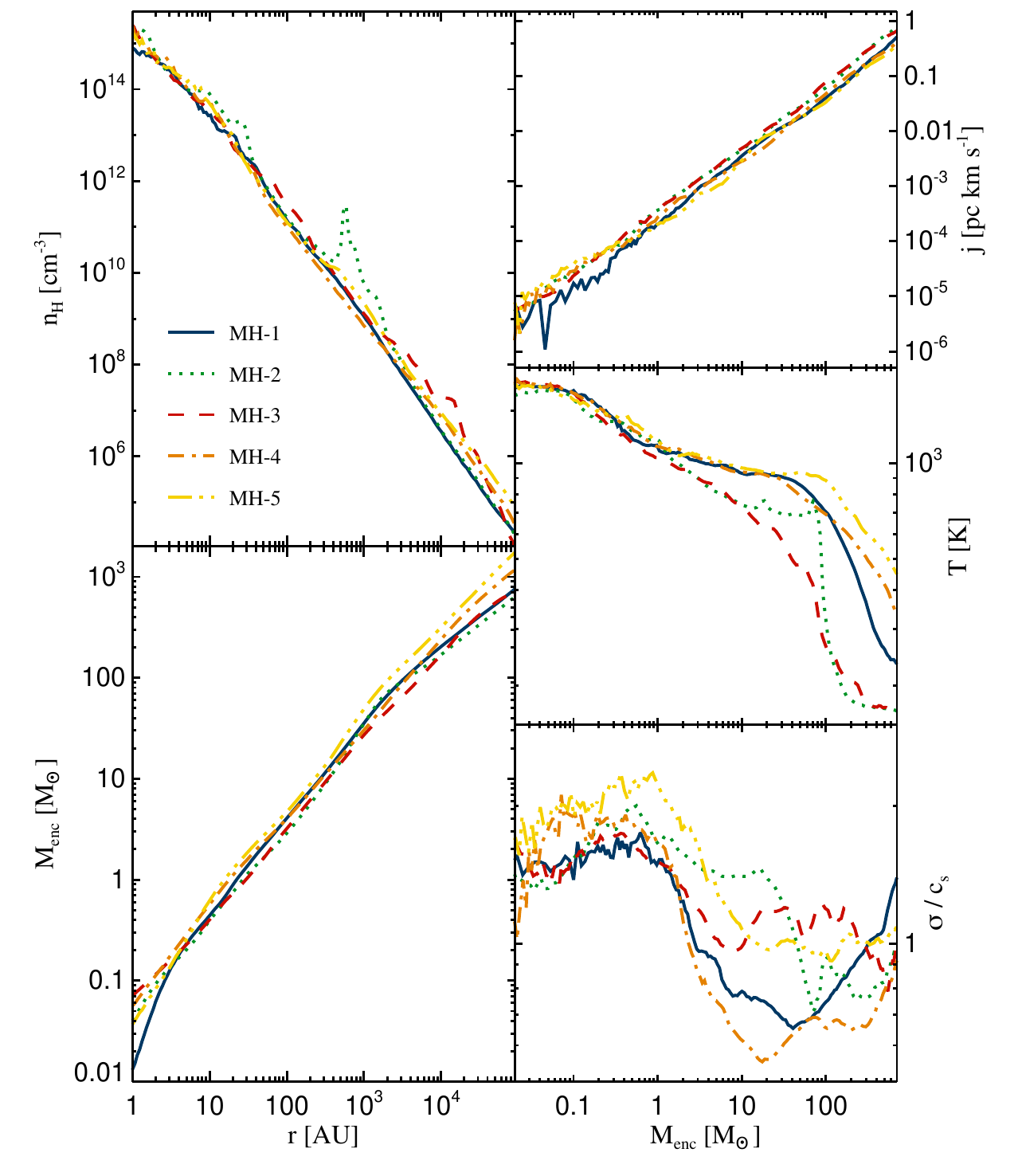}}
\end{picture}
}
\caption{Structural and dynamical properties of three star-forming halos. The left side shows a sequence of zoom-ins focusing on the density-squared weighted hydrogen number density projected along the line of sight. The right plot provides cumulative mass as function of radius (top left) and corresponding number density profile (bottom left) in the inner regions of the halos (plus two additional systems not shown here) shortly before the onset of star formation. Specific angular momentum (top right), gas temperature (middle right), and turbulent velocity dispersion (bottom right) are plotted  as functions of enclosed mass. The figure is adapted from \citet{Greif11} and reproduced with permission of ApJ.}
\label{fig:halo-properties}
\end{figure}

\subsubsection{Disk formation and fragmentation}
\label{sec:disks-and-fragments}
%{\sl (1350 words)}
% 
 The first 3D simulations able to follow the initial cooling and collapse of the gas in high redshift minihalos became available around the year 2000 \citep[e.g.][]{Bromm99, Bromm02, Nakamura01, Abel2000, Abel2002, Yoshida2003}. Although these numerical simulations were fully three-dimensional, the halos considered were relatively round and so assuming spherical symmetry was a very good approximation during the early stages of collapse \citep[see also][]{Yoshida2006, Yoshida2008}. These early calculations typically stopped when the object in the center reached hydrogen number densities of $n \approx 10^{16}\,$cm$^{-3}$, above which the computational timestep became prohibitively small. At this time the hydrostatic object in the center had a mass of only $\sim 10^{-3}\,$M$_{\odot}$. At such an early time, the material that made it to the center carried very little angular momentum, and this object was surrounded by only a small disk-like structure which was more strongly supported by pressure than by rotation. 
 The authors of these studies suggested that this should be true for the entire protostellar accretion history, and hence argued that all the inflowing mass would end up in one single high-mass star \citep[see also][]{Tan2004}. Clearly this supposition needed to be tested, in particular, because the statement that primordial stars only form in isolation is in tension with present-day star formation, where fragmentation is ubiquitous and massive stars are typically found in clusters and aggregates \citep{Lada2003}.
 
 The left side of Figure \ref{fig:halo-properties}, adopted from a high-resolution simulation published by \citet{Greif11} ten years later, gives a visual impression of three star-forming halos at various spatial scales, eventually zooming in on the immediate physical environment where Pop III stars build up. The gas virializes on a scale of 5 kpc (comoving), followed by the runaway collapse of gas in the central $10\,$pc, where it becomes self-gravitating and decouples from the dark matter. In the final stages of the collapse, a fully molecular core forms on scales of $\sim 200\,$AU. The right side of Figure \ref{fig:halo-properties} provides key structural and dynamical parameters, including cumulative mass and the corresponding radial density profile to the left, as well as  specific angular momentum, gas temperature, and turbulent velocity dispersion as a function of enclosed mass to the right. Although the morphological parameters of the halos are very similar on large scales, they differ on small scales. They also have quite distinct dynamical properties. The collapse induces turbulence in the infalling gas, which leads to stochastic variations in the structural appearance of the central regions, demonstrating that small differences in the initial fluctuation spectrum can amplify during the non-linear collapse phase and lead to very different evolutionary pathways.  For example, HD cooling became important in only two of the five halos modeled by \citet{Greif11}, leading to lower temperatures in these halos and a different fragmentation behavior. This suggests that, similar to present-day star formation \citep{Mckee2007, Klessen2016}, stellar birth in the early Universe is also subject to large statistical variations with the outcome sensitively depending on the details of the initial conditions and environmental parameters as well as on various complex non-linear feedback loops.

 \begin{figure}[tp]
 \setlength{\unitlength}{1cm}
 \begin{picture}(15,6.5)(0,0)
  \put(-2.4, 1.0){\includegraphics[width=8.0cm]{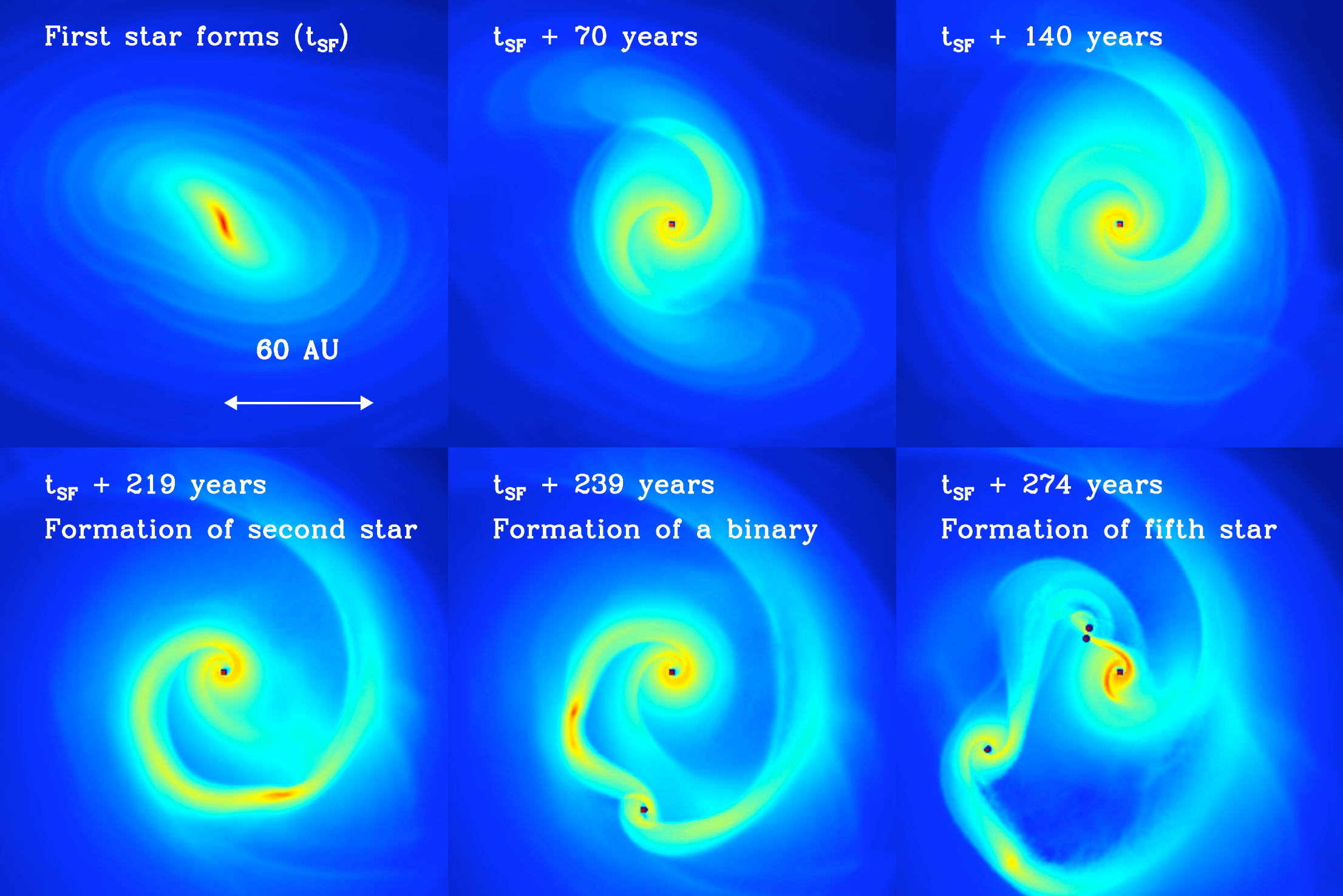}}
  \put(5.7, 0.0){\includegraphics[width=6.6cm]{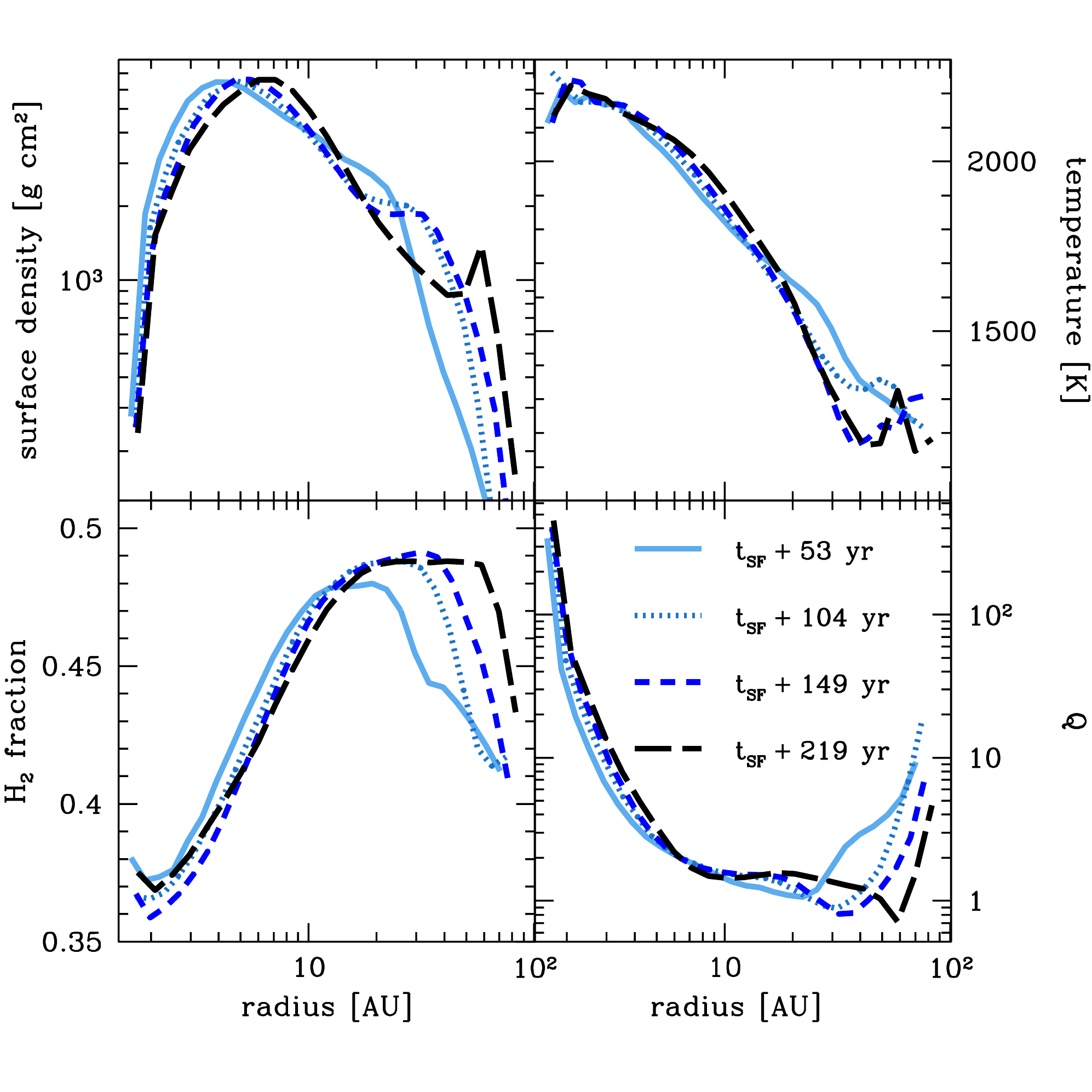}}
%  \put(-2.2, 1.0){\includegraphics[width=8.5cm]{Figures/fig-disk-fragmentation.pdf}}
%  \put(6.3, 0.0){\includegraphics[width=7cm]{Figures/fig-disk-properties.pdf}}
 \end{picture}
\caption{Formation and evolution of the disk around a Pop III protostar, adopted from \citet{Clark11}. The left side gives a visual impression of the build-up of a two-arm spiral structure caused by gravitational instability which subsequently fragments into four protostars within $110\,$yr. The right side shows the radial surface density and temperature profile at different times. It also demonstrates that the disk material is almost fully molecular ($n_{\rm H_{2}}/n = 0.5$) and Toomre unstable ($Q<1$). The figure is reproduced with permission of Science.  
}
\label{fig:disk-fragmentation}
\end{figure}

Current simulations of Pop III star formation acknowledge this complexity and attempt to follow the formation and dynamical evolution of the accretion disk that builds up around the central object, and ideally cover the entire accretion history, until stellar feedback removes the remaining gas and the process of stellar birth is completed. These studies demonstrate that primordial accretion disks are highly prone to fragmentation. They suggest that the standard pathway of Pop III star formation leads to a stellar cluster with a wide range of masses rather than the build-up of one single high-mass object. This is illustrated in Figure \ref{fig:disk-fragmentation}, taken from \citet{Clark11}, showing the early evolution of a typical Pop III accretion disk. It fragments and builds up a multiple system of four protostars within only about one hundred years after the formation of the first object. 

Disk fragmentation across a wide range of spatial and temporal scales is reported in essentially all current studies  \citep[e.g.\ by][which is by no means an exhaustive list]{Machida2008, Machida2008c, Turk09, Clark11, Greif2011, Greif12, Smith2011, Smith2012, Dopcke2013, Susa13, Susa14, Vorobyov13, Stacy2013, Stacy2016, Hirano2014, Hosokawa2016, Takahashi17, Hirano18, Susa19, Wollenberg20, Sugimura20, Sharda21, Jaura22, Chiaki22, Prole22}. They predict the formation of a cluster of multiple Pop III (proto)stars, which grow in mass at rates of $\dot{M} \approx 10^{-3}\,$M$_\odot\,$yr$^{-1}$ with possible brief periods of accretion with rates as high as a few times $10^{-2}\,$M$_\odot\,$yr$^{-1}$. The reason for the high susceptibility of primordial accretion disks to fragmentation is always the same: For typical halo  conditions in the primordial Universe the mass load onto the disk from the infalling envelope exceeds its capability to transport material inwards by gravitational or magnetoviscous torques. As a consequence massive spiral arms build up and speed up the inward transport. Often this is not enough, and the arms become non-linear and interact with each other, leading to run-away collapse in the interaction regions, as clearly visible in Figure \ref{fig:disk-fragmentation}. The disk is fragmenting and forms new protostars. As this happens, a smaller accretion disk forms around the new object, which by itself may become unstable and fragment to form additional protostars. For very quiescent initial conditions with low levels of turbulence, this secondary process may be the prevalent form of fragmentation \citep{Susa19}. As the global accretion disk gains more mass and grows in size by accretion of higher angular momentum material, the Toomre unstable region moves further out. Consequently,  fragmentation and formation of new protostars occurs at larger and larger radii as the evolution progresses.

\subsubsection{Pop III IMF and multiplicity}
\label{sec:IMF-and-multiplicity}
% {\sl (1350 words)}
As argued in the previous Section, it is almost impossible to avoid the fragmentation of primordial accretion disks. The key question then is, what happens to the fragments during their subsequent dynamical evolution? There are three possible outcomes: (1) A fragment grows in mass and survives to become a proper star (or substellar object), or (2) it dies by getting swallowed by the central object, or (3) it merges with another fragment, in which case the same question applies to the new object. A related question is, if the fragment survives, what sets its final stellar mass? Again, there are three answers possible: (1) The system simply runs out of mass, leaving behind relatively massive stars with interesting implications for their detectability at high redshifts (Section \ref{sec:high-z}). (2) Stellar feedback removes material from the system and terminates any subsequent accretion. This is the most complex case and can lead to a wide range of different results, with the currently available simulations being highly inconclusive, as discussed in Section~\ref{sec:small-photoionization}. (3) The fragment gets ejected from the disk by dynamical interactions and its growth is stopped prematurely.  This typically leads to low-mass objects, some of which could potentially have survived until the present day and might be detectable in stellar archeological surveys (see the discussion in Section \ref{sec:low-z}). 
%% MRNT
\begin{marginnote}[]
\entry{IMF}{The stellar initial mass function describes the statistical distribution of stellar masses at birth. Due to mass loss during later phases of stellar evolution, the IMF is different to the mass distribution when stars end their lives. }
\end{marginnote}
%% MRNT
\begin{marginnote}[]
\entry{Hill volume}{Region of influence of a smaller body in the face of gravitational perturbations from a more massive body.}
\end{marginnote}

Let us consider these processes in more detail.  As material flows through the disk towards the center, it first encounters the Hill volume of  protostars that lie further out, and so it preferentially gets swallowed by these objects rather than by those close to the center. 
This reduction of the mass growth rate is more pronounced at smaller radii and can lead to the complete starvation of the  primary object in the very center \citep{Peters2010, Girichidis2011}. In simple binary systems, this process results in the two constituent stars having roughly equal masses \citep[see also][]{Kratter2006}. In the more complex situation of Pop III star formation, with ejections, mergers, and absorption by the central object, all of which are unpredictable and highly stochastic processes, we expect a wide spectrum of stellar masses. 

When considering the question of what fraction of fragments merge or become swallowed by the central object and how many survive, various studies   \citep[e.g.][]{Greif12, Smith2012, Stacy2013} indicate that roughly 2/3 of the fragments quickly disappear again, and about 1/3 remain. These numbers should be taken with caution, as none of the simulations covers the entire duration of disk evolution, with the highest resolution models being able to cover the least amount of time. It therefore could be that protostars that are counted as ejected will eventually fall back again and become accreted then. Also those remaining in the disk may still merge at a later stage. In addition, the result also depends on how mergers are numerically implemented \citep{Wollenberg20}. Still, as long as fragmentation in the disk continues and the appearance and disappearance of new protostars is an ongoing process, adopting a ratio of roughly 1:2 of survivors and mergers is a good estimate. It has also been suggested that the number $N_{\rm frag}$ of fragments, survivors as well as mergers, increases with time $t$ as $N_{\rm frag} \propto t^{0.3}$ \citep[][]{Susa19}.  

Even if Pop III stars form in multiple systems across a wide range of masses, we can nevertheless look for the most bound pairs and investigate their properties. Doing so,  \citet[][]{Stacy2013} and \citet{Stacy2016} report a binary fraction of $\sim 35$\%, with semi-major axes as large as $3000\,$AU and with a wide distribution of orbital periods, ranging from $\sim 30\,$ to $30000$ years, with the smallest period being determined by the minimum numerical resolution achieved. They also find that the  distribution of mass ratios is relatively flat \citep[as also suggested by][]{Susa19}. With these data, it is possible to build large ensembles of Pop III binary systems for the long-term integration in dedicated $N$-body simulations or semi-analytic models \citep[e.g.][]{Liu21, Santoliquido21}. 

\begin{figure}[tp]
\begin{center}
 \vspace*{-1.4cm}
 \includegraphics[width=8cm]{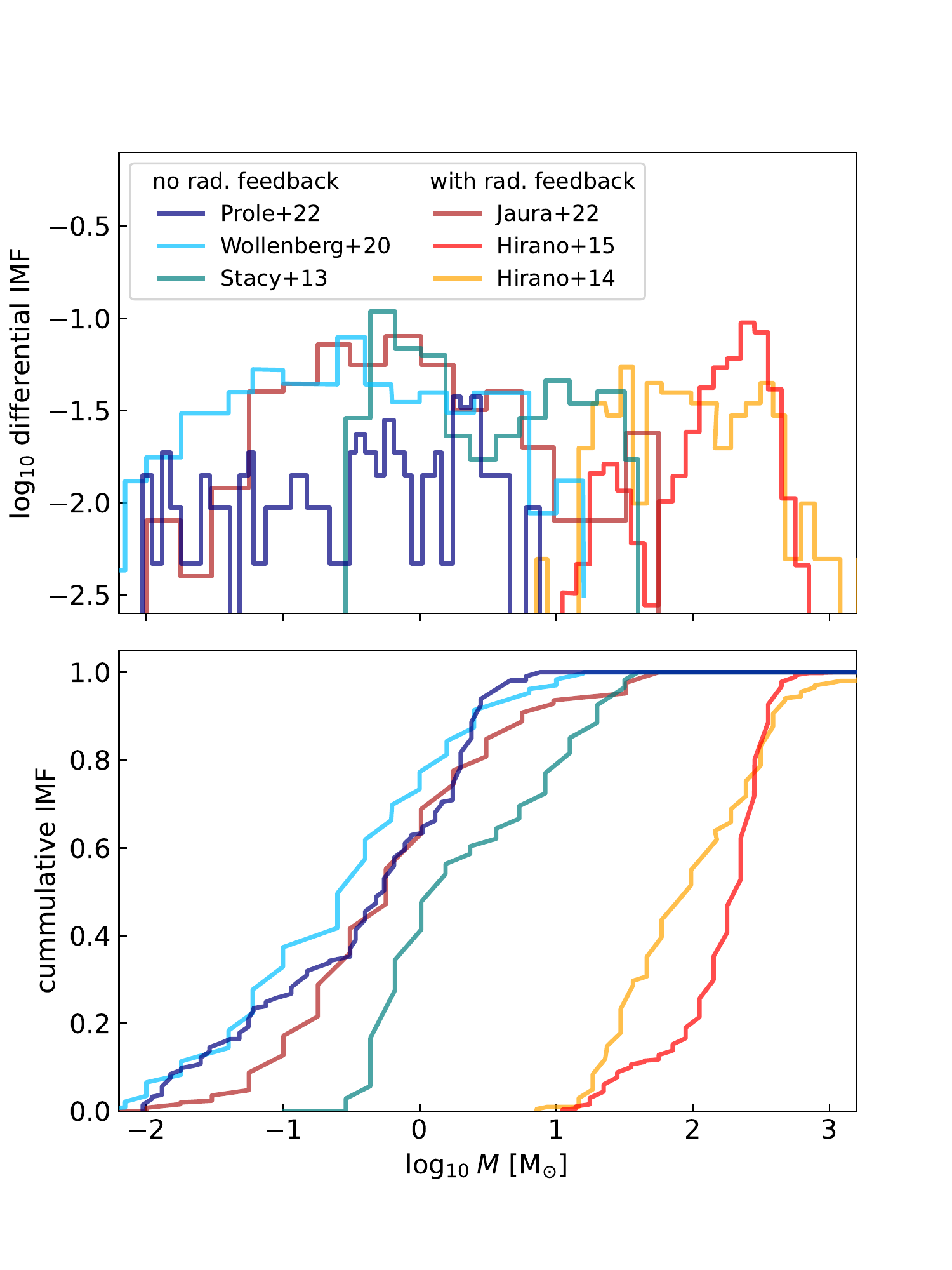}
\end{center}
 \vspace*{-0.3cm}
\caption{Normalized differential (top) and cumulative (bottom) stellar mass distribution obtained from six different studies of Pop III star formation. To represent the current state-of-the art, we select three high-resolution models without stellar feedback (\citealt{Prole2022b}: model with $\rho_{\rm sink} = 10^{-6}\,$g$\,$cm$^{-3}$, \citealt{Wollenberg20}: model ${\alpha}005{\beta}001$, \citealt{Stacy2013}: all models) and three simulations with radiative feedback included (\citealt{Jaura22}: model T$\_$RTP, \citealt{Hirano2015}: model III.1, \citealt{Hirano2014}: all models). Despite the large variation in range, the predicted IMFs are rather flat compared to the present-day IMF.} 
\label{fig:PopIII-IMF}
\end{figure}

\pagebreak
Concerning the resulting distribution of stellar masses, the initial mass function (IMF) of Pop III stars, the ubiquity of fragmentation and the stochasticity of the processes involved lead to a wide range of stellar masses, with all current models suggesting values from the substellar regime up to several hundred solar masses. As mass is the most important parameter in stellar evolution theory \citep{Kippenhahn2012}, predicting the Pop III IMF is a central aspect of many studies. In Figure \ref{fig:PopIII-IMF} we present the results of some of these simulations to give an indication of the diversity of the current state-of-the-art. We select three high-resolution models without stellar feedback: (1) model $\rho_{\rm sink} = 10^{-6}\,$g$\,$cm$^{-3}$ from \citet{Prole22}; (2) model ${\alpha}005{\beta}001$ from \citet{Wollenberg20};  and (3) the combined model from \citet{Stacy2013}. Models (1) and (2) reach very high spatial resolution, but only cover a small fraction of the total accretion time in the halo. Model (3) is not so well resolved, but covers more time. The effect of resolution is noticeable in the resulting mass spectrum. We also chose three simulations with radiative feedback from newly formed stars: (4) T\_RTP from \citet{Jaura22}; (5) III.1 from \citet{Hirano2015}; and (6) the combined models from \citet{Hirano2014}. Model (4) has similar resolution and time coverage as the first two examples (1) and (2)  without feedback, and it predicts a similar mass spectrum. Models (5) and (6)  have somewhat lower resolution, but cover considerably more time until the protostellar accretion has stopped. Whereas (1) -- (4) are fully three-dimensional simulations, (5) and (6) have been performed in  two dimensions assuming axisymmetry. These calculations predict considerably larger masses.  Altogether, the reported stellar mass spectra vary enormously, which is likely a consequence in part of the difference resolutions and periods simulated in the different calculations. However, it is a common feature that they all are approximately logarithmically flat. This is in stark contrast to the present-day IMF, which shows a clear peak around $0.2 - 0.3\,$M$_{\odot}$ followed by a steep power-law fall-off \citep{Kroupa2002, Chabrier2003}. For this reason, the IMF of primordial stars is often called top-heavy in comparison.  

We conclude that our understanding of the IMF of primordial stars is still quite limited. The existing models all have  specific shortcomings, with none being able to reach sufficiently high resolution, cover enough time, and properly include all relevant physical processes. As indicated in Figure \ref{fig:PopIII-IMF}  we find that models that include stellar feedback on average predict larger stellar masses than those without. In addition the outcome strongly depends on the spatial resolution achieved, on the time span covered, and on the details of the numerical implementation. High-resolution simulations tend to yield smaller stellar masses, because they are better able to resolve disk fragmentation (see, e.g., the discussion in \citealt{Stacy2016} or \citealt{Prole22}), but typically only cover a small fraction of the full accretion history of the halo. This raises the question of whether the remaining gas reservoir is mostly used to form new (low-mass) objects, or whether it is largely consumed by existing protostars, which then continue to grow in mass. Similarly, simulations covering a large fraction of the halo collapse timescale do not resolve the inner accretion disk well, and therefore exhibit lower levels of fragmentation. Hence, they are biased towards higher-mass stars. Furthermore, two-dimensional simulations tend to report less fragmentation than full three-dimensional ones. Fragmentation is also influenced by additional physical processes, which we discuss in Section \ref{sec:alternative-PopIII-pathways} below.

\subsubsection{Intermediate conclusions}
\label{sec:intermediate-conclusions}
Altogether, we can draw the following conclusions from the discussion so far: First, once a halo decouples from the cosmic expansion, the question of  whether or not it forms stars depends strongly on its ability to cool. Second, as gas flows in and assembles in an accretion disk it is highly susceptible to fragmentation. Consequently, the formation of binaries or higher-order multiple stellar systems is the norm of Pop III star formation rather than the exception. Third, when focusing on binary stars and the most bound pairs in a hierarchical system, we expect a wide range of separations, a flat mass ratio distribution, and a roughly thermal spread of orbital eccentricities. Fourth, the resulting stellar mass spectrum is very wide, ranging from low-mass stars, which potentially could have survived until the present day with interesting implications for stellar archaeological surveys (Section \ref{sec:arch}), to very massive stars, which produce copious amounts of UV and LW photons in the early Universe. 
Overall the resulting mass spectrum is likely to be rather flat (in the logarithm of mass) and therefore top-heavy compared to the present-day values. Fifth, developing a better understanding the impact of stellar feedback is one of the key challenges of current research into Pop III star formation. Specifically, radiative feedback is very likely to reduce the number of stars within a star-forming halo and  lead to larger stellar masses (see also Section \ref{sec:small-scale-impact}). However, it can also potentially increase the level of (large-scale) fragmentation in externally irradiated halos (Section \ref{sec:large-scale-impact}).

\subsection{Alternative Pop III star-formation pathways}
\label{sec:alternative-PopIII-pathways}
Here we touch upon alternative Pop III star formation scenarios and speculate about the impact of physical processes that are often neglected in numerical and analytical models. We group our discussion into processes that can promote fragmentation in the primordial gas and those that can reduce the level of fragmentation. We defer the discussion of the most extreme physical conditions, which might lead to the formation of supermassive stars and the seeds of supermassive black holes in the Universe, to Section \ref{sec:supermassive}, and we note that a detailed analysis of the impact of stellar feedback is the focus of Section \ref{sec:feedback}. We also  note in passing that adopting different cosmological models compared to standard $\Lambda$CDM (Section \ref{sec:cosmological-model}) can further modify the picture. It is well understood, for example, that in a warm dark matter scenario star formation happens later and in larger halos \citep{Maio15, Dayal15, Magg2016, Mocz20}. A similar result has been reported for fuzzy dark matter models \citep[e.g.][]{Mocz19}. However, it remains to be seen whether these global changes actually influence the properties of the individual stars that form, and we do not follow up further on this topic here.  

\subsubsection{Enhancing fragmentation}
\label{sec:more-fragmentation}
Several numerical studies indicate that fragmentation also occurs on larger scales in the halo, on scales of the star-forming cloud as a whole \citep{Turk09, Stacy2010, Clark11} rather than just at the level of the central disk, as discussed in Section \ref{sec:disks-and-fragments}. It is typically associated with higher levels of turbulence in the halo gas. The initial turbulent flows that are always present with subsonic or transsonic velocities, get amplified during gravitational collapse and induce density fluctuations that can go into run-away growth in their own right. This is very similar to the turbulence-driven mode of star formation that is dominant at the present day \citep{Scalo04, Elmegreen04, Maclow2004, Mckee2007,Krumholz2015, Klessen2016}. There are two systems, for which this effect is thought to be particularly important: atomic cooling halos and halos that are subject to large relative streaming velocities between baryons and dark matter.  

\paragraph{External irradiation and atomic cooling halos}  Atomic cooling halos are halos with virial temperatures high enough to allow cooling by atomic hydrogen, i.e.\ $T_{\rm vir} > 8000\,$K. In general, we expect Pop III stars to form in a halo once its virial temperature exceeds a few thousand K (see Section~\ref{sec:minimum-mass}) and hence that most atomic cooling halos will be associated with metal-enriched gas. However, in the presence of a sufficiently strong LW radiation field, H$_2$ cooling and star formation can be suppressed in halos with $T_{\rm vir} < 8000\,$K \citep[see e.g.][and also Section \ref{sec:LW}]{Oh2002,Agarwal19}. In that case, cooling will only get underway once the halo has grown to the point that $T_{\rm vir} > 8000\,$K. Then the gas will start to cool quasi-isothermally via Lyman-$\alpha$ emission. During this initial period of cooling, the high gas temperature keeps the critical mass for run-away collapse high (see Section \ref{sec:minimum-mass}). This implies that once collapse becomes possible, the associated infall velocities and rates, which we can infer from Equation~\ref{eq:accretion-rate}, are larger than the ones discussed in Section \ref{sec:initial-collapse}. The high gas temperature also keeps the gas partially ionized, enabling it to form H$_{2}$ rapidly once it becomes 
 dense enough to shield itself from the external LW radiation field, resulting in a transition to efficient H$_2$ cooling and a rapid  temperature drop \citep{Oh2002}. The combination of large velocities and low temperatures means that the turbulence associated with the inflow motion is likely to become trans- and supersonic \citep{Greif2008, Wise2007, Wise2008}. The flow is characterized by cold streams that bring dense material rapidly to the center, where it can efficiently fragment and form stars. It has been suggested that this so-called Pop III.2 mode of primordial star formation leads to a different IMF \citep{Mckee2008, Clark11, Maio2011, Stacy2011}, but overall the results are not fully conclusive. 

\paragraph{Streaming velocity between baryons and dark matter} High levels of turbulence are also expected in regions of large relative streaming velocity between baryons and dark matter (Section  \ref{sec:more-complex-models}). Simulations that include this effect suggest that it reduces the gas overdensity in low-mass halos, delays the onset of cooling, and leads to a larger critical mass for collapse to set in \citep[see e.g.][]{Greif2011, Stacy2011, Maio2011, Naoz2012, Naoz2013, Oleary2012, Latif2014Stream, Schauer17a, Schauer2020, Schauer21}. It may also have substantial impact on the resulting $21\,$cm emission \citep[][see also Section \ref{sec:21cm-and-CMB}]{Fialkov12, McQuinn2012,Visbal2012}. Gas in halos that are subject to large streaming velocities is also more turbulent than gas in more quiescent systems, and so we expect more fragmentation and a bias towards smaller stellar masses \citep{Clark2008}. However, this process has not yet been modeled with sufficient resolution, and so no reliable predictions  of its impact on the IMF exist. 

\subsubsection{Reducing fragmentation} 
\label{sec:less-fragmentation}
Besides stellar feedback, 
which we  address in detail in Section \ref{sec:feedback}, two additional physical processes have been suggested to reduce the level of fragmentation and thus to influence stellar IMF and multiplicity. These are magnetic fields and dark matter annihilation, both of which we discuss below.

\paragraph{Magnetic fields} The presence of dynamically important magnetic fields could significantly alter the picture presented so far. We know that the current Universe is highly magnetized on all scales \citep{Beck1996} and that this influences the birth of stars and the evolution of the interstellar medium.\footnote{For the extreme viewpoint of magnetically mediated star formation, see \citet{Shu1987}.} The properties of the magnetic fields observed today are well explained by a combination of small-scale and large-scale dynamo processes \citep{Brandenburg2005}. In contrast, our knowledge of magnetic fields at high redshifts is very sparse. Theoretical models predict that magnetic fields could be produced in various ways, for example via the Biermann battery \citep{Biermann1950}, the Weibel instability \citep{Lazar2009, Medvedev2004}, or thermal plasma fluctuations \citep{Schlickeiser2003}. Other theories place their origin in cosmological phase transitions or during inflation \citep{Sigl1997, Grasso2001,Banerjee2003, Widrow2012}. The resulting fields are thought to be orders of magnitudes too weak to have any dynamical impact, and so magnetohydrodynamic effects have often been neglected in numerical simulations of primordial star formation \citep[however, see the analytic models of][]{Pudritz1989,Tan2004,Silk2006}. 

This situation has changed with the realization that the small-scale turbulent dynamo can efficiently amplify even extremely small primordial seed fields to the saturation level \citep{Kulsrud1997}, and that this process is very fast, acting on timescales much shorter than the free-fall time. An analytic treatment is possible in terms of the Kazantsev model \citep{Kazantsev1968, Subramanian1998, Schobera, Schoberb}. 
This describes how the twisting, stretching, and folding of field lines in turbulent magnetized flows leads to exponential growth of the field. The amplification timescale is comparable to the eddy-turnover time on the viscous or resistive length scale, depending on the value of the magnetic Prandtl number (the ratio of the kinematic viscosity to the magnetic diffusivity).
Once  backreactions become important, the growth rate slows down, and saturation is reached within a few large-scale eddy-turnover times \citep{Schekochihin2004, Schober2015, Liu22a}. Depending on the properties of the turbulent flow the magnetic energy density at saturation is thought to lie between 1\% and a few 10\% of the kinetic energy density  with a field topology that is highly tangled \citep{Federrath2011, Seta21}. 

Magnetic fields with that strength can strongly affect the evolution of protostellar accretion disks. If the field has a strong polodial component, it can efficiently remove angular momentum from the star-forming gas and reduce its level of fragmentation \citep{Machida2008, Machida2008b, Machida2008c, Machida2013, Bovino2013NJP, Latif2013a,LatifMag2014, Hirano22, Saad2022}, which influences the resulting IMF \citep{Turk2011, Peters2014, Sharda21}. A highly ordered small-scale field can also drive protostellar jets and outflows \citep{Machida2006, Sadanari21}. Finally, the presence of a dynamically significant magnetic field can also change the rotational properties of Pop III stars \citep{Machida07, Stacy13b}, which in turn has consequences for their expected lifetimes and overall luminosity (see Section \ref{sec:small-scale-impact}). One important caveat, however, is that many studies of the impact of the magnetic field assume that the field is highly-ordered on small scales. This is not true initially for a field amplified by the small-scale turbulent dynamo, and although we expect the field to become more ordered over time, it remains unclear how rapidly this occurs, with recent simulations yielding contradictory results \citep{Sharda21,Prole2022b,Stacy2022,Saad2022}

Altogether, we expect Pop. III clusters to have fewer members with somewhat higher masses than predicted by purely hydrodynamic  simulations, such as discussed in Section \ref{sec:standard-scenario-of-PopIII-formation}. However, the details depend very much on the adopted field topology, with initially smooth and homogeneous fields leading to more magnetic braking then highly tangled turbulent fields \citep[e.g.][]{Seifried12, Kuruwita19} as expected to result from the small-scale turbulent dynamo at work \citep[e.g.][]{Seta20}. And so the questions of how magnetic fields influence the overall star-formation process in primordial gas and how they affect the resulting IMF remain subject to very active research.

\paragraph{Dark matter annihilation}
Despite its importance for cosmic evolution and structure formation, the true physical nature of dark matter is still unknown. Many models introduce a new class of  weakly interacting massive particles (WIMPs), as they naturally occur in supersymmetry theories \citep[e.g.][]{Jungman96}. The lightest supersymmetric particle is expected to be stable and to have properties consistent with the phenomenological requirements on dark matter  \citep{Bertone2005}. If these particles are self-annihilating, they will act as an additional source of heating.

In most environments the dark matter density is too low for this to be significant. However, this may be different in the very centers of star-forming halos in the early Universe. Here, the collapse of the baryons may lead to  
adiabatic contraction of the dark matter halo \citep{Blumenthal1986},  increasing its central density by several orders of magnitude. As the annihilation rate scales quadratically with density, the corresponding energy input and ionization rate may become large enough to influence gas dynamics. \citet{Spolyar2008} and \citet{Freese2009} suggest that this process may overcome the cooling provided by H$_2$, and speculate that this could halt gravitational collapse and lead to the formation of so-called dark stars. If dark matter particles also scatter weakly on baryons, these dark stars would be stable for a long time without ever becoming dense or hot enough to initiate nuclear fusion. They would be much larger and more massive than normal Pop.~III stars, with sizes of a few AU, lower surface temperatures, and higher luminosities \citep{Freese2008, Iocco2008A,Iocco2008B, Yoon2008,Hirano2011}. 

There are several problems with this scenario. First, it is not clear whether collapse stalls once the energy input from dark matter becomes comparable to the cooling rate. \citet{Ripamonti2010} argue that this is not the case because the larger heating rate catalyzes further formation of H$_2$ and is compensated by the corresponding larger cooling rate. In addition, chemical cooling due to H$_2$ collisional dissociation can help to balance the additional heating after only a moderate increase in the gas temperature \citep{Smith2012}. Second, the implicit assumption of perfect alignment between dark matter cusp and gas collapse is most likely violated in realistic star formation conditions. Three-dimensional simulations \citep{Stacy2012, Stacy2014} clearly demonstrate that the presence of non-axisymmetric perturbations leads to a separation between dark matter cusp and collapsing gas, rendering the annihilation energy input insignificant for dark star formation. However, it is still possible that dark matter annihilation influences the dynamics of the accretion disk in the halo center and that the energy input associated with this process leads to a suppression of disk fragmentation \citep{Smith2012}. We conclude that dark matter annihilation may be able to reduce the multiplicity of metal-free stars and increase their overall mass, but we also note that the existing studies are still premature and it is too early for a reliable assessment of the impact of this process on the IMF of Pop III stars.

\subsection{Supermassive stars and black holes}
% {\sl (1350 words)}
\label{sec:supermassive} 
The existence of extremely bright quasars at redshifts $z \gtrsim 7$ 
\citep[see e.g.][]{Fan06b, Wu15,Banados18,Wang21} implies the presence of supermassive black holes (SMBH) with masses of $M_{\bullet} \sim 10^9\,$M$_{\odot}$ and above. This finding is in tension with physical models in which the seeds of SMBH start out light and then grow gradually by Eddington-limited accretion. The standard Pop III star-formation pathway discussed in Section \ref{sec:standard-scenario-of-PopIII-formation} produces black holes at the end stage of stellar evolution with $M_{\bullet} < 10^2\,$M$_{\odot}$ (see also Section \ref{sec:sne}). These small seeds do not have enough time to grow to the observed large masses over the age of the Universe ($\sim 0.7\,$Gyr at $z=7$, assuming standard $\Lambda$CDM cosmology). In addition, the assumption of persistent Eddington-limited accretion is itself questionable. Stellar feedback from Pop III stars is very efficient at removing gas from their birth sites, leaving little to be accreted by their black hole remnants, and the motion of these black holes within their parent minihalos can also be substantial, further reducing their accretion rate \citep{Smith2018}.
Therefore, more extreme formation scenarios need to be considered. We briefly review a few of the most important models here and refer the reader to \citet{Woods19}, \citet{Latif19}, or \citet{Inayoshi20} for more comprehensive reviews.

\paragraph{Need for high initial infall rates}
If we assume spherical symmetry and if we furthermore assume that the opacity of the material falling onto the black hole is given by Thompson scattering on free electrons, then the released radiation must not exceed the Eddington luminosity, 
\begin{equation}
 L_{\rm Edd} = 4 \pi G M_{\bullet} \mu_{\mathrm{e}} m_{\rm p}c / \sigma_{\rm T} \;, 
 \label{eq:Eddington-luminosity}
\end{equation}
where $m_{\rm p}$ is the proton mass, $G$ is the gravitational constant, $\mu_{\mathrm{e}}$ denotes the effective number of nuclei per free or loosely bound electron and depends on the chemical state of the gas and the spectrum of the radiation field (see also the discussion below Equation~\ref{eq:Jeans-mass}), $c$ is the speed of light, and $\sigma_{\rm T}$ is the Thomson scattering cross section. Otherwise radiation pressure would be too strong to allow for accretion. This can be converted into a characteristic timescale by comparing with the available energy reservoir in the system coming from the rest mass energy of the black hole, $\tau_{\rm Edd} = M_{\bullet}c^2 / L_{\rm Edd} = \sigma_{\rm T}c / (4 \pi G \mu_{\mathrm{e}} m_{\rm p})$,  which is independent of the black hole mass. For typical conditions in primordial gas, $\tau_{\rm Edd} \sim 0.4\,$Gyr. 
The time it takes a black hole to grow via Eddington-limited accretion from an initial mass  $M_{\bullet,i}$ to a final mass $M_{\bullet,f}$ is given by 
\begin{equation}
\tau_{\bullet} = \dfrac{\tau_{\rm Edd}}{f_{\rm Edd}} \dfrac{\epsilon}{1-\epsilon} \ln \left(\dfrac{M_{\bullet,f}}{M_{\bullet,i}} \right)\;,    
\label{eq:Eddington}
\end{equation}
where $\epsilon$ is the accretion efficiency, typically $\sim 0.1$ for accretion through a thin disk \citep{Shakura73}, and where $f_{\rm Edd}$ expresses the fraction of time for which the black hole is able to accrete at the full Eddington rate \citep[for a comprehensive review, see][]{Inayoshi20}.
To build a SMBH of $10^9\,$M$_{\odot}$ at $z=7$ starting from an initial seed mass of $10^2\,$M$_\odot$ requires $\tau_{\bullet} \approx 0.8\,$Gyr, which is longer than the age of the Universe at this redshift, even if we make the unrealistic assumption that $f_{\rm Edd} = 1$ throughout. Reducing $f_{\rm Edd}$ makes this discrepancy even larger.
In principle, accretion at a rate faster than the Eddington limit is possible if the accretion flow strongly deviates from spherical symmetry and material gets delivered to the black hole in a highly filamentary fashion, or alternatively, if the accretion disk is radiatively inefficient or the accreting envelope emits anisotropically \citep[e.g.][]{Mayer19}. However, the inferred small quasar duty cycles would require rates considerably above the Eddington limit during the periods of accretion. This seems unlikely, and so the preferred scenario for the formation of SMBH at high redshifts is to start out with more massive seeds, i.e.\ to increase $M_{\bullet,i}$ to a much larger value.

To build such a massive seed object requires very high accretion rates, which in turn requires extreme environmental conditions. Clearly, the halo in which this happens needs to be massive enough to contain a sufficient amount of gas. To form a seed with a mass of $\sim 10^{5} \, {\rm M_{\odot}}$, we therefore need a halo with a mass $M \gg 10^{7} \, {\rm M_{\odot}}$, unless we assume that an improbably large fraction of the gas ends up in the seed black hole. This is much larger than the critical halo mass required for Pop III star formation and hence points to scenarios in which collapse and star formation is delayed in some fashion. \citet{Inayoshi20} review the different models that have been suggested to explain this, ranging from irradiation of the halo by a high flux of LW photons \citep[e.g.][]{Agarwal2012} to dynamical heating of the gas by repeated major mergers \citep[e.g.][]{Mayer19}.

Once we have physical conditions that allow a large amount of gas to flow into the center of an appropriately massive halo at rates of $\dot{M} \sim 0.1\,$M$_\odot\,$yr$^{-1}$ or higher, possibly reaching up to $1000\,$M$_\odot\,$yr$^{-1}$ \citep{Zwick23}, 
the next question is whether this gas feeds the growth of a single object, or whether the gas fragments and forms multiple objects. We briefly consider both possibilities below.

\begin{figure}[tp]
\begin{center}
     \includegraphics[width=8cm]{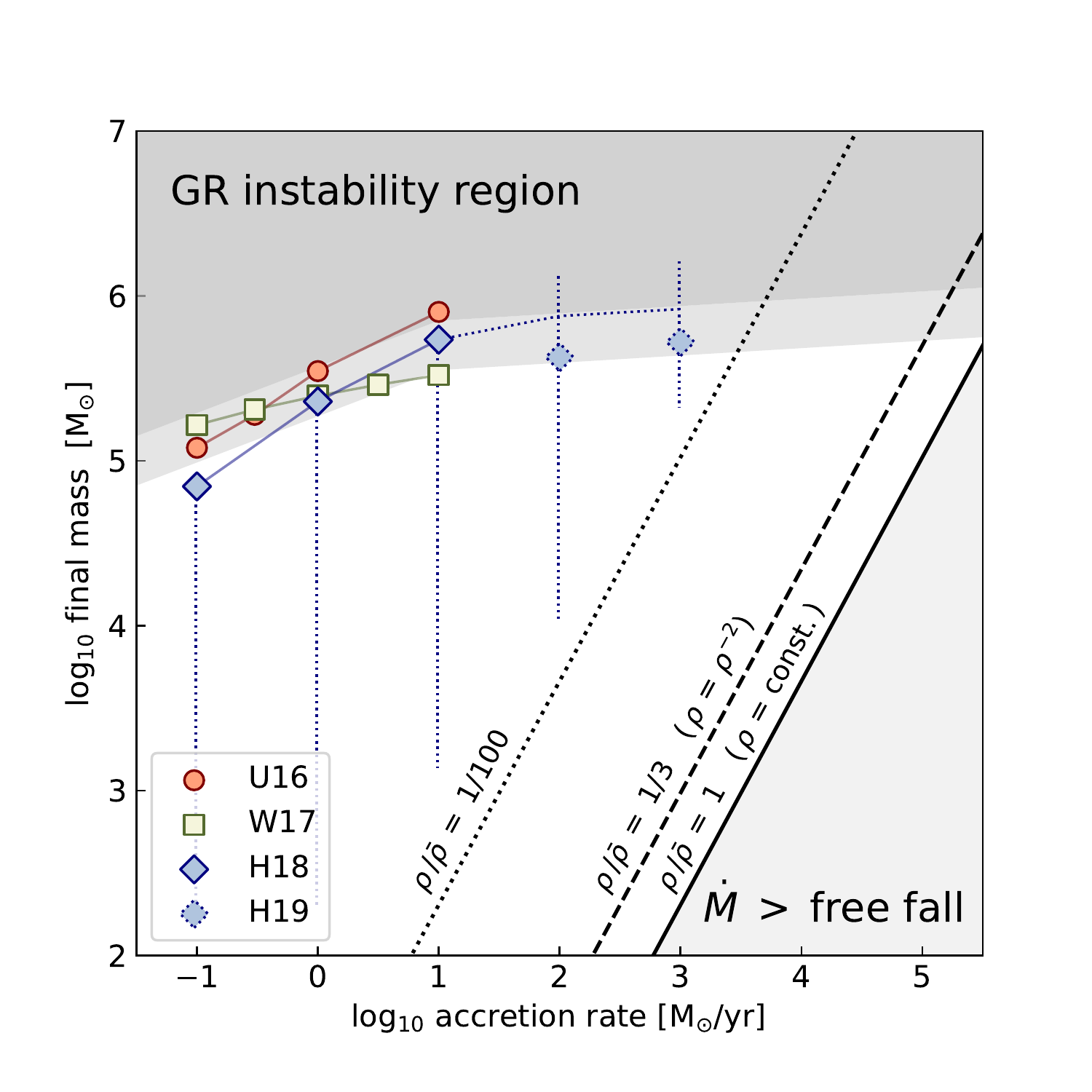}
\end{center}
\caption{Maximum mass reachable by supermassive stars accreting at different rates until the onset of the GR instability triggers collapse to a black hole of the same mass. The colored symbols depict the final masses in the studies of \citet{Umeda2016}, \citet{Woods2017}, and \citet{Haemmerle18} for accretion rates up to $\dot{M} = 10\,$M$_\odot\,$yr$^{-1}$. We also report results from \citet{Haemmerle19} for $100\,$M$_\odot\,$yr$^{-1}$ and $1000\,$M$_\odot\,$yr$^{-1}$, but note that the calculations terminated before the GR instability sets in, and so the expected final masses are higher than shown here, as indicated by the tentative extension of the accretion path (vertical dotted lines). We also indicate the region forbidden by the GR instability as derived from analytic estimates \citep[see][]{Haemmerle20, Haemmerle21c}. The tilted lines show the evolution in $M$ vs.\ $\dot{M}$ space for the free-fall collapse of gas spheres with different initial density profiles, characterized by the contrast $\rho / \bar{\rho}$ between the density of the infalling gas and the mean density of the accreting SMS, following the mass-radius relation for high accretion rates (Equation~\ref{eq:MR-relation-high-Mdot}). The figure is inspired by \citet{Haemmerle21b}.
}
\label{fig:max-SMS-mass}
\end{figure}

\paragraph{Supermassive stars}
If we assume that only a single object forms -- a scenario often referred to as direct collapse -- then 
we can ask about the properties of this object and about the maximum permitted mass. These questions can be addressed with detailed stellar evolution calculations that include a treatment of accretion \citep[e.g.][]{Stahler1986, Behrend01, OmukaiPalla2003, Hosokawa2009, Haemmerle16}. They show that stable supermassive stars (SMS) are possible and can reach maximum masses of up to several $10^5\,$M$_\odot$ \citep{Umeda2016, Woods2017, Haemmerle18}, but always stay below $10^6\,$M$_\odot$ for the accretion rates reachable in atomic cooling halos \citep{Haemmerle19, Haemmerle20, Haemmerle21c, Haemmerle21}. This value is determined by the onset of the general-relativistic (GR) instability, which drives the object into collapse to form a black hole with the same mass (see \citealt{Appenzeller1972a, Appenzeller1972b} or \citealt{Fuller1986} for  early models, or \citealt{Haemmerle21b} for a more recent account). 
In Figure~\ref{fig:max-SMS-mass} we report estimates of the maximum mass from different sets of stellar structure and evolution calculations for different accretion rates 
and from simplified analytic models. 
The stellar structure calculations indicate that primordial SMS evolve as red supergiant protostars \citep{Hosokawa12, Hosokawa2013, Haemmerle18}, with extended radii that follow the relation given by Equation \ref{eq:MR-relation-high-Mdot} and surface temperatures of only $\sim 5000\,$K. Their internal structure consists of a convective core, a radiative zone containing most of the stellar mass, and a convective envelope that covers a dominant fraction of the photospheric radius. To a good approximation, these structures can be described as hylotropes \citep{Begelman2008, Begelman2010}, in particular for accretion rates $\gtrsim10\,$M$_\odot\,$yr$^{-1}$. Although very luminous, their low surface temperatures do not allow them to emit large amounts of ionizing photons. Consequently, they are not able to create extended HII regions (Section \ref{sec:small-photoionization}), which might limit the overall mass growth or affect star formation in neighboring halos (Section \ref{sec:large-photoionization}).

\paragraph{Dense clusters} 
As argued in Section \ref{sec:disks-and-fragments}, it is difficult to prevent primordial gas from fragmenting. This is well established in the standard Pop III formation scenario, but also appears to hold in highly irradiated atomic cooling halos where H$_2$ is suppressed \citep[e.g.][]{Agarwal2012,  Sugimura2014, Agarwal2015, Latif2015, Latif2020, Regan2018}. It becomes even harder to prevent fragmentation in the presence of metals or dust, owing to the additional cooling channels that they provide \citep{omukai2008,Latif2016dust,Chon20}. Therefore, rather than forming a single star, it is plausible that the rapid accretion flows discussed here instead feed the growth of a dense cluster of objects, each 
accreting at its own pace. In this scenario, SMBH seeds result from runaway collisions between stars in this dense environment \citep[e.g.][]{PortegiesZwart00, PortegiesZwart02, Sesana05, Devecchi2009, Devecchi2012,   Katz2015, Sakurai2017, Reinoso18, Reinoso2020, Escala21, Vergara21} or between their black hole remnants \citep{Davies2011, Lupi14, Antonini2019, Kroupa20}. Although many models have treated this as a purely stellar dynamical problem, the difficulty of disrupting these high accretion flows means that in practice one should simultaneously take gas dynamics and stellar dynamics into account. In this context, \citet{Davies2011} and \citet{Lupi14} argue that the inflow of gas into an interacting cluster of stellar mass black holes is needed to steepen the gravitational potential and make mergers more likely than three-body ejections. Focusing on embedded star clusters, \citet{Boekholt2018} demonstrate that the combination of accretion and collisions can lead to masses up to $\sim10^5\,$M$_\odot$. The semi-analytic models of \citet{Tagawa2020} consider the growth of a supermassive object via stellar bombardment in the presence of gas. Other models study the impact of different accretion prescriptions \citep{Das2021a}, or of mass loss occurring in mergers \citep{Alister2020} or associated with stellar winds \citep{Das2021b}. The impact of varying the metallicity of the gas has also been investigated 
\citep{Chon20, Schleicher2022}. 

The emerging picture is that fragmentation leads to the formation of a dense and deeply embedded cluster in which  competitive accretion of individual cluster members in concert with frequent merger events results in the run-away growth of a small number of objects. Depending on the environmental conditions and on the detailed implementation of the physical processes considered, masses of order of $M_\bullet \sim 10^5\,$M$_\odot$ are easily within reach, and it is reasonable to speculate that the internal structure of these very massive run-away objects is similar to that of the SMS discussed above. This is supported by calculations adopting highly time-varying accretion rates \citep{Woods21, Woods2021b} and considering SMS mergers as particularly extreme forms of accretion spikes. They collapse into massive black holes once this run-away growth phase ends or once they reach the mass limit for the general relativistic instability.

\section{Feedback from Pop III and transition to Pop II}
\label{sec:feedback}
The formation of Pop III stars is associated with a range of different feedback processes that affect the gas around them. Radiative and mechanical feedback in various forms influences the star formation efficiency of the gas on both small and large scales, and chemical feedback, i.e.\ the enrichment of the Universe with metals from the first supernovae, drives the transition from Pop III to Pop II star formation. In this section, we review the most important feedback processes. We begin with radiation (Section \ref{sec:radiation-from-PopIII}),  and successively include other forms of stellar feedback,  focussing first on their effects on gas close to the stars (Section~\ref{sec:small-scale-impact}) and later on their effects on much larger scales (Section~\ref{sec:large-scale-impact}). We also briefly review the physics of the Pop III to Pop II transition and the most important open questions associated with this (Section~\ref{sec:transition-PopIII-to-PopII}).

\subsection{Radiation from Pop III stars}
\label{sec:radiation-from-PopIII}
Here we summarize the properties of the radiation produced by primordial stars, which sets the stage for the discussion of the impact of radiative feedback on the star-forming cloud itself and on neighboring halos.  

\subsubsection{Accretion luminosity}
\label{sec:acc-lum}
The first form of feedback to become important in a star-forming minihalo is the accretion luminosity generated by gas accreting onto newly-formed protostars. In principle, this represents a considerable reservoir of energy, owing to the high accretion rates typical in systems forming Pop III stars. A protostar with a mass $M_{*}$ and radius $R_{*}$ that is accreting gas at a rate $\dot{M}$ will produce an accretion luminosity
\begin{equation}
L_{\rm acc} = \alpha \frac{G M_{*} \dot{M}}{R_{*}},
\end{equation}
where $\alpha$ is a dimensionless efficiency factor that depends on the geometry of the accretion flow \citep{Stacy2016}, with high resolution simulations  \citep[e.g.][]{Wollenberg20,Jaura22} suggesting values of $\alpha$ close to unity. Accretion rates onto Pop III protostars can vary substantially from star to star, but at early times, values as high as $10^{-3} - 10^{-2} \, {\rm M_{\odot} \, yr^{-1}}$ are commonly encountered (Section \ref{sec:disks-and-fragments}). For a low mass pre-main sequence Pop III star with $M_{*} = 1 \, {\rm M_{\odot}}$ and $R_{*} = 20 \, {\rm R_{\odot}}$ \citep{OmukaiPalla2003}, this corresponds to accretion luminosities in the range $L_{\rm acc} \sim 2000 - 20000 \, {\rm L_{\odot}}$, which is much higher than the main sequence luminosities of these same stars. Accretion onto more massive Pop III protostars produces even higher luminosities, with $L_{\rm acc}$ scaling approximately as $L_{\rm acc} \propto M_{*}^{0.73}$ for fixed $\dot{M}$ \citep{Smith2012}.

Despite this, the effect of this radiation on the surrounding gas is relatively modest. Rapidly accreting Pop III protostars have photospheric temperatures of at most $\sim 6000$~K \citep{Stahler1986b,OmukaiPalla2003} and hence radiate most of their energy at visible and infrared wavelengths where the continuum opacity of metal-free gas is very small \citep{2005MNRAS.358..614M}. Therefore, only a tiny fraction of the radiated energy is absorbed by the gas in the minihalo, with most escaping into the intergalactic medium (IGM). The heating that this provides to the minihalo gas moderately reduces its propensity to fragment \citep{Smith2011} but otherwise has little impact on its evolution.

\subsubsection{Stellar luminosity}
\label{sec:stellar-lum}
At later times, once the newly formed Pop III stars have joined the main sequence (MS), their intrinsic luminosity becomes more important than their accretion luminosity, specifically when talking about high-mass stars above $\sim 10\,$M$_\odot$. Even on the MS, if the stars continue to accrete the geometry of the flow  has strong impact on the internal structure  \citep{Hosokawa2012}. There are two extreme cases that we can consider. Accretion through a geometrically thin disk-like structure delivers fresh material with relatively low entropy. It settles onto the stellar surface  with the same specific entropy as the photosphere, which allows the Pop III star to retain a relatively compact configuration. The total stellar luminosity $L_{*}$ relates to the  radius $R_{*}$ and effective surface temperature $T_{*}$ as 
\begin{equation}
L_{*} = 4 \pi \sigma_{\rm SB} R_{*}^2 T_{*}^4\;,  
\label{eq:L-R-T}
\end{equation}
with $\sigma_{\rm SB}$ being the Stefan-Boltzmann constant. Consequently,  a star subjected to  cold disk accretion is relatively hot and compact. If the accretion flow is spherically symmetric, the infalling material retains the entropy produced in the accretion shock front, and the star responds to the delivery of high-entropy material by increasing its radius. This implies that the surface temperature must drop, and so stars experiencing spherical hot accretion tend to be relatively cool and bloated \citep{Hosokawa2012}.

\begin{figure}[tp]
\setlength{\unitlength}{1cm}
\begin{picture}(15,7)(0,0)
  \put(-3.5, 0.5){\includegraphics[width=6cm]{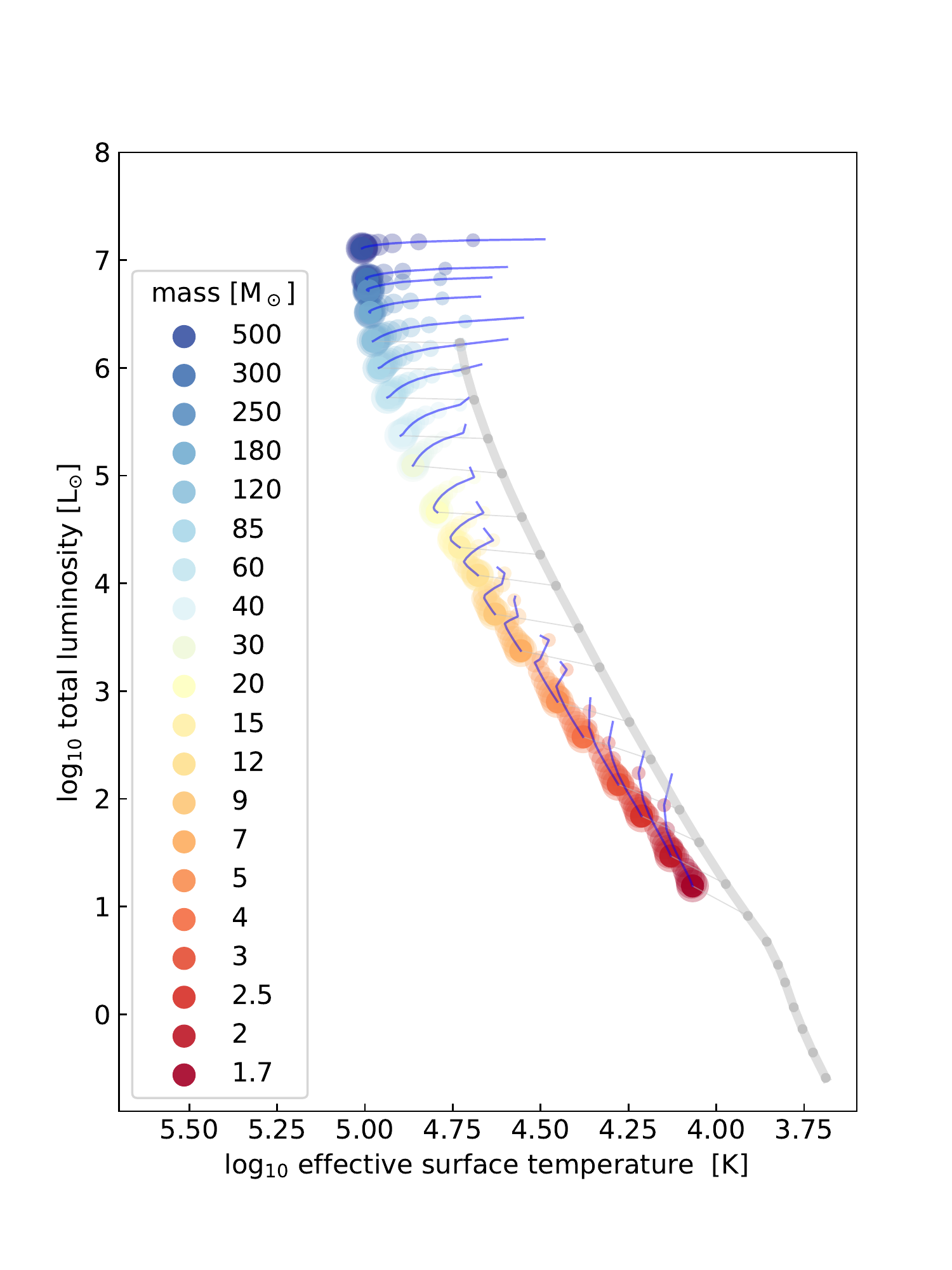}}
  \put( 3.0, 0.5){\includegraphics[width=8cm]{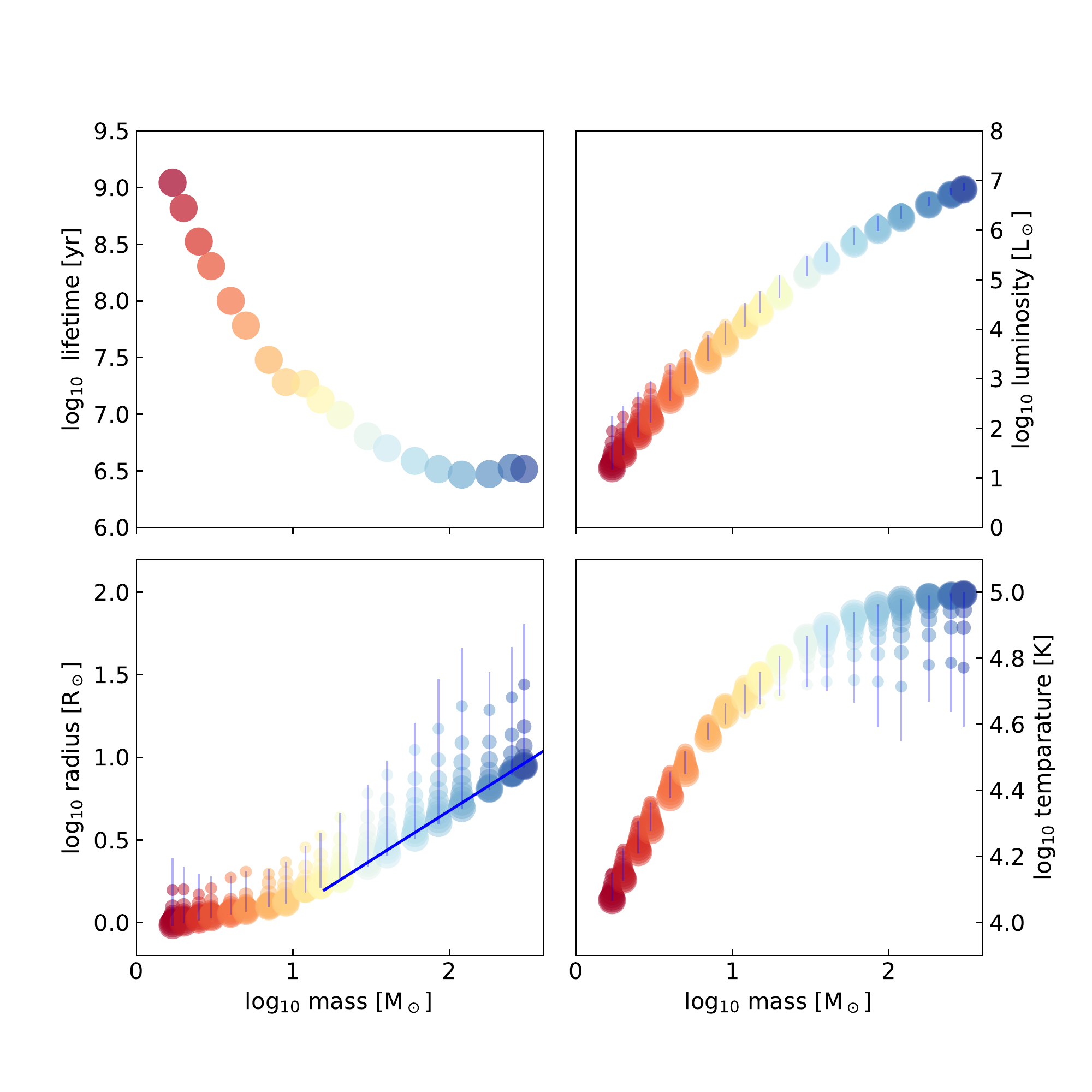}}
%  \put(-3.5, 0.5){\includegraphics[width=6cm]{Figures/fig-PopIII-HRD-v1.pdf}}
%  \put( 3.0, 0.5){\includegraphics[width=8cm]{Figures/fig-PopIII-key-data-v1.pdf}}
\end{picture}
\caption{Summary of key properties of Pop III stars in the mass range  $M_{*} = 1.7\,$M$_\odot$ to $500\,$M$_\odot$. To the left we show the stellar loci in the Hertzsprung-Russell diagram (log$_{10}\,L$ vs.\ log$_{10}\,T$) across their MS and post-MS phases. For comparison, we also show the ZAMS of solar metallicity stars in the range of  $0.8\,$M$_\odot$ to $120\,$M$_\odot$. To the right, we present the numbers for stellar lifetime, stellar radius, total luminosity, and effective surface temperature, each as a function of mass. The symbol size indicates the evolutionary stage, with the largest one indicating the ZAMS  and the smallest the post-MS giant phase, and with a step size that corresponds to 10\% of the total stellar lifetime. The solid blue line illustrates the mass-radius relation for massive main-sequence Pop III stars (Equation \ref{eq:MR-relation-low-Mdot}).  The underlying data are taken from \citet{Murphy21a, Murphy21b} and Martinet et al.\ (in prep.) for zero metallicity, and from \citep[from][]{Ekstroem12} for solar metallicity. More details are provided in the Appendix. 
}
\label{fig:PopIII-HRD}
\end{figure}

%% MRNT
\begin{marginnote}[]
\entry{ZAMS}{The zero-age main sequence of stars indicates the onset of core hydrogen burning. As more hydrogen gets converted into helium the star evolves on the main sequence and becomes somewhat hotter and brighter.  }
\end{marginnote}

The exact values of the stellar radius $R_{*}$, effective surface temperature $T_{*}$ and total luminosity $L = L_{\mathrm{acc}} + L_{*}$ vary with time and strongly depend on the instantaneous accretion rate $\dot{M}$ and mass $M_{*}$  \citep{Hosokawa2009, Hosokawa2010, Hosokawa12, Haemmerle18}.  However, once the stars have finished accreting, they settle onto a well-defined main sequence. Pop III stars on this main sequence are more compact than their present-day counterparts, with a mass-radius relation at the high mass end given roughly by 
 \begin{equation}
R_{*} \sim 0.3\;{\rm R}_\odot \left(\dfrac{M_{*}}{{\rm M}_\odot} \right)^{0.6}\;.
\label{eq:MR-relation-low-Mdot}     
 \end{equation}
They are therefore somewhat hotter than their present-day counterparts
\citep{Maeder2012}. We illustrate key properties of Pop III stars for masses ranging from $M_{*} = 1.7\,$M$_\odot$ to $500\,$M$_\odot$ in Figure \ref{fig:PopIII-HRD}. The values are based on the stellar evolution models computed by \citet{Murphy21a, Murphy21b} and  Martinet et al.\ (in prep.) using the Geneva code \citep{Eggenberger08}. To the left we plot the location of the stars in the log$_{10}\,L$-log$_{10}\,T$ diagram during the main-sequence (MS) and post-MS evolution. For comparison, we also provide the zero-age main sequence \citep[ZAMS, see e.g.][]{Kippenhahn2012} of solar metallicity stars \citep[from][]{Ekstroem12} in the mass range from $0.8\,$M$_\odot$ to $120\,$M$_\odot$. This demonstrates that primordial and present-day stars of the same initial mass have similar total luminosity during the MS evolution, before mass loss becomes important for solar metallicity stars. However, due to their much higher compactness, $T_{*}$ is considerably larger. To the right we visualize the stellar lifetime, as well as the stellar radius (with the blue line justifying Equation~\ref{eq:MR-relation-low-Mdot}), the total luminosity, and the effective surface temperature as function of mass at various evolutionary phases. The largest symbol depicts the stars at the ZAMS and the smallest one in their giant phase. Altogether we provide these  properties in steps of 10\% of the total stellar lifetime. We note that different stellar evolutionary models and codes \citep[e.g. SEVN, see][]{Spera22} lead to very similar numbers. Further information can be found in the  Appendix  together with tables of the fluxes of ionizing and non-ionizing photons, and the total number of photons produced per baryon, including also population-averaged values.  

In summary, the existing models strongly suggest that high-mass Pop III stars are very compact with high  photospheric temperatures \citep{OmukaiPalla2003}, and therefore radiate a large fraction of their energy at ultraviolet wavelengths. Once Pop III stars with masses greater than about $10-20\,$M$_\odot$  have formed, they become powerful sources of ionizing photons. These photons can have a profound effect on the surrounding gas, as we explore below in Sections \ref{sec:small-scale-impact} and \ref{sec:large-scale-impact}.

\subsubsection{Influence of rapid accretion on the stellar spectrum}
\label{sec:rapid-accretion}
The simple picture sketched above assumes that once a Pop III star reaches the main sequence, it remains there in a compact state. This is a reasonable assumption if the accretion rate onto the star is less than $\sim 10^{-2} \, {\rm M_{\odot} \, yr^{-1}}$  \citep[e.g.][]{Hosokawa2009, Hosokawa2010, Maeder2012}.  However, if the accretion rate onto the star becomes larger, this triggers the expansion of the outer envelope of the star to a red supergiant  state, in which the stellar radius is given by \citep{Hosokawa12, Hosokawa2013, Haemmerle18}
\begin{equation}
R_{*} = 2.6 \times 10^{2} \;{\rm R}_\odot \left(\frac{M_{*}}{{\rm M}_{\odot}} \right)^{1/2} \;.
\label{eq:MR-relation-high-Mdot} 
\end{equation}
This can be analytically derived by taking the Eddington luminosity, 
which is realistic for photon-dominated massive stars \citep[e.g.][]{Kippenhahn2012}, adopting an effective surface temperature of $T_{*} \sim 5000\,$K, which is appropriate for red supergiants and which turns out to be roughly constant across a wide range of masses, and inserting both values into Equation~\ref{eq:L-R-T}. For comparison, the radius of a zero-age main sequence Pop III star with $M_{*} = 100 \, {\rm M_{\odot}}$ is $R_{*} \approx 5 \, {\rm R_{\odot}}$ according to Equation~\ref{eq:MR-relation-low-Mdot}. The intrinsic luminosity of the star is not strongly affected by this change in its structure, and Equation~\ref{eq:L-R-T} implies that the increase of $R_{*}$  leads to a substantial drop of $T_{*}$ down to values of $\sim 5000\,$K, remaining relatively constant across a wide range of masses. This leads to a massive drop in the ionizing photon output of the star \citep{Hosokawa2016}. If the Pop III star is located in an environment where an accretion rate of $10^{-2} \, {\rm M_{\odot} \, yr^{-1}}$ or above onto the star can be maintained for an extended period, the star never develops an HII region. Therefore, photoionization feedback never has the opportunity to interfere with accretion onto the star, and it can grow to become extremely massive (as discussed in   Section~\ref{sec:supermassive}).

Furthermore, even if the mean accretion rate onto the star is less than $10^{-2} \, {\rm M_{\odot} \, yr^{-1}}$, it can still grow to reach a supergiant state if the accretion is highly episodic, with periods where the accretion rate exceeds the critical value \citep{Vorobyov13}. In this case, a period of rapid accretion triggers the expansion of the outer envelope of the star, but once the accretion rate falls below the critical value, the star contracts back to its original size. This process can happen repeatedly during the growth of the star \citep{Hosokawa2016} and leads to the repeated appearance and disappearance of the surrounding HII region, in a fashion similar to the ``flickering'' found in simulations \citep[e.g.][]{Peters2010,Galvan2011} and observations \citep[e.g.][]{DuPree2015} of present-day HII regions.

How common this behavior is in real systems depends on how often young Pop III stars encounter such high accretion rates. This is difficult to predict robustly, because it depends not only on the larger-scale environment in the halo, but also on the details of the disk fragmentation, the proportion of fragments that merge, etc. Simulations of this yield numerically converged results only when run with very high resolution \citep{Prole22} and as yet very few such simulations have been run for long enough to also capture the formation of massive stars (see the discussion in Section \ref{sec:IMF-and-multiplicity}).

\subsection{Small-scale impact}
\label{sec:small-scale-impact}
In this section, we summarize the impact that feedback from Pop III stars has on gas close to the stars, which we here take to mean gas within the same dark matter halo. The impact of Pop III stellar feedback on much larger scales is discussed in Section~\ref{sec:large-scale-impact} below.

\subsubsection{Photoionization} 
\label{sec:small-photoionization}
As argued above, high-mass Pop III stars are thought to produce copious amounts of ionizing radiation. Consequently, we begin our discussion of stellar feedback by investigating the impact of photons with energies above $13.6\,$eV. 

\begin{figure}[htp]
\includegraphics[width=\textwidth]{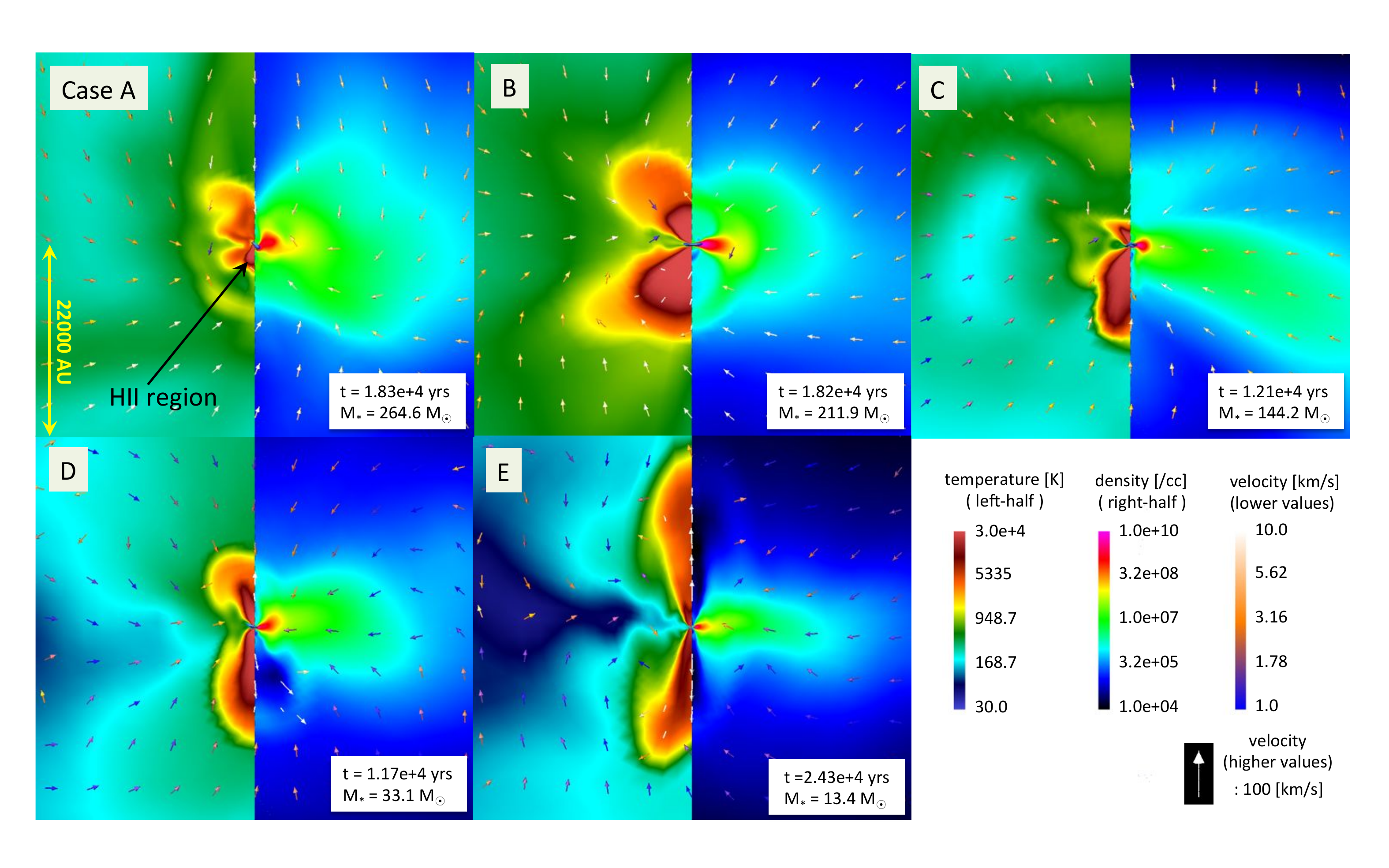}
\caption{Illustration of the bipolar expansion of HII regions around massive Pop III stars. Five examples from the simulations of \citet{Hosokawa2016} are shown. In each case, the Pop III star is located at the center of the double panel and the temperature, density and velocity are shown in the plane containing the polar axis of the disk. Velocities below $10 \, {\rm km \, s^{-1}}$ are denoted by colour-coded arrows of fixed size; those above $10 \, {\rm km \, s^{-1}}$ are denoted by white arrows, with lengths varying in proportion to the velocity. All five HII regions are shown at the time at which their polar extension first reaches $10^{4}$~AU. This time is indicated in the figure for each HII region, along with the mass of the Pop III star at this moment. Figure is from \citet{Hosokawa2016} and reproduced here with permission of the authors and ApJ. 
}
\label{fig:bipolar}
\end{figure}

\paragraph{Impact on surroundings}
The ionizing photons produced by a massive Pop III star are readily absorbed in the surrounding gas, leading to the formation of an HII region. The evolution of this HII region once it breaks out of the dense protostellar accretion disk has been modelled by a number of different groups \citep[e.g.][]{Whalen2004,WiseAbel2008,Stacy2016,Hosokawa2016} and is reasonably well understood. Because of the high densities found in the disk, the gas there resists photoionization and so the HII region initially expands in a bipolar fashion above and below the disk midplane, as illustrated in Figure~\ref{fig:bipolar}. The gas within the HII region is hot -- temperatures of $1 - 2 \times 10^{4}$~K are typical \citep{Whalen2004} -- and highly over-pressured compared to the surrounding neutral gas. It therefore expands outwards, driving a shock in front of it and leaving behind a low density cavity. The expansion of the HII region reduces the flow of gas onto the protostellar accretion disk, since gas can now reach the disk only by moving inwards close to the midplane. In addition, the disk itself begins to be photoevaporated by the radiation from the star \citep{Mckee2008}. Together, these effects lead to the loss of the disk and the termination of accretion onto the star after a few $10^{4} \, {\rm yr}$ \citep{Hosokawa2016,Sugimura20}.
%% MRNT
\begin{marginnote}[]
\entry{HII region}{~~~~~ Bubbles of  ionized gas surrounding high-mass stars or massive star clusters caused by radiative stellar feedback. }
\end{marginnote}

The final fate of the HII region depends upon its location. In a minihalo, the fact that the sound speed of the photoionized gas exceeds the escape velocity of the minihalo means that the HII region will continue to expand for as long as the Pop III star continues to emit ionizing photons. For minihalos with masses close to $M_{\rm crit}$, the end result is that a large fraction of the gas mass is driven out beyond the virial radius \citep{Kitayama2004,Whalen2004,Alvarez2006,Abel2007}. In more massive minihalos, although the ionized gas is too warm to remain gravitationally bound to the minihalo, the larger size of the system means that a single HII region is unlikely to drive out all gas. Finally, in protogalaxies with $v_{\rm esc} \sim 20 \, {\rm km \, s^{-1}}$ or more, the ionized gas remains gravitationally bound to the protogalaxy, which therefore retains most of the gas. For a similar reason, the time-averaged escape fraction of ionizing photons from the HII region, i.e.\ the fraction of photons that can can travel into the IGM, also depends strongly on the location of the HII region. In small minihalos, it can reach values of 50\% or more \citep[see e.g.][]{Kitayama2004,Alvarez2006,wise2014,schauer15}, while in more massive galaxies, a broader range of values is possible \citep[see e.g.][]{Paardekooper2013,Dayal2018}.

\paragraph{HII region trapping}
An additional complication has been highlighted by \citet{Jaura22}. Most numerical studies of the growth of Pop III HII regions implicitly assume that the HII region has already escaped from the dense protostellar accretion disk by choosing an injection region for the ionizing photons that is larger than the disk scale height near the star.\footnote{The studies by \citet{Stacy2016} and \citet{Sugimura20} do not make this assumption. Instead, 
they each adopt a different sub-grid model to treat the behavior of the HII region on scales that are not resolved in their calculation. However, neither of these subgrid models properly captures the behavior of the HII region near the star, as discussed  in \citet{Jaura22}.} \citet{Jaura22} show that if one instead injects the photons at the location of the star itself, then the HII region remains trapped in the dense gas and does not break out of the disk. The physical reason for this is that at the high densities found close to the star, the Str\"omgren radius is much smaller than the Bondi-Parker radius at which the thermal pressure of the ionized gas first starts to dominate over the gravitational attraction of the star. Therefore, the ionized gas remains gravitationally bound to the star, preventing it from flowing outwards and escaping \citep{keto_formation_2007}. This result holds regardless of whether or not one accounts for the radiation pressure exerted on the gas by the ionizing photons, as this is too weak to overcome gravity close to the star. However, \citet{Jaura22} also note that there are two important physical processes that are not included in their model that might change this picture. First, their simulations do not include a magnetic field and hence are unable to capture the formation of a magnetically powered jet or outflow. If such an outflow forms, then this will remove dense gas from the vicinity of the star, potentially allowing the HII region to break out from the disk \citep{Mckee2008}. Second, the high optical depth of the accretion disk to Lyman-$\alpha$ radiation means that any of the photons produced by the star or in the HII region will scatter repeatedly before escaping. The resulting radiation pressure is orders of magnitude larger than the direct radiation pressure of the ionizing photons and hence may be sufficient to overcome the gravitational attraction of the star \citep{Mckee2008,Jaura22}. Modelling this will require a careful treatment of the interaction between the radiation field and the gas in the dense disk environment, and should be a priority for future simulations.

\subsubsection{Photodissociation}
\label{sec:small-pdiss}
As well as producing a large number of ionizing photons, massive Pop III stars also emit many photons in the Lyman-Werner bands of H$_{2}$ ($11.2 -13.6\,$eV). Because of the crucial importance of H$_{2}$ cooling in Pop III star formation,  photodissociation of H$_{2}$ by these photons is 
potentially an important form of negative feedback. 

At high densities (e.g.\ in the immediate vicinity of the disk), the photodissociation region (PDR) created by these photons is restricted to a narrow region just outside of the HII region. This is because as dissociating photons penetrate into the gas ahead of the ionization front (I-front), they build up a dense layer of atomic hydrogen. Once the column density of this layer exceeds $N_{\rm H} \sim 10^{24} \, {\rm cm^{-2}}$, absorption of LW photons in the damping wings of the atomic hydrogen Lyman series lines starts to be highly effective \citep{Wolcott2011}. This shields the gas ahead of the dense atomic layer and limits its maximum column density to $\sim 10^{25} \, {\rm cm^{-2}}$. In the density regime where three-body H$_{2}$ formation is important ($n > 10^{11} \, {\rm cm^{-3}}$; see Figure~\ref{fig:initial-collapse}), the maximum spatial extent of this layer is only $L_{\rm max} \sim 10^{14}\, n_{11}^{-1} \, {\rm cm} \approx 10\, n_{11}^{-1} \, {\rm AU}$, where $n_{11}$ is the gas density in units of $10^{11} \, {\rm cm^{-3}}$. Photodissociation of H$_{2}$ in this narrow PDR leads to a dramatic increase in its temperature, driven primarily by H$_{2}$ formation heating \citep{Susa13}. Unfortunately, the impact of this temperature increase remains unclear. Some authors \citep{Susa13,Susa14,Stacy2016} report that it drives a pressure wave outwards that reduces the gas density near the star and dramatically reduces its accretion rate even in the absence of photoionization feedback. On the other hand, \citet{Hosokawa2016} find it to have a much more limited effect and argue that photoionization is the dominant process. However, none of these studies accounts for the shielding of H$_{2}$ by atomic hydrogen and hence they all overestimate the amount of dense gas affected by photodissociation feedback.

At lower densities, shielding of H$_{2}$ by atomic hydrogen becomes ineffective, as the required shell thickness becomes implausibly large. For example, at a density of $n_{\rm H} = 10^{8} \, {\rm cm^{-3}}$, producing a column density of $10^{25} \, {\rm cm^{-2}}$ requires a shell thickness of $\sim 7000$~AU, several times larger than the entire physical extent of the region with mean density $> 10^{8} \, {\rm cm^{-3}}$ (see Figure~\ref{fig:halo-properties}). Therefore, the PDR is no longer restricted to the gas close to the HII region and instead can expand to large distances within $\sim 10^{4}$~yr. Gas in the PDR is not strongly heated by photodissociation, which is therefore ineffective at driving gas out of the minihalo. However, the loss of the main coolant from the gas leads to it heating up to a temperature of $5000 - 10000\,$K as it collapses \citep{Susa14}. Over time, this heating reduces the flow of gas to the center of the minihalo, limiting the amount of mass available for accretion even in the absence of effective photoionization feedback.

\subsubsection{Outflows}
Protostellar outflows, both in the form of highly-collimated jets and also wide-angle winds, are a ubiquitous feature of present-day star-forming regions 
\citep{Bally2016}. These outflows are magnetocentrifugal in nature. They are powered by the rotation of the protostellar accretion disk and channelled and collimated by the toroidal magnetic field built up by the rotation of the disk 
\citep{Pudritz2019}. Since we now understand that a dynamically significant magnetic field can be present even in the earliest minihalos \citep[][see also Section~\ref{sec:less-fragmentation}]{Schobera}, an obvious question is whether the formation of Pop III stars is also associated with the presence of magnetized outflows.

In practice, the answer to this question appears to depend on what one assumes regarding the initial geometry of the field. \citet{Machida2008}, \citet{Machida2013} and \citet{Sadanari21} simulate the collapse of primordial gas clouds threaded by an initially uniform magnetic field and show that protostellar outflows are a common outcome unless the initial magnetic field strength is very small. On the other hand, the simulations performed by \citet{Sharda2020} and \citet{Prole2022b} that start with a tangled magnetic field of the kind that we would expect to result from the turbulent dynamo find no evidence for the launching of jets or outflows. Whether this means that Pop III protostars do not produce outflows at all or simply that outflows are launched at a later point in the evolution of the system than is followed by these simulations remains to be determined.

The other class of stellar outflows that are potentially of interest are winds driven by the Pop III stars themselves. At solar metallicity, massive stars can produce highly energetic winds that can have a profound impact on the surrounding gas \citep[see e.g.][]{Rahner2017}. However, these winds are driven by the absorption of stellar photons by the lines of highly ionized metals and we  expect the wind strength to be a function of metallicity \citep{Kudritzki2002}. They become weak or non-existent for zero metallicity \citep{Krticka2006,Krticka2009}. Accounting for stellar rotation does not significantly change this picture: Stars rotating very close to their break-up velocity can drive strong winds, but lose so much angular momentum in the process that they do so only very briefly \citep{Ekstrom2008}. Stellar wind feedback therefore does not appear to be important for understanding the formation of Pop III stars.

\subsection{Large-scale impact}
\label{sec:large-scale-impact}
In addition to the profound impact that feedback from the first stars has on the gas in their parent minihalos, feedback from these stars also strongly affects gas on much larger scales, both in the intergalactic medium and in neighboring protogalaxies. In this section, we review the most important of these impacts.
 
 \subsubsection{Photoionization}
 \label{sec:large-photoionization}
 Ultraviolet photons that manage to escape from the halos hosting Pop III star formation will start to ionize the surrounding IGM. Prior to cosmic reionization, the Universe is highly opaque to these types of photons, and so this is essentially a local process: Individual star-forming protogalaxies will produce distinct ionized regions in the IGM, but no large-scale ionizing background will build up until after these individual regions begin to overlap, an event which does not occur until long after Pop II stars have taken over as the dominant source of ionizing photons \citep{Hartwig22}.
 
 The impact of photoionization feedback from Pop III star formation on nearby minihalos and atomic cooling halos depends on their evolutionary state and on the duration of the feedback. Systems in which gas has already cooled and is in the process of forming stars are difficult to disrupt with external photoionization feedback, since the time it takes an ionization front to propagate through the system is much longer than the collapse time of the gas, once the gas density exceeds $n \sim 100 \, {\rm cm^{-3}}$ \citep[see e.g.][]{Whalen2008b}. On the other hand, systems that have not yet gathered together much cool gas can be strongly affected. \citet{Visbal17} show for a constant ionizing flux that photoionization increases the minimum virial temperature required for collapse to $T_{\rm vir} \sim 40000$~K if the ionizing flux exceeds $F_{\rm crit} \sim 10^{6} \, {\rm photons \, s^{-1} \, cm^{-2}}$. For smaller values, photoionization is of limited importance, although cooling is suppressed in halos with $T_{\rm vir} < 10^{4} \, {\rm K}$ by the LW radiation that \citet{Visbal17} assume accompanies the ionizing radiation (see Section~\ref{sec:LW} below). For Pop III minihalos with star formation efficiencies typical of those found in simulations, the ionizing flux exceeds $F_{\rm crit}$ only at distances of a few kpc or less, even when assuming an escape fraction of 100\%. Moreover, the duration of Pop III star formation in individual minihalos is short, owing to the strong effects of photoionization and supernova feedback on small scales. Therefore,  cooling will only be suppressed for a brief period of a few Myr. Over longer timescales, the effect of ionizing feedback within this region can actually be positive, since the residual  fractional ioniztion in the recombining region will promote H$_{2}$ formation  \citep{OShea2005}. Overall,  we do not expect photoionization by Pop III-dominated sources to be the main source of feedback at high redshifts. Feedback from Pop II-dominated protogalaxies on surrounding minihalos is potentially more important, but a thorough discussion of this topic is outside of the scope of this review.

\subsubsection{Photodissociation}
\label{sec:LW}
 As discussed in Section~\ref{sec:small-pdiss}, massive Pop III stars produce large numbers of LW band photons that can photodissociate H$_{2}$. A large fraction of these escape from their parental minihalos  \citep{Kitayama2004,schauer15,schauer17}, and they can propagate to large distances through the IGM owing to its very low opacity below 13.6~eV \citep{haiman00}. The onset of Pop III star formation is therefore quickly followed by the build-up of an extragalactic LW background that pervades the Universe.  This photodissociates H$_{2}$ in 
 newly-assembled minihalos, making it harder for the gas to cool \citep{Haiman1997}. Once the strength of the background exceeds a critical value, H$_{2}$ cooling is almost entirely suppressed, preventing gas in minihalos from collapsing and forming Pop III stars. 

 The strength of the LW background required in order to suppress H$_{2}$ cooling in a minihalo can be estimated by a simple timescale argument \citep{Oh2002}. Models for the build-up of H$_{2}$ in primordial gas demonstrate that most molecules form within the first few recombination times, with the H$_{2}$ formation rate dropping substantially at later times owing to the loss of the required electrons and protons from the gas \citep{Tegmark1997b,Oh2002,Glover2013}. The LW background therefore only significantly suppresses H$_{2}$ formation once the photodissociation timescale, $t_{\rm dis} = 1 / k_{\rm dis}$, becomes comparable to the recombination timescale, $k_{\rm rec} = 1 / (k_{\rm rec} n x_{0})$, where $k_{\rm dis}$ and $k_{\rm rec}$ are the photodissociation and radiative recombination rate coefficients, $n$ is the characteristic number density of the gas and $x_{0}$ is its initial fractional ionization. For the conditions appropriate for a minihalo close to the minimum mass scale for collapse (see Section~\ref{sec:minimum-mass}), that is $T \sim 2000$~K and $x_{0} \sim 2 \times 10^{-4}$, this yields a critical value \citep{Oh2002} of
\begin{equation}
\left(\frac{J_{21}}{n} \right)_{\rm crit} \approx 10^{-4} f_{\rm sh}, 
\end{equation}
 above which we expect the LW background to significantly affect the minihalo. Here, $J_{21}$ is the strength of the background at the Lyman limit in units of $10^{-21} \, {\rm erg \, s^{-1} \, cm^{-2} \, Hz^{-1} \, sr^{-1}}$, $n$ is the density of the gas at the center of the minihalo, and $f_{\rm sh}$ is the self-shielding factor, i.e.\ the ratio of the H$_{2}$ photodissociation rate to its value in the absence of self-shielding. Numerical simulations \citep[e.g.][]{Machacek2001,Yoshida2003,Kulkarni21} indicate that $n \sim 10 \, {\rm cm^{-3}}$ is an appropriate value for the central density in the earliest minihalos, and so in the absence of self-shielding, we expect the LW background to become important once its strength exceeds a critical value of around $J_{\rm 21, crit} \sim 10^{-3}$.
 In this case, the LW background reduces the peak H$_{2}$ abundance by a factor proportional to $t_{\rm dis} / t_{\rm rec} \propto J_{21}^{-1}$, thereby increasing the minimum virial temperature required for efficient H$_{2}$ cooling. At typical minihalo temperatures, the H$_{2}$ cooling rate scales approximately as $\Lambda_{\rm H_{2}} \propto T^{4}$ and the cooling time scales as $t_{\rm cool} \propto (T^{3} x_{\rm H_{2}})^{-1}$. Therefore, for the cooling time to remain approximately constant as $J_{21}$ increases, we must have $T_{\rm min} \propto J_{21}^{1/3}$. Moreover, since $M_{\rm crit} \propto T_{\rm min}^{3/2}$, this implies that the minimum mass scale for collapse must increase as $M_{\rm crit} \propto J_{21}^{1/2}$ for $J_{21} > J_{\rm 21, crit}$. 

\begin{figure}[htp]
\includegraphics[width=8cm]{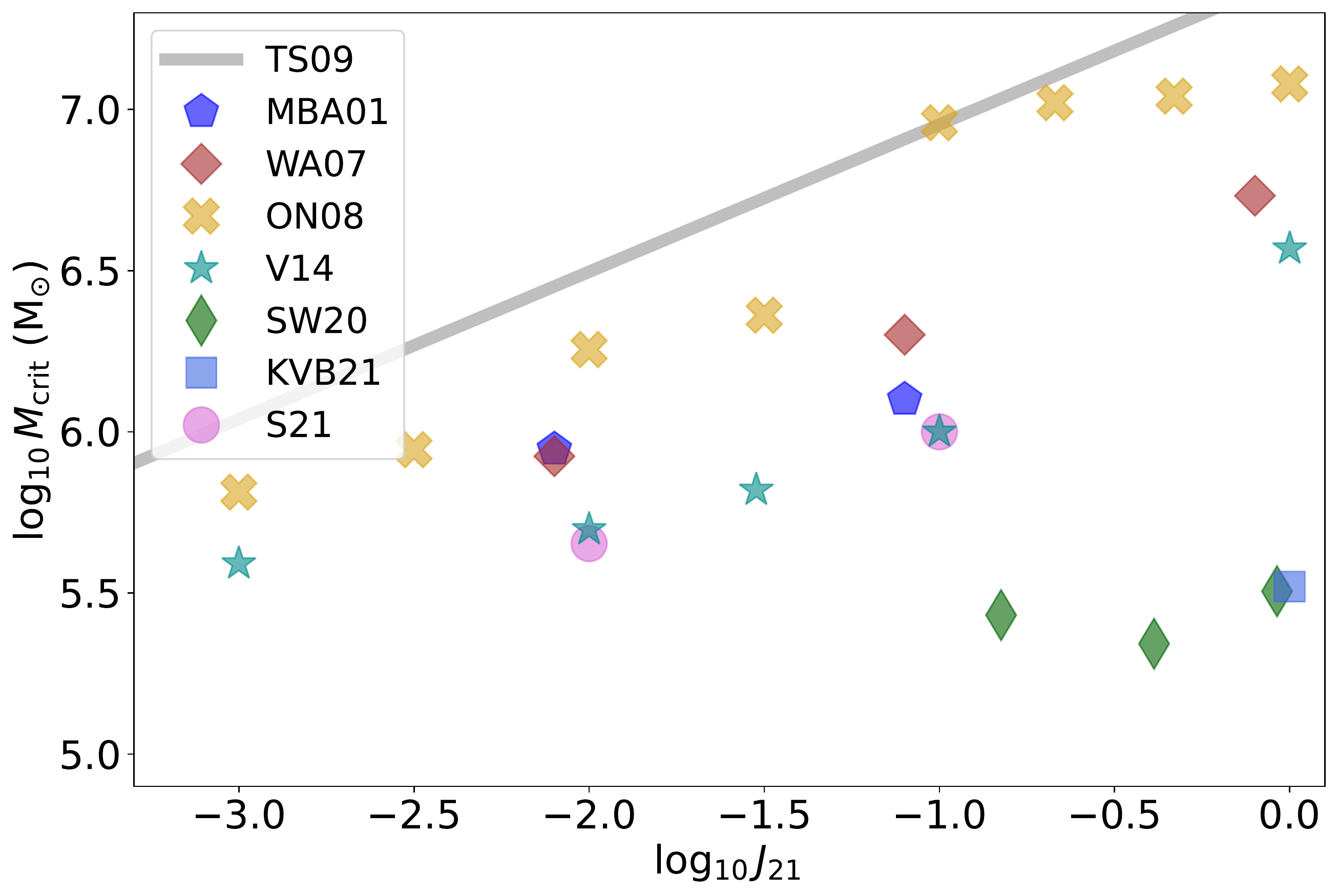}
\includegraphics[width=8cm]{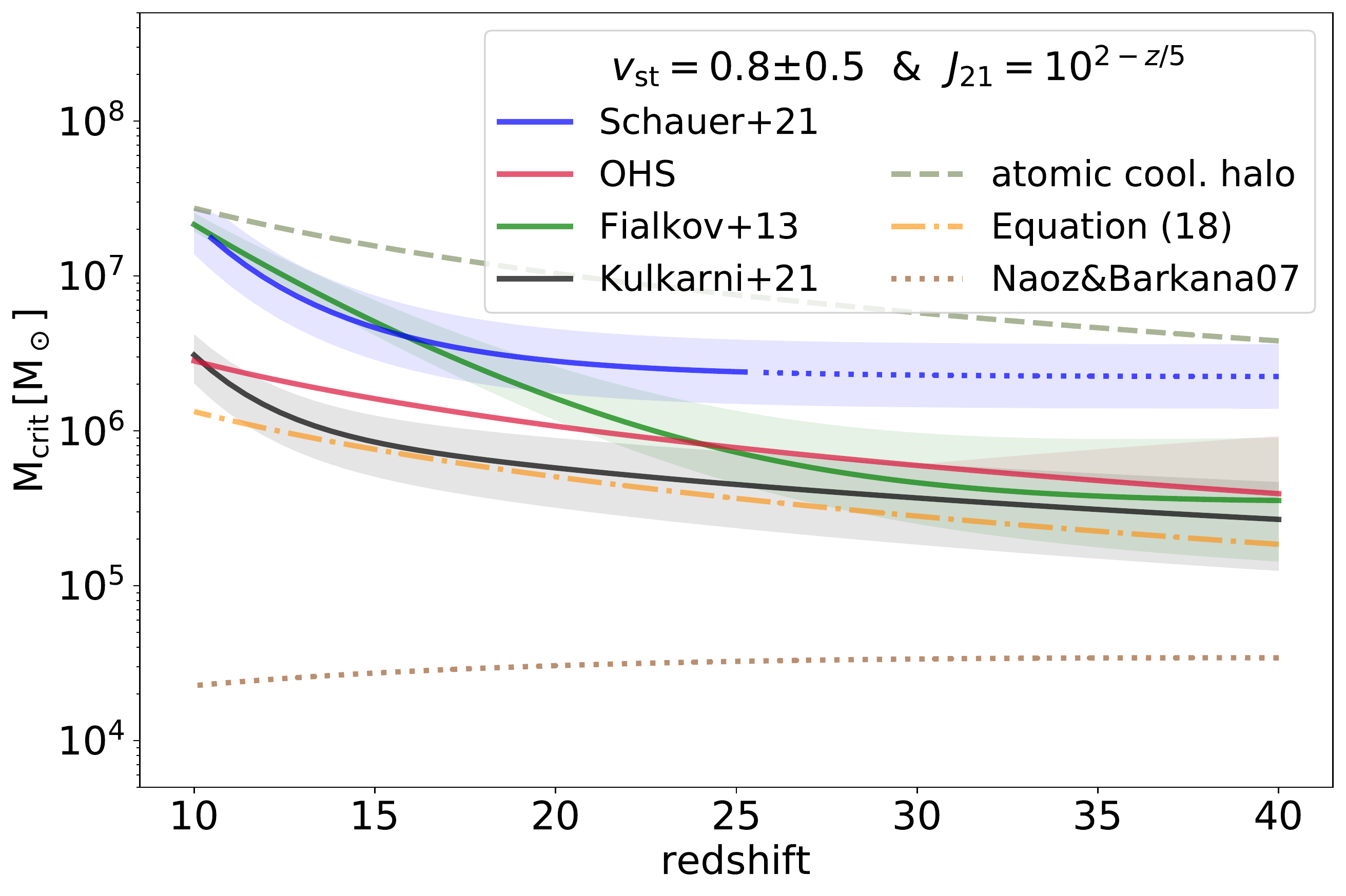}
\caption{{\it Top:} Minimum minihalo mass required for Pop III star formation ($M_{\rm crit}$) plotted as a function of $J_{21}$, the strength of the LW background in units of $10^{-21} \, {\rm erg \, s^{-1} \, cm^{-2} \, Hz^{-1} \, sr^{-1}}$. The values shown are taken from studies by \citet{Machacek2001} (MBA01; blue plus signs), \citet{Wise2007b} (WA07; black crosses), \citet{Oshea08} (ON08; red circles), \citet{Trenti09} (TS09; blue line), \citet{Visbal2014b} (black points), \citet{Skinner2020} (SW20; blue pentagons), \citet{Kulkarni21} (KVB21; green squares) and \citet{Schauer21} (green triangle). For TS09, we show the values computed assuming $z = 20$. SW20 find H$_{2}$ cooling and star formation occurring in minihalos with a broad range of masses for $10^{-1} < J_{21} < 1$; we show only the few least massive points from their study here. Finally, for S21, we show the values they report for the lowest mass at which minihalos form stars. The average mass at which minihalos first form stars (their $M_{\rm ave}$) is about a factor of three larger than this minimum value.
{\it Bottom:} Evolution of $M_{\rm crit}$ with redshift for a selection of models that account for the combined effects of baryonic streaming and the LW background. The lines are computed for a streaming velocity $v_{\rm st} = 0.8$ (in units of the rms value) and the shading indicates the impact of varying this by $\pm 0.5$. For the time evolution of the LW background, we adopt a simple fit to the values computed by \citet{Greif2006}. The meaning of the various lines is the same as in Figure~\ref{fig:Mcrit}.}
\label{fig:LW}
\end{figure}

In the upper panel in Figure~\ref{fig:LW}, we show a compilation of predictions for how $M_{\rm crit}$ varies as a function of $J_{21}$ taken from a number of studies in the literature. These studies range from simple semi-analytical models \citep{Trenti09,Visbal2014b} to detailed numerical simulations both with \citep{Skinner2020,Kulkarni21,Schauer21} and without \citep{Machacek2001,Wise2007b,Oshea08} H$_{2}$ self-shielding. Although there is considerable scatter,\footnote{This scatter likely results from a combination of differences in the microphysical treatment (e.g.\ selection of reactions, choice of reaction rate coefficients and cooling function) and differences in the definition of $M_{\rm crit}$ in the different studies.}  we see that the semi-analytical models and the simulations without H$_{2}$ self-shielding yield results that broadly agree with our expectations based on the discussion above. However, the situation becomes more complicated once we account for H$_{2}$ self-shielding. \citet{Schauer21} include this in their simulations, but do not find it to have a large effect. They report values of $M_{\rm crit}$ that are a factor of $2-3$ smaller than those predicted by simulations that do not account for self-shielding, but whether this difference is due to the inclusion of self-shielding or due to other differences in the chemical or thermal treatment is unclear. On the other hand, \citet{Skinner2020} and \citet{Kulkarni21} find H$_{2}$ self-shielding to have a profound impact on $M_{\rm crit}$, keeping it below $10^{7} \, {\rm M_{\odot}}$ for LW background strengths of up to $J_{21} \sim 10$. If true, these results would imply that the LW background does not play an important role in regulating high redshift star formation, since the mean strength of the background does not exceed $J_{21} = 1$ until after Pop II star formation has already started to dominate \citep[see e.g.][]{Trenti09,ahn09}.

Another view of the current uncertainty in the impact of LW feedback is given in the bottom panel of Figure~\ref{fig:LW}, where we show the redshift evolution of $M_{\rm crit}$ in several different models, assuming a streaming velocity $v_{\rm st} = 0.8$ (in units of the root-mean-squared value) and a LW background that evolves as $J_{21} = 10^{2 - z / 5}$, which is a reasonable fit at $z > 10$ to the values computed by \citet{Greif2006}. At $z > 25$, the LW background is very weak and the evolution of $M_{\rm crit}$ in all of the models is essentially the same as in the $J_{21} = 0$ case discussed earlier (see Section~\ref{sec:more-complex-models} and Figure~\ref{fig:Mcrit}). At $z < 25$, however, the growing background drives a substantial increase in $M_{\rm crit}$ in the \citet{Schauer21} and F13 models, but has little impact on it in the \citet{Kulkarni21} and OHS models.

Figure~\ref{fig:LW} also demonstrates that in halos more massive than $\sim 10^{7} \, {\rm M_{\odot}}$ at $z = 20$, $M_{\rm crit}$ becomes highly insensitive to $J_{21}$ even in the absence of effective H$_{2}$ self-shielding. The reason for this change of behavior is that these halos have high enough virial temperatures to allow them to cool initially via Lyman-$\alpha$ cooling. In these atomic cooling halos, the gas can reach much higher densities without the need for H$_{2}$ cooling, meaning that a much stronger LW background is needed to completely suppress H$_{2}$ formation. The key density in this case is $n_{\rm crit} \sim 10^{4} \, {\rm cm^{-3}}$, the critical density above which the rotational and vibrational level populations of H$_2$ start to approach their local thermodynamical equilibrium (LTE) values. Above $n_{\rm crit}$, the H$_{2}$ cooling time becomes independent of density, while the free-fall collapse time continues to decrease with increasing density. In addition, collisional dissociation of H$_{2}$ also becomes more effective in gas with $n > n_{\rm crit}$. Together, these facts lead to a bifurcation in the behavior of atomic cooling halos depending on the amount of H$_{2}$ they have managed to form by the time their density reaches $n_{\rm crit}$. If they have formed enough H$_{2}$ to provide effective cooling, then the gas remains cold as it collapses, with its subsequent evolution being very much the same as in the standard Pop III formation scenario (Section~\ref{sec:standard-scenario-of-PopIII-formation}). On the other hand, if H$_{2}$ cooling has not become effective by the time the gas reaches $n_{\rm crit}$, it will never become so at higher densities. Instead, the gas will remain warm, with cooling arising primarily from bound-bound transitions in atomic hydrogen and from H$^{-}$ formation \citep{Omukai01,Schleicher10}.

The combined effect of the higher temperatures and higher characteristic densities encountered in atomic cooling halos is a dramatic increase in the value of $J_{\rm 21, crit}$. 
One-zone models of the thermal and chemical evolution of highly irradiated gas in atomic cooling halos typically find $J_{\rm 21, crit} \sim 1000$ if the spectrum of the LW background is dominated by Pop III stars 
\citep[see e.g.][]{Sugimura2014,Glover2015a,Glover2015b}. Three-dimensional models find even larger values, $J_{\rm 21, crit} \sim 10^{4}$ 
\citep{Shang2010,Latif2015,Hartwig2015}, although the scatter between different simulations can be considerable.

To put these numbers into context, we expect the mean strength of the LW background at reionization to be $J_{21} \sim 1$ \citep{haiman00,ahn09}. Therefore, the LW background is unable to suppress H$_{2}$ cooling in the vast majority of atomic cooling halos, although it does produce systematically warmer temperatures and higher accretion rates than would be the case without the background \citep{LatifInflow2015,Regan2018}. Complete suppression of H$_{2}$ cooling occurs only in a small number of atomic cooling halos that happen to lie very close to a strong source of LW photons \citep{Dijkstra2008,Agarwal2012,Dijkstra2014,Visbal2014,Regan2017,Agarwal19} or that are undergoing strong dynamical heating from a succession of mergers \citep{Regan2020}. However, despite their rarity, these systems are of considerable interest as the possible formation sites of supermassive Pop III stars (see Section~\ref{sec:supermassive}).

 \subsubsection{X-rays}
 \label{sec:x-rays}
 Although Pop III stars themselves do not produce significant emission at X-ray wavelengths, we nevertheless expect Pop III star formation to be closely associated with the production of X-rays. We discuss the origin of this emission as well as its impact onto the surrounding gas. 
 
 \paragraph{Stellar sources of high-redshift X-rays}
Three main mechanisms potentially contribute to the X-ray emission from Pop III stars. First, if some fraction of these stars are located in tight binaries, then these systems can become high-mass X-ray binaries 
  \citep[HMXBs, see e.g.][]{Mirabel2011}. Second, hot shocked gas in Pop III supernova remnants (SNRs)  emits strongly at X-ray energies until it cools. Finally, Pop III SNRs may also produce X-rays indirectly, by acting as a source of cosmic ray electrons. Because of the high CMB energy density at high redshift, cosmic ray electrons with energies of up to a few hundred GeV lose most of their energy by inverse Compton scattering CMB photons, producing a power-law inverse Compton spectrum that extends up to X-ray energies \citep{Oh2001}.

A useful metric for comparing the relative importance of these mechanisms is the X-ray luminosity they produce per unit star formation rate, $L_{\rm X} / {\rm SFR}$. In the local Universe, at metallicities close to solar, HMXBs produce a luminosity in the $0.5 - 2\,$keV energy band given by $L_{\rm X} / {\rm SFR} \approx 6 \times 10^{39} \, {\rm erg \, s^{-1} \, (M_{\odot} yr^{-1})^{-1}}$ \citep[see e.g.][]{Glover2003,Kaur2022}, although there is roughly a factor of two scatter between different observational determinations. However, the contribution from SNRs\footnote{We do not consider stellar winds here, as even at solar metallicity, these do not succeed in producing large X-ray luminosities \citep{Wang1991,HarperClark2009}, due to some combination of the escape of hot gas from wind-driven bubbles and efficient cooling of the gas at the bubble boundary \citep{Lancaster2021}. At low or zero metallicity, stellar winds will be much weaker and hence will be even less effective at producing X-rays.} is less well determined. Estimates of the ratio of the X-ray luminosity to the mechanical luminosity range from values as small as $2 \times 10^{-4}$ \citep[e.g.][]{Helfand2001} to ones as large as a few percent \citep[e.g.][]{Smith2005, Franeck2022}, albeit with substantial time variability and scatter from region to region. If we take a value near the midpoint of this range, $L_{\rm X} / L_{\rm mech} = 10^{-3}$ as a representative average  \citep{Rogers2014,Franeck2022}, then we find that $L_{\rm X} / {\rm SFR} \approx 6 \times 10^{38} \, {\rm erg \, s^{-1} \, (M_{\odot} yr^{-1})^{-1}}$ for SNRs \citep{Leitherer1999}, i.e.\ around 10\% of the HMXB contribution. Therefore, on average in the local Universe, X-rays from HMXBs dominate over those produced by hot gas in SNRs. Accounting for the inverse Compton contribution does not substantially change this picture \citep{Oh2001}.

How do these conclusions change as we move to higher redshift? Both models and observations show that HMXBs produce X-rays more efficiently as the metallicity drops, with $L_{\rm X} / {\rm SFR}$ increasing by an order of magnitude or more as the metallicity decreases for $1 \, {\rm Z_{\odot}}$ to $0.03 \, {\rm Z_{\odot}}$ \citep{Kaur2022}. At even lower metallicities, we have no direct observational constraints, and theoretical models disagree on whether $L_{\rm X} / {\rm SFR}$ asymptotes to a constant value \citep{Fragos2013} or continues to increase  \citep{Brorby2016}. It is also unclear whether this finding of a higher $L_{\rm X} / {\rm SFR}$ remains valid if we extrapolate all the way down to ${\rm Z} = 0$, with results from one study indicating that HMXBs formed from Pop III stars may actually be less effective at producing X-rays than in the solar metallicity case \citep{Sartorio2023}, although we note that this result depends sensitively on the choice of the Pop III IMF. Turning our attention to SNRs, we note that we do not expect the fraction of the energy they radiate as X-rays to be particularly sensitive to the metallicity. However, if the Pop III IMF is dominated by massive stars (see Section~\ref{sec:IMF-and-multiplicity}), then the number of SNe that explode per unit star formation rate will be a factor of a few larger than in the local Universe. In addition, if a significant number of Pop III stars explode as hypernovae or pair-instability supernovae (see Section~\ref{sec:sne} below), then the energy available per explosion will also be larger, implying a higher X-ray luminosity even if $L_{\rm X} / L_{\rm mech}$ remains constant. In addition, the inverse Compton contribution is potentially far more important here than at low redshift. For a typical supernova, around $0.1 - 0.2$\% of its initial kinetic energy is used to accelerate relativistic electrons \citep{Lacki2013} and so if the bulk of this is converted to X-rays by inverse Compton scattering, the result is a contribution to $L_{\rm X} / {\rm SFR}$ similar in size to that from hot gas.

In summary, if Pop III HMXBs are as efficient at producing X-rays as present-day HMXBs and if most Pop III SNe are standard core-collapse supernovae with explosion energies $E_{\rm SN} \sim 10^{51} \, {\rm erg}$, then the majority of the X-ray emission associated with Pop III star formation will come from HMXBs. For this reason, it is this scenario that has attracted most attention in the literature, as we discuss in more detail below. However, scenarios in which Pop III supernovae dominate the production of X-rays at high redshift cannot be completely ruled out \citep[see e.g.][for an example]{Ricotti2016}.

\paragraph{Impact of the high-redshift X-ray background}
The X-rays produced by Pop III HMXBs and other sources can propagate to large distances through the IGM. The Universe is optically thin to X-ray photons with energies $E > 1.8 [(1 + z) / 15]^{1/2} \, {\rm keV}$ and the comoving photon mean free path exceeds 1~Mpc for photon energies $E > 0.25$~keV \citep{Furlanetto2006}. Therefore, once Pop III star formation begins, a pervasive X-ray background develops. It affects gas in high redshift minihalos in two ways. First, it partially ionizes it, thereby catalyzing the formation of additional H$_2$ \citep{haiman00}. Second, it also heats the gas. The first of these effects acts as positive feedback, making cooling and star formation more likely, while the second one acts as negative feedback. Their relative importance depends on the density of the gas and the strength and spectral shape of the background. Heating dominates at low densities and can completely suppress the collapse of gas into low-mass minihalos if it is able to heat the IGM to a temperature above the virial temperature of the minihalo. The strength of the background required to accomplish this depends sensitively on the adopted low-energy cutoff, since the heating rate due to photons of energy $E$ scales approximately as $E^{-2}$ owing to the strong energy dependence of the photionization cross-section. For example, \citet{Hummel2015} take a minimum photon energy of 1~keV and estimate that a critical value of $J_{\rm X, crit} \approx 10^{-21} \, {\rm erg \, s^{-1} \, cm^{-2} \, Hz^{-1} \, sr^{-1}}$ at 1~keV is required to suppress the collapse of gas into a $10^{6} \, {\rm M_{\odot}}$ halo. On the other hand, \citet{Park2021} adopt a minimum photon energy of 200~eV and find that a much weaker background is sufficient to suppress collapse, $J_{\rm X, crit} \approx 10^{-23} \, {\rm erg \, s^{-1} \, cm^{-2} \, Hz^{-1} \, sr^{-1}}$.

At higher densities, the heating provided by the X-rays becomes unimportant in comparison to the increased cooling provided by the additional H$_{2}$. Consequently,  the gas cools to  lower temperatures than in the case without X-rays \citep{Machacek2003,Hummel2015,Park2021}. The density at which the behavior changes  lies in the range $n \sim 1 - 100\,$cm$^{-3}$, with the value depending on the strength of the background. However, if the X-ray background is very strong, there is so much heating at low densities that the gas never reaches the positive feedback regime, and the feedback in this case is wholly negative.

At even higher densities, the gas becomes optically thick to the X-ray background, which therefore plays no direct role on scales comparable to the size of the protostellar accretion disk. However, its presence indirectly affects the behavior of the gas on these scales, as the lower temperatures that are reached on larger scales lead to a slower inflow of mass to the center of the halo. This may result in the  formation of a less massive, more stable accretion disk \citep{Hummel2015,Park2021,Park2021b}, which means fewer fragments and implies a more top-heavy IMF as well as lower stellar multiplicity.

Another interesting question is whether or not positive feedback from X-rays dominates over negative feedback from LW photons. The answer obviously depends on the relative strength of the two backgrounds \citep{Park2021}, but models that self-consistently follow their build-up typically find that LW feedback dominates \citep[see e.g.][]{Glover2003}.

Finally, the heating of the IGM produced by even a weak X-ray background has an important impact on its visibility in the 21~cm line of atomic hydrogen \citep{Furlanetto2006}. Prior to the onset of Pop III star formation, the IGM is colder than the CMB and hence the 21~cm signal is visible in absorption (see Section~\ref{sec:21cm-and-CMB}). However, once X-rays heat the IGM above the CMB temperature, the absorption signal vanishes and the line instead becomes visible in emission. Measuring the redshift at which the transition from absorption to emission occurs therefore allows us to constrain the strength of the high redshift X-ray background, and may also allow us to distinguish between the contributions made to it by Pop III and Pop II sources \citep{Mirocha18}.
 
 \subsubsection{Supernovae}
 \label{sec:sne}
In this section, we discuss how the final fate of a Pop III star depends on its initial mass, with a particular focus on identifying the range of masses that lead to supernova explosions. We also briefly outline the impact that these explosions have on the surrounding gas.
 
 \paragraph{The fate of massive Pop III stars} 
 How massive Pop III stars end their lives depends on the mass of their helium core at the moment that the star dies \citep{Heger2003}. For non-rotating stars with negligible mass loss, there is a simple mapping between helium core mass and initial mass, and so it is common to discuss the fates of these stars as a function of their initial mass. Below $M \sim 9 \, {\rm M_{\odot}}$, Pop III stars end their lives as white dwarfs, supported by electron degeneracy pressure. Above this mass, electron degeneracy pressure is not sufficient to support the star. Stars in the narrow mass range $9\, {\rm M_{\odot}} < M < 10 \, {\rm M_{\odot}}$ form degenerate oxygen-neon cores that collapse, while those with $10\, {\rm M_{\odot}} < M < 100 \, {\rm M_{\odot}}$ form iron cores that also undergo collapse. In stars with $10\, {\rm M_{\odot}} < M < 25 \, {\rm M_{\odot}}$, this core collapse triggers a supernova explosion (hereafter referred to as a core-collapse supernova or CCSN) with an energy of around $10^{51} \, {\rm erg}$ that leaves behind a neutron star remnant. For $25\, {\rm M_{\odot}} < M < 40 \, {\rm M_{\odot}}$, a core-collapse explosion also occurs, but is unable to completely unbind the stellar envelope from the neutron star remnant. Instead, the inner portions of the envelope fall back onto the neutron star, causing it to collapse and form a black hole. The energy released into the surrounding environment depends on the extent of this ``fallback'' and can potentially be much smaller than $10^{51} \, {\rm erg}$, in which case we refer to the CCSN in question as a faint supernova. For masses between $40 \, {\rm M_{\odot}}$ and a somewhat uncertain upper limit in the range $70 - 100\,$M$_{\odot}$, no supernova explosion occurs. Instead, the star simply collapses directly into a black hole at the end of its life. 
 
 For more massive stars, a new physical effect becomes important. These stars have such high central temperatures during the later stages of their evolution that they become susceptible to the so-called pair instability \citep{Fowler1964, ElEid1983, Ober1983, Bond1984, Heger2002}. Briefly, as the temperature increases, so too does the fraction of photons in the core that have energies greater than 1.022~MeV, the rest mass of an electron-positron pair. Interaction of these photons with the surrounding nuclei can lead to their conversion into  electron-positron pairs. Since the core is supported by radiation pressure, the loss of photons via this mechanism results in a reduction in the pressure and hence contraction of the core. This further increases the core temperature, making pair production more likely, and in sufficiently massive stars, this can become a runaway process triggering the dynamical implosion of the core \citep{Woosley2017}. This implosion is terminated by the onset of rapid O and Si burning. Early calculations predicted an onset of the pair instability above a mass of around $100 \, {\rm M_{\odot}}$ \citep{Heger2003}, but more recent calculations find lower values of $\sim 70 \, {\rm M_{\odot}}$ \citep{Woosley2017}, albeit with some remaining uncertainty due to the uncertainty in the rate of the $^{12}$C($\alpha, \gamma)^{16}$O nuclear reaction \citep{Farmer19}. In stars above this mass, but below $140 \, {\rm M_{\odot}}$, this leads to a pulsational instability in which the star undergoes a series of implosions, nuclear flashes and re-expansions that disrupt the stellar envelope and part of the core. This process terminates once the core mass becomes too small for it to be susceptible to the pair instability, following which it evolves in the same fashion as a lower mass star. The energy released by this instability ranges from $\ll 10^{51} \, {\rm erg}$ to a few times $10^{51} \, {\rm erg}$, and the ejected mass also spans a large range of values \citep{Woosley2017}.
 
 Above $140 \, {\rm M_{\odot}}$,  pulsations do not occur, as the energy released in the initial nuclear flash completely disrupts the star, leaving no remnant behind. The resulting explosion is termed a pair-instability supernova (PISN). It has a typical energy of around $10^{53} \, {\rm erg}$, two orders of magnitude larger than the value of a typical CCSN. Because all of the mass of the star is expelled in the ejecta, PISNe typically have very large metal yields. Finally, for Pop III stars more massive than $260 \, {\rm M_{\odot}}$, the pair instability occurs, but the energy released by the resulting rapid nuclear burning is not sufficient to disrupt the star, which instead simply collapses to a black hole 
 \citep{Heger2002}.
 
 Accounting for the effects of rotation has two main effects on the picture sketched above. First, rotation-driven mixing leads to rapidly rotating Pop III stars having larger helium core masses than non-rotating stars of the same mass  \citep{Yoon2012}. Since the onset of the pair instability depends on the core mass, an important consequence of this is a reduction in the minimum mass required for a PISN, which can reach values as low as $M \sim 100 \, {\rm M_{\odot}}$ for rapidly rotating stars. Second, rapidly rotating Pop III stars can potentially explode as jet-driven supernovae \citep[see e.g.][and references therein]{Grimmett2021}. These are widely believed to be responsible for the unusually bright class of SNe known as ``hypernovae''  \citep[HNe, see][]{Umeda2002}, which have typical energies of $10^{52} \, {\rm erg}$. Rapid rotation is also a necessary (but not sufficient) condition for a supernova to produce a gamma-ray burst \citep[GRB, see][]{Toma2016}.
 
 Unfortunately, the rotational properties of Pop III stars are poorly constrained. In order to measure the angular momentum of Pop III protostars reliably in simulations, extremely high resolution is required in order to distinguish between tight binaries with separations of a few AU and single stars. 
 One of the few sets of simulations that do have sufficient resolution are those presented by \citet{Greif12}. \citet{Stacy13b} analyzed the rotational properties of the protostars formed in these simulations and showed that most were rotating very rapidly, with rotational velocities between $\sim 50$\% and 100\% of the Keplerian velocity. However, these simulations cover only a very short period ($\sim 10\,$yr) at the start of the evolution of the protostellar system and do not account for the effects of the magnetic field, which may provide strong magnetic breaking on protostellar scales \citep{Hirano18}. In view of these uncertainties, most studies of the impact of Pop III SNe on their surroundings have focussed on the
non-rotating case. 

 \paragraph{Impact of Pop III supernovae}
 \label{sec:PopIIIsn}
 Pop III supernovae affect their surroundings via both mechanical feedback (the injection of energy and momentum) and chemical feedback (the enrichment of the gas with metals). We first consider the role played by mechnical feedback. The gravitational binding energy of a minihalo with a mass close to $M_{\rm crit}$ is
 $\sim 10^{51} \, {\rm erg}$, comparable to the energy of a single CCSN. We would therefore expect a single CCSN to be able to eject a considerable fraction of the gas from a low-mass minihalo, provided that the supernova remnant does not lose too much energy via radiative cooling. 
 
 In practice, the main factor that determines whether or not radiative cooling of the supernova remnant is effective is the density of the region in which the supernova explodes. If prior photoionization feedback has cleared away most of the gas, then cooling is ineffective and even a single supernova can eject a large fraction of the gas from the halo \citep{Kitayama2005,Ritter2012,Jeon2014b,Chiaki2019}. On the other hand, if prior photoionization feedback is ineffective, as would be the case for e.g.\ a $10 - 15\,$M$_{\odot}$ CCSN, then little gas is ejected \citep{Jeon2014b,Chiaki2019}. Highly energetic SNe, such as hypernovae or PISNe, are less sensitive to the local environment and typically clear almost all gas out of the halo \citep{Greif2010,Magg2022}. This picture changes in  massive minihalos, due to their larger gravitational binding energies. For minihalo masses larger than a few times $10^{6} \, {\rm M_{\odot}}$, a single CCSN is no longer sufficient to eject the gas, although a single PISN remains effective \citep{Chiaki2019,Magg2022}.
 
 The time taken for gas to become available again for star formation after the explosion of a Pop III SN, the so-called fallback or recovery time,  varies considerably with halo mass and SN progenitor mass, ranging from a few Myr to 100 Myr or more. Smaller values are typically associated with lower mass progenitors or higher mass minihalos, but there is  considerable scatter from halo to halo even for fixed progenitor mass. \citet{Magg2022} argues that this is a consequence of the small scale structure of the gas prior to the supernova explosion. The greater the amount of gas located in dense clumps, the better it is able to survive the impact of the SN remnant and the shorter the resulting fallback time.
 
Supernova remnants that successfully expel gas from their parent minihalo will subsequently expand into the surrounding IGM, reaching final sizes of $500 - 1500\,$pc by the time that they finish expanding \citep{Greif2010,Jaacks18b}, depending on their initial energy and the density of the surrounding gas. These sizes are large enough to reach neighboring minihalos. If these minihalos are already in the process of forming stars, then the interaction with the expanding supernova remnant will have little effect on them. It will strip away the low density gas in the outskirts of the minihalo, but will not penetrate into the dense core \citep{Chen2017}. Minihalos that are still assembling their gas are more strongly affected, as in this case the expanding remnant can strip most of the gas from the halo, preventing it from cooling and collapsing \citep{Sigward2005}. However, these minihalos will already have been irradiated by photoionizing radiation from the supernova progenitor \citep{Jaacks18b}, which has a similar suppressive effect, and so this direct mechanical feedback is generally considered to be of limited importance. 
 
Turning to chemical feedback, we note that this proceeds in two qualitatively different ways: internal enrichment (i.e.\ enrichment of gas in the halo where the supernova exploded) and external enrichment (i.e.\ enrichment of nearby halos that interact with the supernova remnant). As we might expect from the discussion above, the relative importance of these mechanisms depends on how effectively supernovae can drive gas and metals out of their host halos, with internal enrichment dominating in more massive halos and external enrichment dominating in lower-mass halos \citep{Chiaki2019}. The likelihood of external enrichment occurring also depends on the local number density of minihalos capable of forming stars, as this determines how many are likely to be located within the region of the IGM enriched by the supernova. Therefore, we would expect the importance of external enrichment to also depend on $M_{\rm crit}$, with this process becoming less important as $M_{\rm crit}$ increases. As an illustration, consider the studies of early metal enrichment by \citet{Hicks2021} and \citet{Magg2022}.  \citet{Hicks2021} find that $\sim 60$\% of the metal-enriched minihalos in their simulation gain their metals via external enrichment. However, these halos are small, with most having masses $M < 10^{6} \, {\rm M_{\odot}}$, and it is unclear how many will actually form stars. In comparison, \citet{Magg2022} examine only the enrichment of halos that successfully form stars and find that in this case, only around 35\% are externally enriched. Although these studies are a good start, more work along these lines is necessary to fully pin down the relative importance of external versus internal enrichment and how this depends on the local extragalactic environment, the strength of the LW background etc.\
 
Finally, a quantity of great interest for the interpretation of observations of extremely metal-poor stars in our own galaxy (see Section~\ref{sec:arch}) is the metallicity of the gas enriched by a single Pop III supernova. This depends on the efficiency with which metals are mixed into the surrounding environment: the greater the amount of mixing, the lower the resulting metallicity. However, we can put an upper limit on this metallicity by assuming that metals only mix within the volume occupied by the supernova remnant at the time that it finishes expanding. \citet{Magg2020} show that in this case, the minimum mass of gas into which the metals are mixed is given by
\begin{equation}
 M_{\rm dil, min} = 1.9 \times 10^{4} E_{51}^{0.96} n^{-0.11} \, {\rm M_{\odot}}\;,
\end{equation}
where $E_{51}$ is the kinetic energy of the SN explosion in units of $10^{51} \, {\rm erg}$ and $n$ is the number density of the gas in which the explosion takes place in units of cm$^{-3}$. For explosions taking place within a pre-existing HII region, $n^{-0.11} \sim 1$ and so a CCSN will mix its metals into a minimum of $2 \times 10^{4} \, {\rm M_{\odot}}$ of gas, while a PISN will mix its metals into a minimum of $2 \times 10^{6} \, {\rm M_{\odot}}$ of gas. The total mass of metals ejected by a Pop III CCSN ranges from around $0.1$ to a few M$_{\odot}$ \citep{Nomoto2006}, and taking a value at the top end of this range then yields a maximum metallicity of $Z_{\rm max} \sim 0.01 \, {\rm Z_{\odot}}$ for enrichment by a single CCSN. For a PISN, the mass of metals ejected is much larger, $\sim 100 \, {\rm M_{\odot}}$ \citep{Heger2002}, but since this is mixed into a much larger gas mass, the resulting maximum metallicity is actually smaller, $Z_{\rm max} \sim 3 \times 10^{-3} \, {\rm Z_{\odot}}$. \citet{Magg2020,Magg2022} show that in practice, individual Pop III supernovae mix their metals into gas masses ranging from a few times this minimum value to as much as $10^{4}$ times more. Large values typically correspond to cases of external enrichment, when a large mass of pristine gas associated with an existing minihalo is enriched by a small mass of metals from an external source. It should also be stressed that this mixing is generally not uniform, there can be substantial metallicity variations within the enriched volume \citep{Ritter2014,Tarumi2020}.
Individual Pop III SNe can therefore produce metal-enriched gas with a broad range of metallicities, making it plausible that at least some of the extremely metal-poor stars observed in the Milky Way will have formed from gas enriched by only a single Pop III SN (see also Section~\ref{sec:arch}).
 
\subsection{Transition to Pop II star formation}
\label{sec:transition-PopIII-to-PopII}
Second generation stars, sometimes termed early Pop II stars, formed from material that was enriched by metals from the first generation of stars. Unlike Pop III stars, which have not yet been directly detected (Section~\ref{sec:direct-detection}), members of the second generation have already been found in surveys looking for extremely metal-poor stars in our Milky Way and neighboring satellite galaxies (Section~\ref{sec:arch}). As discussed in Section \ref{sec:initial-collapse}, the chemical composition of the gas governs its thermodynamic response and therefore its fragmentation behavior under different environmental conditions. In turn, this determines the stellar mass spectrum  (Section \ref{sec:disks-and-fragments}), which for genuine Pop III stars is predicted to be very different compared to what we find in the Universe today (Section~\ref{sec:IMF-and-multiplicity}). Unlike purely primordial gas, where cooling is largely provided by hydrogen (see top right of Figure \ref{fig:initial-collapse}), metal-enriched gas has access to a much wider range of  cooling mechanisms, allowing it to reach lower temperatures at various points in its evolution. This enhances the fragmentation of the gas and reduces the rate at which the fragments accrete additional material. Together, these effects are widely believed to lead to an IMF dominated by low-mass stars, similar to the present-day IMF \citep{Kroupa2002, Chabrier2003}. However, there remains the question of when and where this transition occurs, whether it is sudden or gradual, and what is the underlying physical cause for it. Indeed, there are two competing models based on two different low-metallicity cooling channels, each associated with a different critical metallicity. 

The first model is based on atomic fine-structure lines from alpha elements such as carbon or oxygen. These start to dominate the coolingat number densities around $10^4\,$cm$^{-3}$ and above a critical metallicity of about $10^{-3}\,Z_{\odot}$ \citep[see, e.g.][]{Chon21}. This leads to a markedly different evolution in the log~$T-$log~$n$ diagram as indicated in the upper right part of Figure \ref{fig:initial-collapse}. The required metallicity (or, more precisely, the required carbon and oxygen abundances) can be estimated by comparing the cooling rate due to metals with the compressional heating rate at a characteristic temperature of 200~K and density of $10^{4} \, {\rm cm^{-3}}$ \citep[see e.g.][]{Bromm03, Santoro2006}, corresponding to the point at which H$_{2}$ cooling becomes inefficient in the Pop III case (see Figure \ref{fig:initial-collapse}).
To quantify this, \citet{Frebel2007} introduced the 
transition discriminant, $D_{\rm trans} = \log_{10} \left( 10^{\rm [C/H]} + 0.3 \times 10^{\rm [O/H]}\right)$ and argued that low-mass stars can only form for metallicities $D_{\rm trans} \gtrsim -3.5$.

The second line of reasoning considers dust cooling as the primary agent for fragmentation, which leads to a considerably lower critical metallicity in the range $10^{-5}$ to $10^{-6}\,Z_{\odot}$. This value was first estimated using analytical or simple one-zone numerical models that compared the strength of various cooling and heating processes in a metal-poor environment
\citep[e.g.][]{Omukai2005, Schneider2006, Schneider12, Chiaki2013, Chiaki2013B}. However, it has subsequently received strong support from detailed 3D numerical simulations of dust cooling in dense, metal-poor gas \citep[e.g.][]{Tsuribe2006, Tsuribe2008, Clark2008, Dopcke2013, Safranek-Shrader2014, Chiaki2016, Chiaki2019, Chon21, Chiaki22}. Together, these models suggest that dust is the driving agent for the transition in the IMF. It occurs at a metallicity of about $10^{-5}$ solar and leads to an IMF peaking below  $1\,$M$_{\odot}$ with  a functional form similar to what is found in the solar neighborhood \citep{Kroupa2002, Chabrier2003}.

\begin{figure}[ht]
\begin{center}
\includegraphics[width=10.0cm]{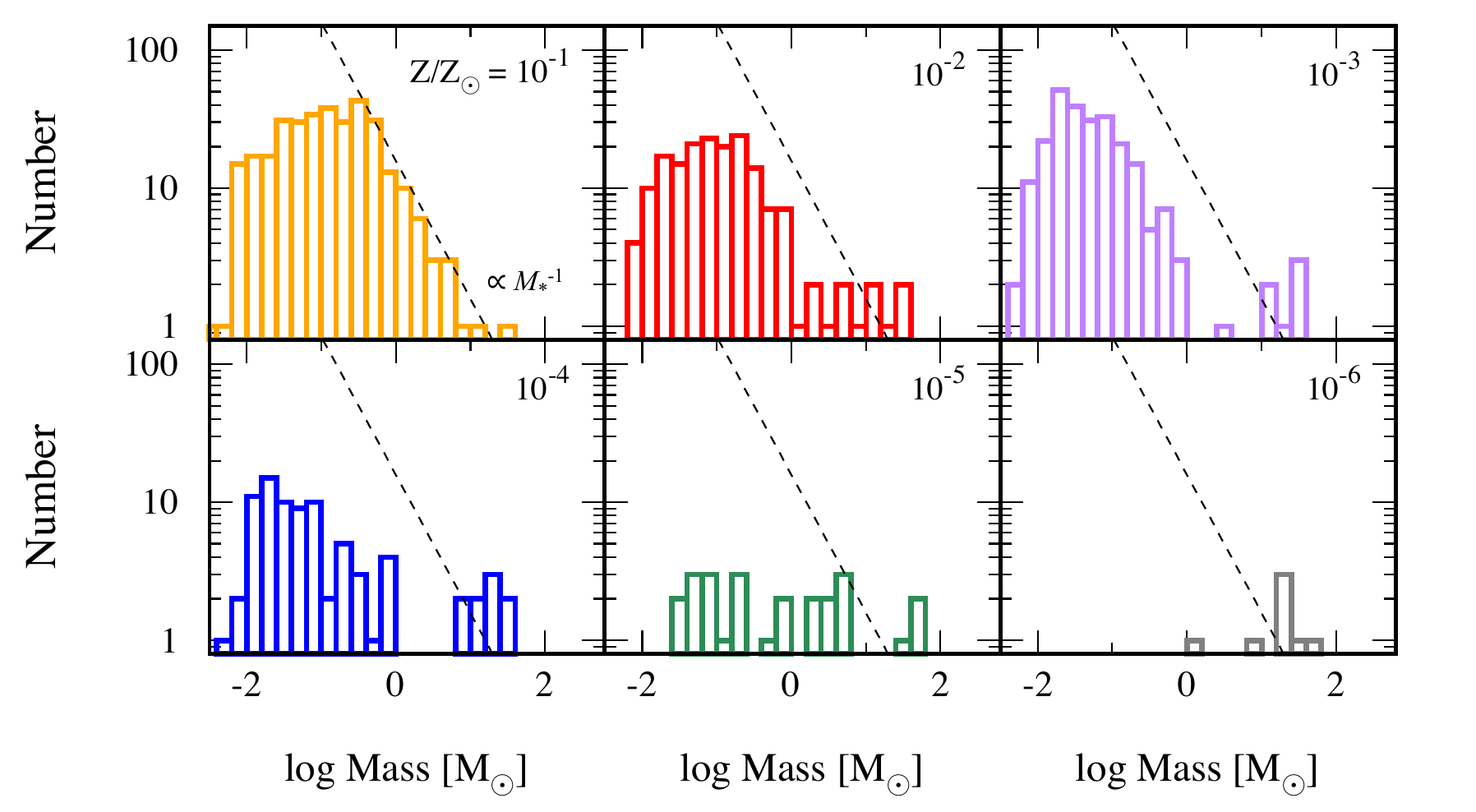}
\end{center}
 \caption{Illustration of the change of fragmentation behavior and resulting stellar mass spectrum for different metallicities, ranging from $10^{-6}\,Z_{\odot}$ to $10^{-1}\,Z_{\odot}$. For high metallicities the resulting IMF is bottom heavy with a peak below one ${\rm M}_{\odot}$ and a power-law decline above, similar to the present-day IMF. At low $Z$ the mass spectrum becomes logarithmically flat and top heavy similar to what we expect for the Pop III IMF (cf.\ Figure \ref{fig:PopIII-IMF}). This is a reproduction of Figure 10 of \citet{Chon21} and used here with permission of the authors and MNRAS.   }
\label{fig:IMF-transition}
\end{figure}

The existence of the star SDSSJ1029151+172927 \citep{Caffau2011Nat} provides strong evidence for the validity of the dust-induced fragmentation model \citep[e.g.][]{Schneider2012, Klessen2012, Chiaki2014, Bovino2016} and is a challenge to models based on fine-structure cooling. SDSS J1029151+172927 is  a truly primitive star in the constellation of Leo with a metallicity that falls well below the suggested critical value of $D_{\rm trans}$ for atomic fine-structure lines to be relevant. The star has elemental abundances in the range $10^{-5}$ to $10^{-4}$ of the solar value for all of the elements measured in its spectrum \citep{Caffau2012}, setting it apart from other extremely metal-poor stars which typically have enhanced CNO abundances despite being very iron poor \citep{Beers2005}. 

What is the physical origin of this behavior? Simple numerical experiments \citep{Li03} indicate that turbulent gas in a cooling regime (i.e.\ one in which $T$ drops with increasing density and the effective adiabatic index is $\gamma_{\rm eff} < 1$) can fragment very efficiently to build up a cluster of low mass stars. Conversely, star formation in a heating regime ($T$ increasing with density, $\gamma_{\rm eff} >1$) is biased towards forming very few and very massive objects. It has been posited \citep{Larson05} that the transition between these regimes provides ideal conditions for efficient fragmentation and introduces a characteristic mass scale to the system.  This mass scale is approximately the Jeans mass at the density and temperature of the transition point \citep{Jappsen2005}. It can be understood as a consequence of the fact that fragmentation in turbulent, self-gravitating gas often occurs in filaments, which are unstable against fragmentation for $\gamma_{\rm eff} < 1$ and stable for  $\gamma_{\rm eff} > 1$ \citep{Larson85}, with isothermal gas constituting a critical case \citep{Kawachi98}. As all current models of the assembly of the first galaxies indicate that the gas develops a filamentary morphology, 
it is reasonable to assume that filament fragmentation will dominate in the earliest galaxies just as it does in the local ISM of the Milky Way \citep{Hacar22}. 

The transition from $\gamma_{\rm eff} < 1$ to $\gamma_{\rm eff} > 1$ corresponds to a downward ``kink" in the effective equation of state. Inspection of Figure \ref{fig:initial-collapse} reveals two sets of such kinks, one associated with atomic fine-structure line cooling (for metallicities $-3 \lesssim \log_{10} Z/Z_{\odot} \leq 0$) and one with dust cooling (in the range $-5 \lesssim \log_{10} Z/Z_{\odot} \leq 0$). 
The associated characteristic masses are $\sim 100\,$M$_{\odot}$ for fine structure cooling and $< 1\,$M$_{\odot}$ for dust cooling.
This implies that atomic fine-structure cooling is not able to produce the extremely metal-poor low-mass stars we detect in stellar archeological surveys (see Section \ref{sec:low-z}) in sufficiently large numbers, providing additional evidence for dust cooling being the dominant process driving the transition from the Pop III to the Pop II IMF

To summarize the above, in Figure \ref{fig:IMF-transition} we show the stellar mass spectrum that results from systematically varying the metallicity of the star-forming gas \citep[for further details, see][]{Chon21}. Although similar caveats apply here as for our previous discussion of the Pop III IMF, we nevertheless see a clear influence of metallicity. Values close to solar lead to a bottom-heavy IMF which peaks around $0.2 - 0.3\,$M$_{\odot}$ with a power-law drop towards larger masses, similar to the present-day values \citep{Kroupa2002, Chabrier2003}. As the metallicity gets smaller, deviations from the present-day IMF become more pronounced and the mass spectrum becomes logarithmically flat and top heavy, more comparable to the Pop III IMF.  

\section{Observational probes of Pop III stars}
\label{sec:obs-probes}
Characterizing the properties of Pop III stars is difficult, as most of what we know about them comes from theoretical models and numerical simulations rather than observational constraints, which are hard to obtain and highly indirect at best. In this section, we provide an overview of the available data from the high-redshift and the  present-day Universe.

\subsection{High-redshift Universe}
\label{sec:high-z} 
There are several different approaches that we can take when trying to use observations of the high-redshift Universe to constrain the properties of Pop III stars. First, we can try to detect the stars themselves. Second, we can observe the bright explosive transients that the most massive Pop III stars produce at the end of their lives (supernovae and gamma-ray bursts), Finally, we can look for larger-scale signatures of the energetic feedback produced by massive Pop III stars, particularly through the imprint they leave on the cosmic 21-cm background.

\subsubsection{Direct detection of Pop III stars}
\label{sec:direct-detection}
Many of our uncertainties regarding the formation of Pop III stars could be resolved if we could observe them directly in the high-redshift Universe. The possibility of doing so with current or near-future facilities has been investigated by a number of authors \citep[see e.g.][or \citealt{Carr1984} for some early work]{Oh2001b,Scannapieco2003,Greif2009,Zackrisson2011,Zackrisson2012,Rydberg2013,MasRibas2016,Riaz2022}. Unfortunately, the general conclusion of these studies is that this is not possible, at least if we are interested in Pop III stars with masses $< 1000 \, {\rm M_{\odot}}$ forming in minihalos or atomic cooling halos with efficiencies similar to those predicted by numerical simulations. A recent calculation by \citet{Schauer2020} highlights the problem. They show that a single $1000 \, {\rm M_{\odot}}$ Pop III star forming at a redshift $z > 10$ will produce a flux in the NIRCAM wide-field filters on JWST of between $5 \times 10^{-4}$ and $2 \times 10^{-3}$~nJy, considering only the stellar emission. Accounting for nebular emission increases these numbers by about a factor of ten, but they are still far below the limiting values of a few nJy reachable with deep JWST imaging \citep[][]{merlin2022}, although they would be in reach of a hypothetical 100~m extraterrestrial telescope \citep{Schauer2020}.

In order to be detectable by JWST, our source therefore needs to be about 100 times brighter than we expect for a single very massive Pop III star. One way to reach the necessary brightness would be to have a cluster of such stars. However, the required stellar mass -- around $10^{5} \, {\rm M_{\odot}}$ in massive Pop III stars \citep{Zackrisson2011} -- does not seem achievable in a minihalo, given what we know about Pop III star formation. For instance, simply building up this much mass in a central cluster given the typical mass inflow rates we measure in simulations would require $\sim 10$~Myr, but feedback from the assembling cluster will disrupt the inflow long before this. Moreover, the required star formation efficiencies are unreasonably large, for instance, a minihalo with a mass close to $M_{\rm crit}$ would have to convert close to 100\% of its baryonic mass into stars. Clusters of the required size could conceivably build up in more massive galaxies, but the availability of sufficient metal-free gas in these systems is unclear. Models for the evolution of the Pop III star formation rate density agree that some level of Pop III star formation persists down to $z \sim 5-6$ (see Figure~\ref{fig:SFRD}), but disagree greatly on the average mass in young Pop III stars per galaxy. For example, \citet{Xu16a}, \citet{Mebane18}, \citet{Liu2020} and \citet{Skinner2020} argue that close to reionization, the typical mass in young Pop III stars per galaxy is $\sim 10^{3} \, {\rm M_{\odot}}$, while \citet{Pallottini2014} and \citet{Sarmento2018} find much larger values of $\sim 10^{5} \, {\rm M_{\odot}}$. This difference stems from differences in how Pop III star formation in minihalos is treated in the models. If the process is strongly suppressed by LW radiation, baryonic streaming or the impact of limited numerical resolution, then there is less pre-enrichment accreted by more massive galaxies, making it more likely to find large clusters of Pop III stars in these systems. On the other hand, if Pop III star formation in minihalos is barely affected by early feedback, as in e.g.\ \citet{Skinner2020}, then there is greater pre-enrichment of the gas in massive galaxies, resulting in smaller Pop III cluster masses. Intriguingly, this suggests that a non-detection by JWST of any Pop~III-dominated high redshift galaxies may allow us to constrain the efficiency of star formation in minihalos.

Another way to reach the brightness necessary for JWST to detect small clusters of massive Pop III stars is gravitational lensing. For the example discussed above, we would need a magnification factor of around 100 in order to increase the brightness sufficiently. To detect individual Pop III stars at high redshift, even higher magnifications would be required. \citet{Rydberg2013} estimate that a magnification of around 3000 is required for JWST to detect a lensed $60 \, {\rm M_{\odot}}$ star at $z \sim 10$. Such high magnifications are achievable close to lensing caustics  \citep{Windhorst2018,Diego2019}, but the disadvantage is that only a very small fraction of the high redshift Universe is magnified by such a large amount. The chance of actually detecting a highly lensed Pop III star therefore depends on the number of massive Pop III stars present in the high redshift Universe and hence on the Pop III SFRD. Unfortunately, even for values of this at the upper range of predictions from the literature (see Section~\ref{sec:csfrd}), the probability of actually detecting a highly lensed Pop III star or star cluster is small \citep{Rydberg2013,Diego2019}. Despite this, two possible candidate objects have recently been reported in the literature: a highly lensed star cluster at $z = 6.629$ \citep{Vanzella2020} and {\em Earendel}, an extremely magnified single star at $z = 6.2$ \citep{Welch2022}, which \citet{Schauer2022} argue has a non-negligible chance of being a Pop III star. Whether either of these objects is truly primordial is currently unclear but will likely be resolved by JWST in the near future.

Finally, if some fraction of Pop III stars are supermassive, with masses of $M \sim 10^{5} \, {\rm M_{\odot}}$ and luminosities of $L \sim 10^{9} \, {\rm L_{\odot}}$, then these may be directly detectable by JWST with no or minimal lensing \citep{Surace2018, Surace2019,Vikaeus2022}. They may also be detectable by Euclid and the Nancy Grace Roman Space Telescope (RST), given magnifications of $10 - 1000$ \citep{Vikaeus2022}. However, we expect these stars to be rare objects (see discussion in Section~\ref{sec:supermassive}), which strongly limits the number we expect to detect in a typical survey. For example, \citet{Vikaeus2022} show that with reasonable survey parameters  RST will only detect $\sim 10$ supermassive stars, even if their  number density at high redshift is close to the predicted upper limits, while Euclid will be unlikely to detect any. On the other hand, JWST should detect $\sim 30$ or more in the redshift range $7 < z< 10$.

It is also important to note that detecting Pop III stars in the early Universe is only part of the problem. Equally important is our ability to distinguish them from other high-redshift objects. In the observer's frame, Pop III stars will have very red colours owing to their high redshifts and the almost complete absorption of radiation shortwards of Lyman-$\alpha$ by the neutral IGM. This is true regardless of their rest-frame colours, but will be exacerbated in cases where pre-main sequence Pop III stars or highly-inflated massive stars contribute significantly to the observed flux \citep{Mitani2019,Woods2021}. However, Pop II-dominated galaxies at the same redshift will also be very red, as will dusty starbursts at lower redshifts \citep{Naidu2022, Zavala23}. Pop III-dominated systems can potentially be distinguished from Pop II-dominated systems based on their broad-band colours, but doing this purely with the NIRCAM filter set requires an almost complete absence of nebular emission (i.e.\ a very high escape fraction of ionizing photons, $f_{\rm esc} > 0.95$; see \citealt{Zackrisson2011}). For galaxies with lower escape fractions, detection in the MIRI 560W and 770W filters is also necessary for distinguishing between Pop III and Pop II stars, limiting this method to only the brightest sources \citep{Zackrisson2011,Trussler2022}. 

A more promising route for distinguishing Pop III stars from other high redshift objects involves spectroscopy. Massive Pop III stars are unusually hot compared to metal-enriched stars \citep{Schaerer2002} and hence produce a much larger flux of photons with energies above 54~eV that are capable of ionizing He$^{+}$ to He$^{++}$. It was therefore quickly realized that the recombination lines of HeII, particularly the 1640\AA\ line, could act as an important diagnostic of Pop III star formation \citep{Oh2001b,Tumlinson2001,Schaerer2002}. Unfortunately, detection of this line by itself does not prove that the source is a Pop III star, as Wolf-Rayet stars, X-ray binaries and black holes can also produce strong HeII emission \citep[see e.g.][]{Erb2010,Schaerer2019}. A salutary example is provided by the $z=6.6$ Lyman-$\alpha$ emitter CR7 \citep{Sobral2015}. It shows a strong HeII 1640\AA\ emission line that was interpreted by some authors as being due to Pop III star formation \citep[e.g.][]{Pallottini2015,Visbal2016}. However, subsequent observations detected [OIII] emission from this source \citep{Bowler2017}, which is inconsistent with this explanation, and it is now thought that CR7 is either a low luminosity AGN or a metal-poor starburst. A Pop III origin for HeII 1640\AA\ emission becomes more plausible if there is no associated emission from highly ionized metals, but even in this case a Pop II origin cannot be  excluded
\citep{Katz2022}. Altogether, the equivalent width of  HeII recombination lines appears to be a promising diagnostic parameter, with Pop III stars producing larger widths than other astrophysical sources \citep{Nakajima2022}, although the requirement to measure it accurately may again restrict the use to only the brightest sources.

\subsubsection{Pop III transients: supernovae and gamma-ray bursts}
\label{sec:transients}
The high peak brightness of Pop III CCSN and PISN potentially make them much easier targets to observe than their massive star progenitors. The predicted peak luminosity of a Pop III CCSN can reach values as high as a few times $10^{11} \, {\rm L_{\odot}}$ \citep{Whalen2013b}, around $10^4$ times larger than the luminosity of the single $1000 \, {\rm M_{\odot}}$ star. Pop III PISNe are even brighter, with peak luminosities a factor of ten or more greater than for CCSNe \citep{Whalen2013c}. Although these peak luminosities persist for only brief periods, Pop III SNe remain brighter than their progenitors over a period of weeks to months.

In a series of papers, Whalen and collaborators have examined the detectability of Pop III CCSNe, hypernovae, pulsational pair instability SNe and PISNe by JWST and other future facilities 
\citep{Whalen2013b,Whalen2013c, Whalen2014,Smidt2014,Smidt2015}. They show that Pop III PISNe will be bright enough to detect with JWST at redshifts $7 < z < 15$ for hundreds of days. The brighter explosions may also be marginally detectable with the RST, but will likely be too faint to detect with Euclid or in the Pan-STARRS or LSST sky surveys. Similar results are found for pulsational pair-instability SNe and hypernovae. CCSNe are harder to detect: Some explosion models are bright enough to be detectable with JWST at high redshift but others are not, and all are likely too faint to detect with the RST.

Unfortunately, although many Pop III SNe will be bright enough for JWST to detect, we expect them to be rare because of their short duration, and so the expected number of detections is small. The Pop III SNe rate depends on the star formation rate and IMF of Pop III stars and hence is subject to the considerable uncertainties in both of these quantities that we have discussed previously. However, several recent studies broadly agree that we should expect around $3 -7 \times 10^{-4}$ Pop III PISNe per JWST NIRCAM field of view per year \citep{Hummel12,Johnson2013,Magg2016,Hartwig2018}. The numbers for CCSNe are larger, but not by more than an order of magnitude \citep{Magg2016}. The chance of detecting any Pop III SNe in a dedicated survey is therefore tiny. Serendipitous detections remain plausible, although unambiguously identifying the detected objects as Pop III SNe without multi-epoch imaging may be problematic \citep{Hartwig2018}. At redshifts close to the epoch of reionization, confusion with SNe produced by low metallicity Pop II stars will also be a major issue \citep{Moriya2022}. 

The other transients related to Pop III star formation that are detectable at cosmological distances are long-duration gamma-ray bursts (GRBs). The extremely high luminosities associated with these events makes them readily visible even at high redshifts, as demonstrated by GRB 090423, which has a spectroscopically-confirmed redshift of $z \approx 8.2$ \citep{Tanvir2009} and GRB 090429B, which has a photometric redshift of $z \approx 9.4$ \citep{Cucchiara2011}. Current gamma-ray telescopes such as Swift \citep{Gehrels2004} are unlikely to detect GRBs at redshifts much higher than $z \sim 9$, because of limited sensitivity, and hence are mostly sensitive to GRBs produced by metal-enriched stars \citep{Ghirlanda2015}. However, future gamma-ray facilities such as THESEUS will expand the redshift range in which GRBs and their afterglows can be readily detected up to $z \sim 20$ \citep{Amati2018}, allowing us to probe redshifts at which Pop III sources will dominate. In view of this, it is clearly of interest to be able to make predictions for the number of Pop III GRBs that we expect these future facilities to detect. Unfortunately, this turns out to be even less well constrained than the number of Pop III SNe we expect at high redshift. As well as uncertainties due to the Pop III SFR density and IMF, the number of GRBs that we expect to detect also depends on several other highly uncertain numbers \citep{BrommLoeb2006}: the fraction of massive Pop III stars that are found in close binary systems, the fraction of these stars that are rapidly rotating, and the beaming factor (i.e.\ the probability that a randomly selected GRB happens to beam its radiation in our direction). Estimates of the number of Pop III GRBs that we should detect given a perfect telescope therefore range from values as high $\sim 100 \, {\rm yr^{-1}}$ \citep{Kinugawa2019,Lazar2022} to ones as small as $0.1 \, {\rm yr^{-1}}$ \citep{Toma2016}.

%% MRNT
\begin{marginnote}[]
\entry{Gamma-ray bursts}{GRBs are highly energetic bursts of electromagnetic radiation that can last from tens of milliseconds to several hours.}
\end{marginnote}

\subsubsection{$21\,$cm global signal}
\label{sec:21cm-and-CMB}
The ground state of atomic hydrogen exhibits a hyperfine splitting owing to the interaction between the proton and electron spins, with an energy difference of $5.9 \times 10^{-6}\,$eV (corresponding to $\lambda \approx 21\,$cm) between the two states. The radiative transition between these states -- the $21\,$cm line -- is frequently used to determine the properties of atomic hydrogen from the local Universe to very high redshifts.
The global (i.e.\ sky-averaged)  $21\,$cm signal is an important probe of the Universe at the redshifts when the first stars formed. When averaged on very large scales, the contribution of density fluctuations cancel out and the strength of the line is determined solely by the spin temperature $T_{\rm s}$ of the transition, yielding an emission line if this is larger than the CMB temperature $T_{\rm CMB}$, and an absorption line if it is smaller. The evolution of $T_{\rm s}$ is controlled by three physical processes: collisions with H, H$^{+}$ and e$^{-}$, the absorption and emission of 21~cm photons, and the Wouthuysen-Field effect (i.e.\ hyperfine transitions brought about by the absorption and emission of Lyman-$\alpha$ photons;  see \citealt{Wouthuysen52}, \citealt{Field58}). 
%% MRNT
\begin{marginnote}[]
\entry{Spin temperature}{Effective temperature that would produce the observed ratio of parallel to antiparallel spins if the atomic hydrogen gas were in thermal equilibrium.}
\end{marginnote}

At redshifts immediately prior to the formation of the first stars, the interaction with the CMB dominates and $T_{\rm s} \approx T_{\rm CMB}$. However, once Pop III stars start to form, the Lyman-$\alpha$ photons that they produce allow the Wouthuysen-Field effect to become dominant, driving $T_{\rm s}$ towards the kinetic temperature of the gas. The timing of this transition therefore lets us learn something about when (and how quickly) Pop III stars started to form, and the information that it provides us on the temperature of the gas constrains the rate at which the high-redshift IGM is heated by X-ray sources \citep[e.g.][]{Fialkov14, Pacucci14}, its degree of ionization \citep[which becomes more important at lower redshift; ][]{Park20}, and possibly also the coupling between dark matter and gas \citep[e.g.][]{Barkana18, Munoz18, Liu19}. A measurement of the global 21~cm signal could also in principle  test the existence of any radio background in addition to the CMB \citep[e.g.][]{Feng18, Ewall18, Fialkov19, Reis20}. Measurement of the high-$z$ $21\,$cm line therefore provides stringent constraints on the early evolution of the cosmic star formation history and the properties of the first stars \citep{Madau14-review}, as reviewed in  \citet{Furlanetto2006}, \citet{Barkana2001}, \citet{Fan06_Review}, \citet{Morales10}, \citet{Pritchard12}, or \citet{Mesinger19}.

Experiments aiming at measuring the global signal at high redshifts include EDGES \citep{Bowman13}, LEDA \citep{Price18}, SARAS \citep{Singh18},  PRIZM \citep{Philip19}, or REACH \citep{deLeraAcede22}.  
The first tentative detection of this signal was reported by the EDGES collaboration \citep{Bowman18}. Their signal is much stronger than expected and efforts to explain this have led to considerable theoretical speculation but few firm conclusions. In addition, we note that the true nature of this signal is highly debated \citep[e.g.\ see][]{Hills18, Sims20}. If it is of cosmological origin it represents one of our first direct constraints on the star formation process at a redshift of $z\sim 17$ as reflected by the impact of Lyman-$\alpha$ coupling and X-ray heating.

The characteristics of early star formation can influence the global $21\,$cm signal in a variety of additional ways. 
For example, the  recovery time (i.e.\ the average duration for metal-enriched material, which has been expelled from the Pop III star-forming halo, to mix with the ambient material and recollapse to form Pop II stars; see Section~\ref{sec:PopIIIsn}) can influence the observable $21\,$cm signal.  \citet{Mirocha18} and \citet{Magg2022a} find that short recovery times, resulting from weak supernova feedback, possibly due to Pop III star formation occurring in more massive halos or with low efficiency, result in a noticeably steeper global $21\,$cm signal than models with long recovery times from strong feedback, characteristic of scenarios in which stars build up in low-mass halos or form in groups leading to mechanical energy and momentum input from multiple supernovae. The global $21\,$cm signal is also sensitive to the Pop III IMF.
In a study focusing on the impact of Lyman-$\alpha$ radiation at high redshift and neglecting ionization and X-ray emission,  
\citet{GesseyJones22} show that current and future observations can probe the characteristic stellar mass if the IMF is dominated by stars lighter than $\sim 20\,$M$_\odot$. However, the expected signatures are relatively weak and comparable to the instrumental precision. If the Pop III IMF is dominated by very massive stars, we can no longer discriminate between different IMF models. This may change once X-rays are included, which are thought to have a different dependence on the IMF compared to the Lyman-$\alpha$ emission.  Detailed measurements of the $21\,$cm emission at high redshifts, once available, will open up many interesting pathways to study the birth of Pop III stars and the transition to Pop II star-formation.

\subsection{Low-redshift Universe}
\label{sec:low-z}
The most stringent constraints on the properties and mass distribution of Pop III stars are likely to come from the study of  extremely metal-poor (EMP) stars in the Local Group. These stars have a metallicity of $\mathrm{[Fe/H]} < -3$ as expressed by the iron abundance relative to the solar value \citep[for a review see][]{Beers2005}. A common approach is to use large-area  photometric or low-resolution spectroscopic surveys\footnote{For example, the  Sloan Digital Sky Survey provides spectral data for several $10^5$ stars (for DR12, see \citealt{SDSS-DR12-2016}), and Gaia for more than $10^8$ sources (for DR3, see \citealt{Gaia-DR3-2022}). 
}
for a first estimate of the stellar metallicity \citep[e.g.][]{Ludwig08, Caffau2011}. This information can then be used to put together a list of objects for high-resolution spectroscopic follow-up observations. Good targets are main-sequence turn-off stars, because they are numerous and bright, and they have chemical abundances that are essentially unaltered since the time of their formation \citep{Pinsonneault97}. The resulting stellar archeological surveys, such as those aiming at the Galactic halo and/or bulge \citep{Christlieb08, Schoerck09, Frebel2010, Salvadori2010, Caffau2013, Starkenburg17, Arentsen2020, Ishigaki21, Li22} or nearby satellite dwarf galaxies \citep[e.g.][]{Koch2013, Salvadori2015, Kirby2015, Roederer2016, Ji2016A, Ji2016B, Ji23},  can contribute to our knowledge of primordial stars in two key ways. 
%% MRNT
\begin{marginnote}[]
\entry{Abundances}{We adopt the notation ${\rm [X/H]} = \log_{10}(m_{\rm X}/m_{\rm H}) - \log_{10}(m_{{\rm X},\odot}/m_{{\rm H},\odot})$ where $m_{\rm X}$  and $m_{\rm H}$ are the mass abundances of any element X and hydrogen, and $m_{{\rm X},\odot}$ and $m_{{\rm H},\odot}$ are the corresponding solar abundances.}
\end{marginnote}

\subsubsection{Galactic archaeology and the high-mass end of the Pop III IMF}
\label{sec:arch}
We can use the  abundance pattern determined in EMP stars to infer the properties of the progenitor stars which provided the observed chemical enrichment.  It is common practice to employ stellar evolution calculations combined with explosive nucleosynthesis models for this comparison, both for individual stars \citep[e.g.][]{Ishigaki14b, Bessell15,  Placco16, Aguado18b, Aguado19} as well as for large samples \citep{Fraser17, Ishigaki18}. This is facilitated by libraries of modelled supernova yields  \citep{Heger10, Nomoto2006, Umeda08, Ishigaki18},  which typically depend on the stellar mass of the exploding star, the explosion energy and one or several parameters that quantify the mixing-and-fallback process. The latter cannot be simulated self-consistently in 1D supernova models and in principle require detailed 3D radiation-hydrodynamic calculations of the star-forming halo to determine \citep{Magg2022}. An additional constraint on this process is provided by the fact that the supernova yields need to be diluted with metal-free gas in the surrounding of the exploding star in order to match the absolute metallicity of the observed star \citep{Magg19}. 

The observational data indicate that most EMP stars are rich in carbon in relative terms \citep{Beers2005, Frebel15}. Of specific interest amongst these carbon-enhanced metal-poor (CEMP) stars are the so-called CEMP-no stars. Despite an excess of carbon relative to iron of more than ${\rm [C/Fe]} = 1$ they show no enhancement in neutron-capture elements, i.e.\ they have ${\rm [Ba/Fe]} < 0$ \citep{Frebel2005, Komiya2007, Norris2013, Keller2014, Aguado18a, Nordlander19}. The origin of the elemental abundance patterns in these stars is one of the key questions of early chemical evolution. Their dynamics appear to be inconsistent with them having gained their metals via mass transfer from a binary companion \citep{Zepeda2022}. \citet{Umeda2002} propose that the abundance pattern of CEMP-no stars is the fingerprint of a so-called mixing-and-fallback or faint supernova \citep[see also][]{Ishigaki14,Chan20}. Such events have relatively low explosion energies, and consequently the inner layers of the star, containing in particular iron, are suggested to fall back on the compact remnant, such that only the outer  layers, containing carbon and other light elements, are ejected and become mixed with ambient material. 

Assuming that the oldest and most metal-poor stars in the Galaxy have been supplied with heavy elements by only one or at most two supernova explosions 
-- a common assumption but one that may not be true for all EMP stars \citep{Hartwig2018b,Hartwig2019,Welch2021} -- one finds that the relative abundances of heavy elements are most consistent with core collapse supernovae from Pop III stars in the mass range $20 - 40\,$M$_\odot$ 
\citep[e.g.][]{Bonifacio2015, Caffau2011Nat, Caffau2012, Cooke2014, Frebel2005, Iwamoto2005, Joggerst2009, Joggerst2010, Norris2013, Keller2014, Lai2008}, although a few examples exist of stars that are more consistent with enrichment by hypernovae \citep{Placco2021,Skuladottir2021} or PISNe \citep{Aoki2014,Salvadori2019}. The conclusion that most Pop III SNe were CCSNe is consistent with the work of \citet{Mapelli2006}, who compute the expected density of intermediate-mass black holes in the Galaxy by assuming they are the relics of higher-mass Pop III stars. This allows them to derive an upper limit from the non-detection of ultra-luminous X-ray sources in the Milky Way. The approach has been extended by \citet{deBennassuti14, deBennassuti17} who also include the Galactic metallicity function and the abundances of EMP stars. Again these data appear inconsistent with pair-instability supernovae being important (although see \citealt{Salvadori2019} for a contrary view). Taken together, these studies suggest that most Pop III stars forming through the standard pathway outlined in Section \ref{sec:standard-scenario-of-PopIII-formation} are not likely to reach masses significantly above $\sim 40\,$M$_\odot$, with more massive objects being extremely rare and possibly being the result of collisions in very dense clusters (see Section \ref{sec:GW}). 

Finally, we note that complementary information on the metal yields produced by the first generation of stars is potentially provided by observations of extremely metal poor foreground absorption systems in the spectra of distant quasars. The most metal-poor of these systems have metallicities comparable to EMP stars \citep[e.g.][]{Cooke2017,Robert2022,Welsh2022}. If we assume that each of these systems has been enriched by either a single Pop III supernova or a small number of such SNe, then we can use the observed metal abundances to infer their properties. The results we obtain from this procedure are consistent with those coming from EMP stars. Most of the systems show elemental abundances consistent with enrichment primarily by CCSNe \citep{Welsh2019}, although there is an example of a cloud in the broad-line region of a $z = 7.6$ quasar with an abundance pattern that might indicate a PISN origin \citep{Yoshii2022}. This supports the idea that most Pop III supernovae are CCSNe, with PISNe occurring only rarely.

\subsubsection{Potential for direct detection of Pop III stars and low-mass end of the IMF}
\begin{figure}
\begin{center}
\includegraphics[width=12.2cm]{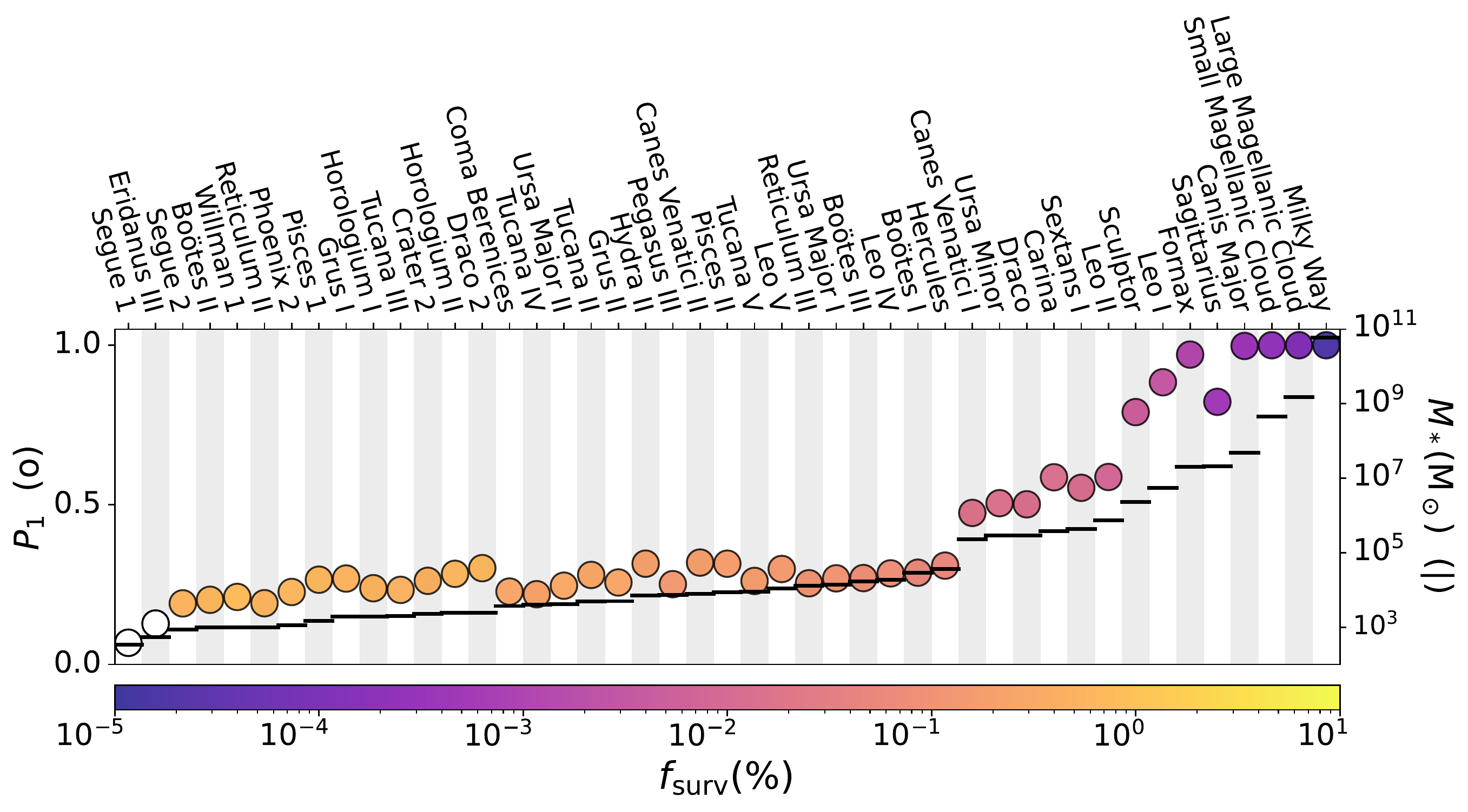}
\end{center}
 \caption{Properties of the Milky Way and its satellite galaxies. The horizontal line in each column indicates the stellar mass as inferred from observations. The filled circles indicate the probability that the systems contains at least one low-mass Pop III survivor, with the color encoding the median fraction. The figure is based on data from \citet{Magg18}.   }
\label{fig:PopIII-survivors}
\end{figure}

Stellar evolution models \citep[e.g.][]{Kippenhahn2012, Maeder2012} tell us that  low-mass Pop III stars with $M < 0.8\,${M$_\odot$} must have survived until the present day. Furthermore, models of the surface pollution of Pop III stars by metals from the ISM suggest that this is not an important effect and that Pop III stars around at the present day would still be spectroscopically identifiable as such \citep{Frebel2009,Tanaka2017}. Therefore, if these stars exist, there is a chance to directly detect some of them in current or future stellar archeological surveys.  One can use semi-analytic models to predict the most likely locations to find these Pop III survivors. Whereas the predicted numbers of survivors are highest in the Galaxy, specifically in the bulge, they are very difficult to find there because of the presence of so many higher metallicity stars. It is therefore better to search in environments that did not have continuous star formation over secular timescales, but instead only experienced an initial burst of star formation at high redshift and then remained quenched during subsequent cosmic evolution. Consequently, the lowest-mass  satellites of the Milky Way are expected to contain the largest fraction of Pop III survivors compared to their Pop II stellar content, and so they are the most promising targets for future attempts to find Pop III survivors with high-resolution, high signal-to-noise spectroscopic observations \citep{Magg18}.  Figure \ref{fig:PopIII-survivors} shows predictions of the expected fraction of genuine low-mass Pop III stars in Local Group galaxies together with their total stellar mass. We see that the lowest-mass satellites are the best places to look for these stars, although it is possible that some of these systems will be too small to contain any Pop III survivors.

Even non-detections allow us to put stringent limits on the low-mass end of the Pop III IMF. For example, \citet{Hartwig2015} estimate the expected numbers of low-mass Pop III stars in the Galactic halo based on semi-analytic models of the early star formation history in Milky Way-like halos. They conclude that if no such object is found in surveys with sample sizes of 4 million stars then we can exclude the existence of low-mass Pop III stars with masses below $0.8\,${M$_\odot$} with high significance. On similar grounds, \citet{Salvadori2007, Salvadori2010} develop a detailed model of the metallicity distribution function of metal-poor stars in the Galactic halo and suggest that Pop III stars should be more massive than $0.9\,${M$_\odot$} when matching their predictions with the observational data. A different conclusion was reached by  \citet{Tumlinson2006, Tumlinson2010}, who argues that the current abundance measurements are not precise enough to distinguish between different models of the primordial IMF based on studying the chemical evolution during the early build-up of the Galaxy. However, he also suggests that characteristic masses of a few $10\,${M$_\odot$} provide a better fit to the available data than masses of $100\,${M$_\odot$} or above, consistent with the supernova yields mentioned above.  Finally, \citet{Rossi2021} argue that our failure to find any Pop III stars in nearby ultra-faint dwarf galaxies already allows us to exclude Pop III stars with $M < 0.8\,${M$_\odot$}. Altogether, the prospects of finding surviving low mass first stars in our immediate neighborhood are highly exciting, but at present no firm conclusions about the low-mass end of the Pop III IMF can be drawn. 
This may change with the availability of new generations of multi-object spectrographs, such as 4MOST or WEAVE \citep[e.g.][]{Dalton2012, Feltzing2016, Feltzing2018}, which will allow for galactic archeological surveys of unprecedented size and depth that can help to resolve this issue.

\subsection{Gravitational waves}
\label{sec:GW} 
The detection of gravitational waves with extremely precise interferometers (advanced LIGO, \citealt{Aasi15},  advanced Virgo, \citealt{Acernese15}, or KAGRA, \citealt{Aso13}, all of which are currently operational, and the upcoming  LISA, \citealt{AmaroSeoane23}, DECIGO, \citealt{Kawamura11}, and TianQin \citealt{Luo16} experiments in space or the Einstein Telescope on the ground, \citealt{Punturo10}) opens up a new observational window to the Universe. The signals detected so far come from the gravitational inspiralling and merging of two stellar-mass black holes, two neutron stars, or pairs of a black hole and a neutron star. Whereas LIGO/Virgo/KAGRA can detect the mergers of compact objects  with masses up to several hundred solar masses out to $z \sim 1$ \citep[e.g.][]{Abbott19}, the planned observatories (such as LISA, TianQin, and DECIGO, or the  Einstein Telescope) have the potential to detect  considerably more massive objects out to higher redshifts. 

There are two primary ways to infer information about primordial stars from these data. On one hand, we can look at the properties of individual measurements, which gives us information about the total mass of the system, the mass ratio of the two constituents, and their spin. This provides constraints on the mass spectrum and multiplicity of Pop III stars and on possible formation and evolution pathways. On the other hand, we can study the rate of these events (typically normalized to a cosmic volume of $1\,$Gpc$^{3}$) as function of redshift, and compare the results with our understanding of the cosmic star-formation history. However, the astrophysical interpretation of these data is highly challenging for several reasons. First, the detected gravitational wave signal from the merger of two compact objects can occur many Gyr after the formation of the original stellar binary system. To link these two events requires complex binary evolution calculations that reproduce the correct stellar structure in the MS and post-MS phases and that reliably model mass transfer and the change of orbital parameters over secular timescales \citep[see e.g.][]{Belczynski02, Belczynski16, Eldridge16, deMink16, Mandel16, Marchant16, Marchant21, Stevenson17, Mapelli17,  Spera19}. Second, to explain the current data, the mergers of stars in a dense cluster environment may be necessary. The corresponding stellar dynamical processes contribute additional stochasticity and uncertainty to the problem \citep[e.g.][]{PortegiesZwart02, PortegiesZwart22}. And finally, the signal coming from genuine Pop III stars is much weaker than the rates expected from metal-enriched binaries, simply because the total number of metal-free stars is many orders of magnitude smaller, which can be see for example by comparing the corresponding cosmic star formation rate densities in Figure \ref{fig:SFRD}, and which cannot be compensated by differences in the IMF (Figure \ref{fig:IMF-transition}).     

The combined data of the first three observing runs of LIGO, Virgo, and KAGRA \citep{Abbottb-2021-LIGOVirgo, Abbott2021a-LIGO-Virgo} demonstrate that the mass spectrum inferred for the primary and secondary merger component is very wide and contains structure beyond a simple power law with a sharp high-mass cut-off. The mass distribution of the primary shows a first peak at $\sim 10$M$_\odot$ and a secondary one at $30 - 35\,$M$_\odot$. Although the majority of the systems detected have primary masses smaller than $45\,$M$_\odot$, the mass distribution clearly extends beyond $\sim 65\,$M$_\odot$ and reaches into the so-called pair-instability gap, at least for metal-rich stars (e.g.\ \citealt{Farmer19, Woosley21}; see also Section \ref{sec:sne}). Pair-instability and pulsational pair-instability supernovae lead to the complete disruption of the star and leave no remnant behind. Because metal-rich stars experience considerable mass loss due to strong winds or pulsational instabilities prior to the explosion, we expect that no black hole in the mass range between about $50\,$M$_\odot$ to $130\,$M$_\odot$ should form, whereas for zero-metallicity stars, which experience no mass loss, the pair-instability gap  covers masses from $\sim 90\,$M$_\odot$ to $\sim 260\,$M$_\odot$ \citep[however, with considerable uncertainties, see e.g.][]{Heger2002, Woosley15, Farmer20, Costa21, Vink21}. 

As we expect that most of the  observed gravitational wave detections come from the remnants of metal-enriched rather than zero-metallicity stars \citep[e.g.][]{Hartwig2016b, Belczynski2017, Tanikawa22}, these findings cast doubts on the widespread occurrence of pair-instability supernovae, and it has been argued that stellar collisions in a dense cluster environment provide an alternative pathway to produce the high-mass end of the observed black hole mass spectrum  \citep[e.g.][]{PortegiesZwart00, PortegiesZwart02, PortegiesZwart04, Giersz15, Mapelli16, DiCarlo19, DiCarlo21, Liu20a, DallAmico21, Wang22, Costa22}. At the low mass end of the distribution, the data indicate a lack of low-mass black holes below $\sim 6\,$M$_\odot$ with high significance. This is larger than the masses reported for black hole candidates in Galactic binaries \citep[reaching down to $\sim 3\,$M$_\odot$, see][]{Thompson19} and potentially hints towards a more top-heavy IMF of the progenitor stars. Most systems have mass ratios $q$ close to unity, but there are some for which $q<1$ is inferred with high significance. 

Another feature of these reported gravitational wave detections is that about one third of the systems exhibit spins which are possibly inclined with respect to the orbital angular momentum and which show signs of precession \citep[see also][]{Callister22}. This provides additional support for a collisional origin of these systems. The inferred merger rates for binary black holes have a 90\% credibility interval of $18 - 44\,$Gpc$^{-3}\,$yr$^{-1}$; those for binary neutron stars  lie in the range of $10$ to $1700\,$Gpc$^{-3}\,$yr$^{-1}$ \citep{Abbott2021a-LIGO-Virgo}. These values are still highly uncertain, but the precision will increase with time  as the number of detected gravitational wave events continues to go up. The data furthermore indicate a merger rate that grows with redshift as $(1+z)^\kappa$, with $\kappa$ only marginally constrained in the range from $1.1$ to $4.6$, at least for $z \lesssim 1$.  Theoretical predictions of gravitational rate event rates, specifically from the remnants of Pop III stars \citep[][]{Kinugawa2014, Hartwig2016b,  Schneider17, Belczynski2017, Belczynski22,  Dayal19, Neijssel19, Tang20, Ng21, Tanikawa21, Tanikawa22}, are roughly consistent with these numbers, but more stringent constraints clearly require larger statistics \citep[for a comprehensive overview of this aspect, see][]{Mandel22}.

\section{Summary}
\label{sec:summary}

\begin{summary}[Summary points]
\begin{enumerate}
\item The formation of the first stars begins at redshifts of $z \sim 30$ and is likely to continue to $z \sim 5$, dominating the cosmic star formation rate density until $z \sim 15$, at which point the build-up of metal-enriched Pop II stars takes over (Section \ref{sec:minimum-mass}).
\item Stellar birth in the high-redshift Universe shares the same complexity as present-day star formation. Primordial gas is highly susceptible to fragmentation and we expect the first stars to form in binary or higher-order multiple stellar systems with a wide range of separations, a flat mass ratio distribution, and a roughly thermal spread of orbital eccentricities (Section \ref{sec:disks-and-fragments}).
\item Most primordial stars build up in isolated high-redshift halos, which have not been affected by stellar feedback. We call this the standard pathway of Pop III star formation. We expect a wide distribution of masses, potentially ranging from the sub-solar regime up to several hundred solar masses. The initial mass function (IMF) is approximately logarithmically flat, which makes it top-heavy compared to the Milky Way values (Section \ref{sec:IMF-and-multiplicity}).
\item Feedback from the first massive Pop III stars plays a major role in regulating subsequent star formation. Photoionization and supernovae dominate within individual halos, whereas the Lyman-Werner and X-ray backgrounds built up by Pop III stars and their remnants govern the evolution on cosmic scales (Sections~\ref{sec:small-scale-impact} and \ref{sec:large-scale-impact}).
\item Pop III supernovae produce metals that enrich gas associated with their host halo (``internal enrichment'') but can also reach neighboring halos (``external enrichment''). The time it takes Pop II stars to form from metal-enriched gas remains poorly constrained and depends on the halo mass, the supernova energy, and details of the local environment. It is plausible that some fraction of the observed extremely metal-poor stars in the Milky Way and its satellites were enriched by single supernovae (Section~\ref{sec:transition-PopIII-to-PopII}).
\item If they exist, Pop III stars with masses below $\sim 0.8\,$M$_\odot$ will have survived until today and should be detectable in surveys of extremely metal-poor stars. However, none has yet been found. At the high-mass end, observations point towards Pop III stars that end their lives in core-collapse supernovae, implying a mass range of $20\ - 40\,$M$_\odot$. Evidence for more massive stars is sparse (Sections \ref{sec:high-z} and \ref{sec:low-z}). 
\item Under very rare and extreme conditions, the formation of supermassive stars is possible. Subjected to accretion rates of $\dot{M} \sim 0.1\,$M$_\odot\,$yr$^{-1}$ or higher, they can grow to masses of several $10^5\,$M$_\odot$, before the general-relativistic instability kicks in and triggers the collapse to a black hole. These objects can be the seeds for the supermassive black holes observed in high-redshift quasars. Because the infalling gas is expected to fragment, run-away collisions in dense and deeply embedded stellar system are thought to be important in this process (Section \ref{sec:supermassive}).
 \end{enumerate}
\end{summary} 
\begin{issues}[Future Issues]
\begin{enumerate}
\item From a theoretical point of view, better understanding the impact of stellar feedback is a key challenge of current research into Pop III star formation. Specifically, the question of how photoionizing radiation escapes from the immediate vicinity of the star remains uncertain. On large scales, the relative importance of Lyman-Werner feedback, X-rays and streaming velocities remains unclear, with several recent studies coming to highly different conclusions. Overall, it is likely that feedback has a negative effect on the Pop III star formation rate, but its influence on the mass distribution and the multiplicity of Pop III stars remains undetermined. 
\item On the observational side, no genuine low-mass Pop III star has been found so far in stellar archeological surveys. With the increasing sample size provided by future surveys, it will become possible to either confirm or rule out the existence of Pop III stars with masses below $\sim 0.8\,$M$_\odot$ with high statistical significance. 
\item Halos forming Pop III stars are unlikely to be observable with JWST without the aid of gravitational lensing, unless they have higher masses and star formation efficiencies than current models predict. Supermassive Pop III stars or extremely strongly lensed systems should be bright enough to see, but will be very rare, and so it is plausible that none will be detected (Section~\ref{sec:direct-detection}). Pop III supernovae should be visible with JWST, but are also very rare. Nevertheless, serendipitous discoveries remain a realistic possibility. Pop III gamma-ray bursts (GRB) are unlikely to be detected using current facilities, but will become accessible by next-generation GRB telescopes (Section~\ref{sec:transients}). Much work still needs to be done to understand how to relate detections (or non-detections) of high-redshift Pop III stars to constraints on models of Pop III star formation.
%\item Direct detection of Pop III stars or their associated transients (supernovae, gamma-ray bursts) is an exciting but unlikely prospect. We need to better understand what we learn about Pop III star formation in the case we do not observe any.
\item Detection and characterization of the global $21\,$cm signal at high redshift will allow us to put stringent constraints on the properties of high-mass Pop III stars as well as the early evolution of the cosmic star formation rate density. 
%Similarly, increasing the number of well characterized gravitational wave signatures from the mergers of black holes and neutron stars will reveal information about the formation and dynamical evolution of high-redshift star clusters.
\item The detection of gravitational waves from merging binary black holes in the mass range forbidden by pair-instability models provides additional support for the importance of stellar collisions, possibly occurring in the star clusters that are a natural consequence of the standard Pop III formation pathway (Section \ref{sec:GW}). Our ability to use these events to constrain the physics of Pop III star formation will improve as our sample size grows.
\end{enumerate}
\end{issues}

\section*{Acknowledgements}
This review would not have been possible without the input and help of many colleagues. Specifically, we mention Li-Hsin Chen, Sunmyon Chon, Paul Clark, Anastasia Fialkov, Lionel Haemmerl\'{e}, Tilman Hartwig, Takashi Hosokawa, Mattis Magg, S\'{e}bastien Martinet, Laura Murphy, Kazuyuki Omukai, Lewis Prole, and Raffaella Schneider for making some of their data available to us for some of the figures and tables provided here. In addition, we thank Tim Beers, Volker Bromm, Norbert Christlieb, Pratika Dayal, Anna Frebel, Melanie Habouzit, Alexander Heger,  Joseph Lewis, Mordecai-Mark Mac~Low, Michela Mapelli, Lucio Mayer, Dylan Nelson, Annalisa Pillepich, John Regan, Anna Schauer, Dominik Schleicher, Jennifer Schober, Matthew Smith, Marta Volonteri, Tyrone Woods, Naoki Yoshida, and many others for highly stimulating scientific discussions and valuable input.

We acknowledge funding from the European Research Council via the ERC Synergy Grant ``ECOGAL -- Understanding our Galactic ecosystem: From the disk of the Milky Way to the formation sites of stars and planets'' (project ID 855130) and via the ERC Advanced Grant ``STARLIGHT: Formation of the First Stars'' (project ID 339177). We also acknowledge funding from the Deutsche Forschungsgemeinschaft (DFG) via the Collaborative Research Center (SFB 881 -- 138713538) ``The Milky Way System'' (subprojects A1, B1, B2 and B8) and from the Heidelberg Cluster of Excellence (EXC 2181 - 390900948) ``STRUCTURES: A unifying approach to emergent phenomena in the physical world, mathematics, and complex data'', funded by the German Excellence Strategy. We also thank the German Ministry for Economic Affairs and Climate Action for funding in the project ``MAINN -- Machine learning in astronomy: understanding the physics of stellar birth with invertible neural networks'' (funding ID 50OO2206). We also thank for computing resources provided by the Ministry of Science, Research and the Arts (MWK) of \textit{The L\"{a}nd} through bwHPC and DFG through grant INST 35/1134-1 FUGG. Data are stored at SDS@hd supported by MWK and DFG through grant INST 35/1314-1 FUGG.

\pagebreak
 \appendix

\section*{Appendix: Average stellar properties and production rates of ionizing and non-ionizing photons}
\addcontentsline{toc}{section}{Appendix: Average stellar properties and photon production rates}
\label{sec:photon-production-rates}

The stellar properties discussed in Section \ref{sec:stellar-lum} and summarized in Figure \ref{fig:PopIII-HRD} can be used to estimate the number of ionizing and non-ionizing photons emitted by stars of different mass and metallicity. We focus our discussion on the standard pathway of Pop III formation (Section \ref{sec:standard-scenario-of-PopIII-formation}), that is on stars which have experienced small to moderate accretion rates, and which therefore enter the main sequence (MS) compact and hot. Our approach is based on evolutionary tracks of individual stars that are computed by self-consistently solving the equations of stellar structure, the production and transport of energy in the interior of the star, and the loss of mass and radiation through its surface as a function of time in spherical symmetry \citep[for further details, see the book of][]{Kippenhahn2012}. We consider both non-rotating stars as well as stars that spin with 40\% of the break-up velocity \citep[e.g.][]{Maeder2012}. The data presented here are obtained with the Geneva  code \citep{Eggenberger08}, and they are derived from stellar evolution models provided by \citet{Ekstroem12}, \citet{Georgy13}, \citet{Groh19}, \citet{Murphy21a, Murphy21b}, and Martinet et al. (in prep.).  We note that other approaches lead to very similar results. Specifically, we have compared with stellar evolution models obtained with the SEVN code \citep{Spera22} and found differences of at most a few percent (M. Mapelli \& T. Hartwig, private communication). 

\begin{figure}[htp]
 \setlength{\unitlength}{1cm}
 \begin{picture}(16,4)(0,0)
  \put( 0.0, 0.5){\includegraphics[width=16.1cm]{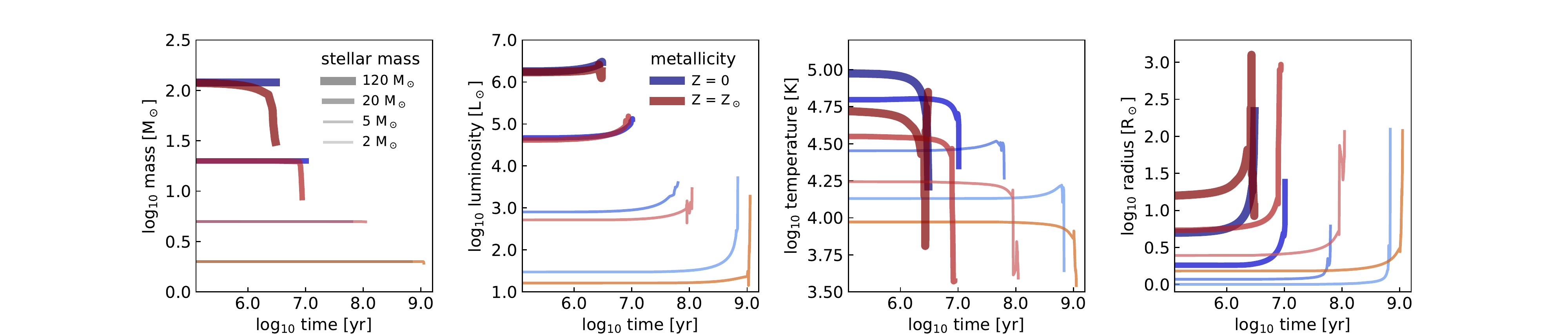}}
 \end{picture}
\caption{Stellar parameters as function of time for selected non-rotating stars of metallicity $Z=0$ (blue) and $Z=Z_\odot$ (red) with  masses of $2\,$M$_\odot$, $5\,$M$_\odot$, $20\,$M$_\odot$ and $120\,$M$_\odot$ (different line strength). From left to right: stellar mass, bolometric luminosity, effective surface temperature, and stellar radius. The tracks have been computed with the Geneva stellar evolution code by \citet{Murphy21a, Murphy21b} and \citet{Ekstroem12}.
}
\label{fig:stellar-evolution-2-5-20-120-Msun}
\end{figure}

It is clear that key stellar parameters, such as mass, bolometric luminosity, effective  temperature and radius, change during the main-sequence (MS) and post-MS evolutionary phases. We illustrate this in Figure \ref{fig:stellar-evolution-2-5-20-120-Msun}, which shows these parameters as function of time for non-rotating stars of mass $2\,$M$_\odot$, $5\,$M$_\odot$, $20\,$M$_\odot$ and $120\,$M$_\odot$ and zero as well as solar metallicity. When looking at the left-most panel, we notice that none of the zero metallicity Pop III stars experiences mass loss, whereas high-mass solar metallicity Pop I stars shed significant amounts of material in the post-MS phase due to strong line-driven winds and/or pulsational instability. The bolometric luminosity remains roughly constant during the main sequence, but can easily rise by a factor of ten or more in the giant phase, which is most noticeable in the second panel for stars of intermediate masses ($M = 2\,$M$_\odot$ and $M = 5\,$M$_\odot$). Independent of metallicity the effective surface temperature drops sharply (third panel), while the stellar radius increases enormously (right-most panel) in the giant phase of post-MS evolution.     

\begin{figure}[htp]
 \setlength{\unitlength}{1cm}
 \begin{picture}(10,11)(0,0)
  \put( 0.0, -0.5){\includegraphics[width=16.1cm]{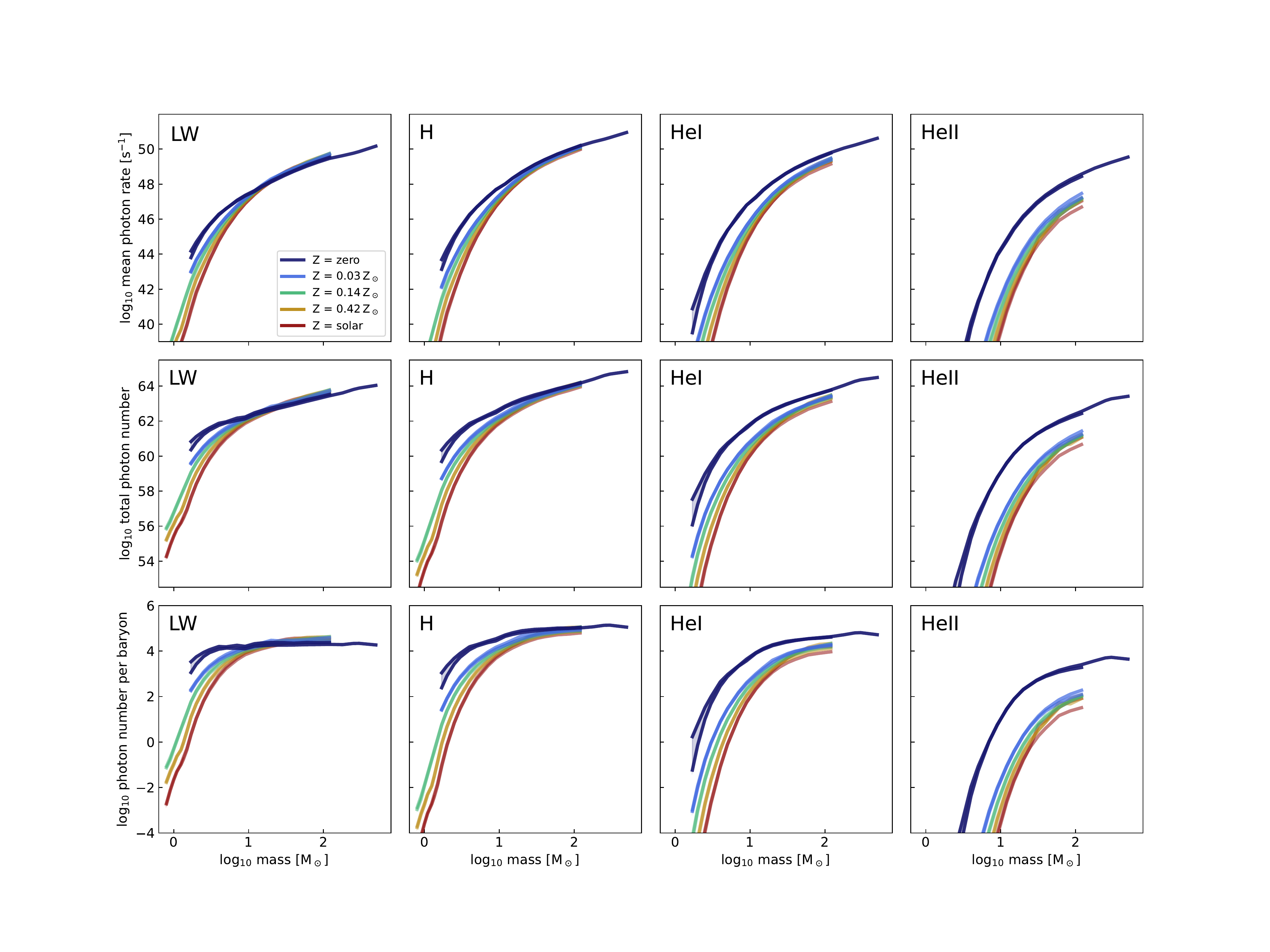}}
%  \put( 0.2, -0.5){\includegraphics[width=16.1cm]{Figures/fig-PopIII-photon-rates.pdf}}
 \end{picture}
\caption{Stellar emission properties for five different metallicities (dark blue:  $Z=0$, light blue: $Z = 0.03\,Z_\odot$, green:  $Z = 0.14\,Z_\odot$, orange: $Z = 0.42\,Z_\odot$, red: solar metallicity) and masses covering the range $0.8\,$M$_\odot \le M \le 120\,$M$_\odot$ for $Z \ge 0.14\,Z_\odot$,   $1.7\,$M$_\odot \le M \le 120\,$M$_\odot$ for $Z \le 0.03\,Z_\odot$, plus four additional models $
180\,$M$_\odot \le M \le 500\,$M$_\odot$ for $Z = 0$. We focus on the Lyman-Werner band ($11.2 {\rm eV} \le h\nu < 13.6\,{\rm eV}$) and on photons that can ionize H ($h\nu \ge 13.6\,{\rm eV}$), He ($h\nu \ge 24.6\,{\rm eV}$), and He$^{+}$ ($h\nu \ge 54.4\,{\rm eV}$). The top row depicts the number of photons emitted per second averaged over the entire MS and post-MS lifetime. The middle row provides the total number of photons produced, and the bottom row gives the total number of photons per baryon in the star.  The figure is constructed from data obtained from  \citet{Ekstroem12}, \citet{Georgy13}, \citet{Groh19}, \citet{Murphy21a, Murphy21b} and  Martinet et al.\ (in prep.). The tabulated values are available as machine readable table from ARAA or directly from the authors at \href{https://heibox.uni-heidelberg.de/f/6b5b3fcbd3974fb98d50/}{heibox.uni-heidelberg.de/f/6b5b3fcbd3974fb98d50/}. 
}
\label{fig:PopIII-photon-rates}
\end{figure}

We use these stellar evolution calculations to compute the spectral energy distribution at each point in time. For simplicity, we assume pure blackbody radiation and neglect individual line emission and absorption features (see \citealt{Schaerer2002} or \citealt{GesseyJones22} for a critical assessment of this approach). We then integrate or take the average of the corresponding radiative flux over the entire MS and post-MS evolutionary phase. This allows us to ask how many ionizing and non-ionizing photons Pop III stars of different masses produce over their lifetime. Specifically, we are interested in photons in the Lyman-Werner bands (with energies in the range $11.2\,{\rm eV} \le h\nu < 13.6\,{\rm eV}$), as these are important for the dissociation of H$_2$ (Sections \ref{sec:small-pdiss} and \ref{sec:LW}), and in  photons that can ionize hydrogen (H: $h\nu \ge 13.6\,{\rm eV}$), neutral helium (HeI: $h\nu \ge 24.6\,{\rm eV}$), and singly ionized helium (HeII: $h\nu \ge 54.4\,{\rm eV}$). The results are presented in Figure \ref{fig:PopIII-photon-rates}, which not only considers the zero and solar metallicity case, but also values in between, roughly corresponding to what we expect for the Large and Small Magellanic Clouds ($Z=0.42\,Z_\odot$ and  $Z=0.14\,Z_\odot$, respectively) and for low-mass low-metallicity dwarf galaxies, such as I~Zwicky~18 ($Z=0.03\,Z_\odot$).  

We consider models without and with stellar rotation at the 40\% break-up level. The non-rotating models tend to give slightly larger fluxes at all frequencies, but this is only noticeable in the low-mass regime. At the high-mass end, the difference between non-rotating and rotating stellar evolution models is small \citep[see also][]{Maeder2012}. In the top row we provide the photon flux averaged over the stellar lifetime, in the middle row we present the total number of photons produced by each star during the MS and post-MS evolution, and in the bottom row the corresponding total number of photons per stellar baryon. The data entering Figure \ref{fig:PopIII-photon-rates} are made available for further use in Table \ref{tab:stellar-data}, which we provide for all metallicities and both levels of stellar rotation in electronic form.\footnote{The full dataset is available from ARAA or from the authors (at \href{https://heibox.uni-heidelberg.de/f/6b5b3fcbd3974fb98d50/}{heibox.uni-heidelberg.de/f/6b5b3fcbd3974fb98d50/}) as a space-separated ASCII table, which can be read with the python {\tt astropy.io.ascii.read} command.  } Here we only reproduce the entries for zero metallicity and non-rotating stellar models. Note that the lifetime-averaged data provided in the Table are close to the zero-age main sequence (ZAMS) values. This has two reasons. First, the changes of mass, luminosity, effective surface temperature, and radius are small as the star evolves through the MS. Second, the time the star spends on the post-MS (during which these parameter can vary by orders of magnitude, as illustrated in Figure \ref{fig:stellar-evolution-2-5-20-120-Msun}) is comparatively short, and so this evolutionary phase does not contribute much to the overall photon budget. This is especially true for the high frequency range relevant for ionization, as the effective stellar temperature drops significantly on the post-MS and with it the corresponding UV flux in the blackbody approximation adopted here.

We see that for high-mass stars above $\sim 10\,$M$_\odot$ the average fluxes in the LW band and for H ionization at different metallicities are comparable to each other within a factor of order unity. This may seem surprising at first glance, given the differences in effective surface temperatures (see Figures \ref{fig:PopIII-HRD} and \ref{fig:stellar-evolution-2-5-20-120-Msun}), but can be explained by subtle differences in stellar lifetime and post-MS evolution largely off-setting each other. When considering the hard UV emission necessary for helium ionization, the impact of the much higher effective surface temperatures of Pop III stars becomes clearly noticeable. These stars produce an order of magnitude more photons than their low-metallicity or solar metallicity counterparts, thus opening up a potential observational window to discriminate between the contributions of Pop III and Pop II stars to early reionization by looking for differences in the volume-averaged rate of H and He ionization in the early Universe (Section \ref{sec:21cm-and-CMB}).

\renewcommand{\thetable}{A1:} 
\begin{landscape}
\begin{table}[h]
\tabcolsep4.5pt
\caption{Key stellar properties and production rates of ionizing and non-ionizing photons as function of stellar mass and metallicity}
\label{tab:stellar-data}
\begin{center}
\begin{tabular}{@{\hskip 1pt} c @{\hskip 3pt} c  @{\hskip 3pt} r r r r r r r r r r r r r r r r r r   @{\hskip 3pt} r   @{\hskip 3pt} r @{\hskip 1pt}}
\hline
$Z$ & rot & $M$ &  age & $T_*$ & $L_*$ & $R_*$ & $Q_\mathrm{tot}$ & $Q_\mathrm{LW}$ & $  Q_\mathrm{H}$ & $Q_\mathrm{HeI}$ & $Q_\mathrm{HeII}$ & $N_\mathrm{tot}$ & $N_{\mathrm{LW}}$ & $N_{\mathrm{H}}$ & $N_{\mathrm{HeI}}$ & $N_{\mathrm{HeII}}$ & ${N^b}$ & $N^b_\mathrm{LW}$ & $N^b_\mathrm{H}$ & $N^b_\mathrm{HeI}$ & $N^b_\mathrm{HeII}$ \\
$[Z_{\odot}]$ & [$v_\mathrm{max}$] & [$M_\odot$] & [yr] & [K] & [$L_\odot$] & [$R_\odot$]  & [s$^{-1}$] & [s$^{-1}$] & [s$^{-1}$] & [s$^{-1}$] & [s$^{-1}$] & -- & -- & -- & -- & -- & -- & -- & -- & -- & -- \\[0.1cm]
\hline
0.0 & 0.0 &  1.7  & 9.06 & 4.10 & 1.48 &  1.16 & 46.51 & 43.80 & 43.13 & 39.51 & 29.26 & 63.07 & 60.36 & 59.69 & 56.07 & 45.82 & 5.76 & 3.06 & 2.39 & 0    &    0 \\
0.0 & 0.0 &  2.0  & 8.84 & 4.17 & 1.76 &  1.18 & 46.70 & 44.46 & 43.95 & 40.90 & 32.19 & 63.04 & 60.80 & 60.28 & 57.24 & 48.53 & 5.66 & 3.42 & 2.91 & 0    &    0 \\
0.0 & 0.0 &  2.5  & 8.54 & 4.25 & 2.13 &  1.20 & 46.96 & 45.19 & 44.86 & 42.43 & 35.37 & 63.00 & 61.23 & 60.90 & 58.47 & 51.41 & 5.53 & 3.76 & 3.43 & 1.0  &    0 \\
0.0 & 0.0 &  3.0  & 8.33 & 4.32 & 2.41 &  1.23 & 47.17 & 45.65 & 45.46 & 43.40 & 37.36 & 62.99 & 61.48 & 61.28 & 59.23 & 53.19 & 5.44 & 3.93 & 3.73 & 1.68 &    0 \\
0.0 & 0.0 &  4.0  & 8.02 & 4.42 & 2.84 &  1.29 & 47.49 & 46.24 & 46.22 & 44.61 & 39.78 & 63.00 & 61.76 & 61.73 & 60.13 & 55.29 & 5.33 & 4.08 & 4.06 & 2.45 &    0 \\
0.0 & 0.0 &  5.0  & 7.80 & 4.48 & 3.15 &  1.34 & 47.71 & 46.60 & 46.69 & 45.35 & 41.21 & 63.01 & 61.90 & 61.99 & 60.65 & 56.51 & 5.24 & 4.13 & 4.22 & 2.87 &    0 \\
0.0 & 0.0 &  7.0  & 7.50 & 4.58 & 3.59 &  1.45 & 48.04 & 47.05 & 47.29 & 46.25 & 42.94 & 63.04 & 62.04 & 62.29 & 61.25 & 57.94 & 5.12 & 4.12 & 4.37 & 3.33 & 0.02 \\
0.0 & 0.0 &  9.0  & 7.30 & 4.64 & 3.90 &  1.58 & 48.29 & 47.33 & 47.67 & 46.80 & 43.98 & 63.09 & 62.13 & 62.47 & 61.60 & 58.78 & 5.06 & 4.10 & 4.44 & 3.57 & 0.75 \\
0.0 & 0.0 & 12.0  & 7.29 & 4.69 & 4.24 &  1.84 & 48.53 & 47.58 & 48.01 & 47.27 & 44.83 & 63.31 & 62.37 & 62.80 & 62.06 & 59.62 & 5.16 & 4.21 & 4.65 & 3.90 & 1.46 \\
0.0 & 0.0 & 15.0  & 7.15 & 4.73 & 4.48 &  2.03 & 48.79 & 47.84 & 48.34 & 47.68 & 45.49 & 63.43 & 62.49 & 62.98 & 62.32 & 60.14 & 5.18 & 4.23 & 4.73 & 4.07 & 1.89 \\
0.0 & 0.0 & 20.0  & 7.01 & 4.77 & 4.81 &  2.43 & 49.08 & 48.12 & 48.68 & 48.10 & 46.18 & 63.59 & 62.63 & 63.19 & 62.61 & 60.69 & 5.21 & 4.26 & 4.82 & 4.24 & 2.32 \\
0.0 & 0.0 & 30.0  & 6.82 & 4.82 & 5.23 &  3.22 & 49.46 & 48.49 & 49.12 & 48.61 & 46.95 & 63.78 & 62.81 & 63.44 & 62.93 & 61.27 & 5.23 & 4.26 & 4.88 & 4.38 & 2.72 \\
0.0 & 0.0 & 40.0  & 6.72 & 4.84 & 5.50 &  3.96 & 49.71 & 48.73 & 49.39 & 48.92 & 47.39 & 63.92 & 62.94 & 63.60 & 63.14 & 61.60 & 5.25 & 4.26 & 4.93 & 4.46 & 2.93 \\
0.0 & 0.0 & 60.0  & 6.61 & 4.86 & 5.85 &  5.26 & 50.03 & 49.03 & 49.72 & 49.29 & 47.90 & 64.13 & 63.13 & 63.82 & 63.40 & 62.01 & 5.28 & 4.28 & 4.97 & 4.54 & 3.15 \\
0.0 & 0.0 & 85.0  & 6.53 & 4.87 & 6.10 &  6.77 & 50.29 & 49.26 & 49.97 & 49.56 & 48.26 & 64.32 & 63.29 & 64.00 & 63.59 & 62.30 & 5.31 & 4.28 & 4.99 & 4.59 & 3.29 \\
0.0 & 0.0 & 120.0 & 6.48 & 4.88 & 6.34 &  8.55 & 50.51 & 49.47 & 50.19 & 49.80 & 48.57 & 64.50 & 63.45 & 64.17 & 63.79 & 62.55 & 5.34 & 4.29 & 5.02 & 4.63 & 3.40 \\
0.0 & 0.0 & 180.0 & 6.49 & 4.92 & 6.57 &  9.44 & 50.66 & 49.62 & 50.41 & 50.06 & 48.91 & 64.65 & 63.61 & 64.40 & 64.05 & 62.90 & 5.32 & 4.28 & 5.07 & 4.72 & 3.57 \\
0.0 & 0.0 & 250.0 & 6.54 & 4.92 & 6.76 & 11.37 & 50.83 & 49.76 & 50.56 & 50.23 & 49.13 & 64.87 & 63.81 & 64.60 & 64.27 & 63.17 & 5.40 & 4.33 & 5.13 & 4.80 & 3.70 \\
0.0 & 0.0 & 300.0 & 6.53 & 4.92 & 6.86 & 13.06 & 50.95 & 49.86 & 50.66 & 50.33 & 49.24 & 64.98 & 63.89 & 64.69 & 64.36 & 63.28 & 5.43 & 4.34 & 5.14 & 4.81 & 3.72 \\
0.0 & 0.0 & 500.0 & 6.38 & 4.90 & 7.14 & 19.61 & 51.29 & 50.17 & 50.95 & 50.62 & 49.54 & 65.17 & 64.04 & 64.82 & 64.49 & 63.42 & 5.39 & 4.27 & 5.05 & 4.72 & 3.64 \\
\hline
\end{tabular}
\end{center}
\begin{tabnote}
{\em General comment:} Note that the part of the table depicted here only covers models of non-rotating zero-metallicity Pop III stars. The complete table is available electronically in space-separated ASCII format from ARAA or from the authors at  \href{https://heibox.uni-heidelberg.de/f/6b5b3fcbd3974fb98d50/}{heibox.uni-heidelberg.de/f/6b5b3fcbd3974fb98d50/}. It can be easily read, for example, using the {\tt astropy.io.ascii.read} command. \\[0.3cm]
{\em Column description:} $Z$ = metallicity in solar units (here only Pop III stars) $\blacksquare$ rot $=$ rotational velocity in units of the break-up velocity $v_\mathrm{max}$ (here only non-rotating stellar models) $\blacksquare$ $M$ = stellar mass in solar units $\blacksquare$ age = stellar lifetime in years $\blacksquare$  $T_*$ = decadic logarithm of the lifetime averaged effective temperature in Kelvin $\blacksquare$ $L_*$ = decadic logarithm of the lifetime averaged bolometric luminosity in solar units $\blacksquare$ $R_*$ = lifetime averaged stellar radius in solar units $\blacksquare$ $Q_\mathrm{tot}$ = decadic logarithm of the lifetime averaged total photon flux per second $\blacksquare$ $Q_\mathrm{LW}$ =  decadic logarithm of the lifetime averaged flux of photons in the Lyman-Werner band (with energies in the range $11.2 {\rm eV} \le h\nu < 13.6\,{\rm eV}$) per second $\blacksquare$ $Q_\mathrm{H}$ =  decadic logarithm of the lifetime averaged flux of hydrogen ionizing photons (with energies  $h\nu \ge 13.6\,{\rm eV}$) per second $\blacksquare$ $Q_\mathrm{HeI}$ =  decadic logarithm of the lifetime averaged flux of helium ionizing photons (with energies  $h\nu \ge 24.6\,{\rm eV}$) per second $\blacksquare$ $Q_\mathrm{HeII}$ =  decadic logarithm of the lifetime averaged flux of He$^{+}$ ionizing photons (with energies  $h\nu \ge 54.4\,{\rm eV}$) per second $\blacksquare$ $N_\mathrm{tot}$ =  decadic logarithm of the total number of photons emitted during the lifetime of the star $\blacksquare$ $N_{\mathrm{LW}}$ = decadic logarithm of the total number of LW photons emitted during the lifetime of the star $\blacksquare$ $N_{\mathrm{H}}$ = decadic logarithm of the total number of H ionizing photons emitted during the lifetime of the star $\blacksquare$ $N_{\mathrm{HeI}}$ = decadic logarithm of the total number of He ionizing photons emitted during the lifetime of the star $\blacksquare$ $N_{\mathrm{HeII}}$ = decadic logarithm of the total number of He$^+$ ionizing photons emitted during the lifetime of the star $\blacksquare$ ${N^b}$ = decadic logarithm of the total number of photons emitted during the stellar lifetime per baryon $\blacksquare$  $N^b_\mathrm{LW}$ = decadic logarithm of the total number of LW photons emitted during the stellar lifetime per baryon $\blacksquare$ $N^b_\mathrm{H}$ = decadic logarithm of the total number of H ionizing photons emitted during the stellar lifetime per baryon $\blacksquare$ $N^b_\mathrm{HeI}$ = decadic logarithm of the total number of He ionizing photons emitted during the stellar lifetime per baryon $\blacksquare$ $N^b_\mathrm{HeII}$  = decadic logarithm of the total number of He$^+$ ionizing photons emitted during the stellar lifetime per baryon \\
\end{tabnote}
\end{table}
\end{landscape}

The total number of photons per stellar baryon (bottom row of Figure \ref{fig:PopIII-photon-rates}) is a highly useful quantity when estimating the impact of star formation on the galactic and intergalactic environment in numerical simulations and theoretical models. Combined with information about the stellar initial mass function (IMF) and linked to the cosmic star-formation rate density (see Figure \ref{fig:SFRD}) it gives an indication of the volume-averaged photon flux as function of redshift, which in turn is a prerequisite for estimating the evolution of the cosmic radiation background in different frequency bins. We compute the numbers for a representative population of zero-metallicy stars and compare to solar-metallicity stars in the Milky Way today. For Pop III we assume a logarithmically flat mass spectrum in the range $0.08 \le M/\rm{M}_\odot \le 500$ (see Section \ref{sec:IMF-and-multiplicity}), whereas we adopt a \citep{Kroupa2002} multi-component power-law IMF with $0.08 \le M/\rm{M}_\odot \le 120$ for the solar neighborhood (see Section \ref{sec:transition-PopIII-to-PopII}). The resulting population averaged photon numbers per baryon are listed in Table \ref{tab:pop-avg-photon-numbers} in log units. We find that the more top-heavy IMF of Pop III stars and their larger mass range makes the difference in number of ionizing photons emitted per baryon for the population average even larger than was already noticeable when comparing individual stars. Altogether, we expect over 52000 hydrogen ionizing photons per baryon for the typical stellar population in the early Universe, and the numbers are about 19000 and 1100 for the ionization of He and He$^+$. In contrast, at the present day we have on average fewer than 500 photons per stellar baryon that can ionize H,  30 for He, and none for He$^+$. This enormous discrepancy explains why primordial stars are so efficient in ionizing their environment and why they are key drivers of early cosmic reionization.

\renewcommand{\thetable}{A2:} 
\begin{table}[h]
\tabcolsep7.5pt
\caption{Population averaged number of photons per baryon (in log units)}
\label{tab:pop-avg-photon-numbers}
\begin{center}
\begin{tabular}{@{\hskip 1pt} r  r  r r r r r @{\hskip 1pt}}
\hline
population & IMF & $\langle{N^b}\rangle$ & $\langle{N^b_\mathrm{LW}}\rangle$ & $\langle{N^b_\mathrm{H}}\rangle$ & $\langle{N^b_\mathrm{HeI}}\rangle$ & $\langle{N^b_\mathrm{HeII}}\rangle$\\[0.1cm]
\hline
Pop~III & log-flat & 5.31 &   4.11~ &  4.72~ &  4.29~~ &   3.05~~ \\
Pop~I  & Kroupa &    6.11 &   2.67~ &  2.69~ &  1.50~~ &      0~~ \\
\hline
\end{tabular}
\end{center}
\begin{tabnote}
{\em Column description:}  population = metal-free Pop III or solar-metallicity Pop I stars $\blacksquare$ IMF = stellar initial mass function, either logarithmically flat or \citep{Kroupa2002} multi-component power-law model $\blacksquare$
$\langle{N^b}\rangle$ = decadic logarithm of the population averaged total number of photons emitted per stellar baryon $\blacksquare$  $\langle{N^b_\mathrm{LW}}\rangle$ = decadic logarithm of the population averaged number of LW photons emitted per stellar baryon $\blacksquare$ $\langle{N^b_\mathrm{H}}\rangle$ = decadic logarithm of the population averaged number of H ionizing photons emitted per stellar baryon $\blacksquare$ $\langle{N^b_\mathrm{HeI}}\rangle$ = decadic logarithm of the population averaged number of He ionizing photons emitted per stellar baryon $\blacksquare$ $\langle{N^b_\mathrm{HeII}}\rangle$  = decadic logarithm of the population averaged number of He$^+$ ionizing photons emitted per stellar baryon \\
\end{tabnote}
\end{table}

 \pagebreak

\section*{}
\addcontentsline{toc}{section}{Bibliography}
 
\bibliographystyle{ar-style2}

\pagebreak

\end{document}